# Operator Scaling via Geodesically Convex Optimization, Invariant Theory and Polynomial Identity Testing

Zeyuan Allen-Zhu* Ankit Garg† Yuanzhi Li‡ Rafael Oliveira§ Avi Wigderson¶

November 3, 2017
(version 1)‖

**Abstract**

We propose a new second-order method for geodesically convex optimization on the natural hyperbolic metric over positive definite matrices. We apply it to solve the *operator scaling* problem in time polynomial in the input size and *logarithmic* in the error. This is an exponential improvement over previous algorithms which were analyzed in the usual Euclidean, "commutative" metric (for which the above problem is not convex). Our method is general and applicable to other settings.

As a consequence, we solve the *equivalence* problem for the left-right group action underlying the operator scaling problem. This yields a *deterministic* polynomial-time algorithm for a new class of Polynomial Identity Testing (PIT) problems, which was the original motivation for studying operator scaling.

*Microsoft Research Redmond, email: zeyuan@csail.mit.edu.
†Microsoft Research New England, email: garga@microsoft.com.
‡Princeton University, email: yuanzhil@cs.princeton.edu.
§University of Toronto, email: rafael@cs.toronto.edu.
¶Institute for Advanced Study, Princeton, email: avi@math.ias.edu.

# 1 Introduction

Group orbits and their closures capture natural notions of *equivalence* and are studied in several fields of mathematics like group theory, invariant theory and algebraic geometry. They also come up naturally in theoretical computer science. For example, graph isomorphism, the $\mathcal{VP}$ vs $\mathcal{VNP}$ question and lower bounds on tensor rank are all questions about such notions of equivalence.

In this paper, we focus on the *orbit-closure intersection* problem, which is the most natural way to define equivalence for continuous group actions. We explore a general approach to the problem via geodesically convex optimization. As a testbed for our techniques, we design a deterministic polynomial-time algorithm for the orbit-closure intersection problem for the left-right group action. Recent results by [24, 49] have reduced this problem to polynomial identity testing (PIT), which yields a randomized polynomial-time algorithm. We derandomize this special case of PIT, perhaps surprisingly, by continuous optimization.

On the optimization side, we propose a new second-order method for geodesically convex optimization and use it to get an algorithm for *operator scaling* with time polynomial in the input bit size and *poly-logarithmic* in $1/\varepsilon$ ($\varepsilon$ is the error). In contrast, prior work [34] gives an operator-scaling algorithm that runs in time only *polynomial* in $1/\varepsilon$, which is not sufficient for an application to the general orbit-closure intersection problem.

On the PIT side, we have continued the line of research initiated by Mulmuley [63] to the study of problems in algebraic geometry and invariant theory from an algorithmic perspective in order to develop and sharpen tools to attack the PIT problem. Our result adds to the growing list of this agenda [32, 34, 48, 49], and continues the paper [34] in building optimization tools for PIT problems (at least for those arising from invariant theory). Could it be possible that the eventual solution to PIT will lie in optimization (perhaps very wishful thinking)?

Below is an outline of the rest of the introduction. We review geodesically convex optimization and explain its application to the operator-scaling problem in Section 1.1. We discuss the basics of invariant theory in Section 1.2 and an optimization approach to invariant-theoretic problems in Section 1.2.1. We discuss the left-right group action in detail and explain how the orbit-closure intersection problem for this action (for which we give the first deterministic poly-time algorithm) is a special case of PIT in Section 1.3. In Section 1.4, we discuss the unitary equivalence problem for the left-right action. Section 2 contains an overview of the techniques we develop.

## 1.1 Geodesically convex optimization and operator scaling

Convex optimization provides the basis for efficient algorithms in a large number of scientific disciplines. For instance, to find an $\varepsilon$-approximate minimizer, the interior point method runs in time polynomial in the input and *logarithmic* in $1/\varepsilon$. Unfortunately, many problems (especially machine-learning ones) cannot be phrased in terms of convex formulations. A body of general-purpose non-convex algorithms have been recently designed with theoretical guarantees (see [3–6, 16, 66, 69]). However, their guarantees are not as good as in the convex case: they only converge to approximate local minima (some only to stationary points) and run in time *polynomial* in $1/\varepsilon$.

So one might wonder, for what generalizations of convex optimization problems, can one design optimization algorithms with guarantees comparable to convex optimization? One avenue for such a generalization is given by *geodesically convex* problems. Geodesic convexity generalizes Euclidean convexity to Riemannian manifolds [15, 38]. While there have been works on developing algorithms for optimizing geodesically convex functions [1, 67, 76, 81, 83, 84], the theory is still incomplete in terms of what is the best computational complexity.

We focus on geodesically convex optimization over the space of positive definite (PD) matrices endowed with a different geometry than the Euclidean one. This specific geometry on PD matrices is well studied, see [42, 62, 74, 81].

An $n \times n$ (complex) matrix $M$ is positive definite (PD) if it is Hermitian (i.e., $M = M^\dagger$) and all of its eigenvalues are strictly positive. We write $M \succ 0$ to denote that $M$ is PD. The *geodesic path* from any matrix $A \succ 0$ to matrix $B \succ 0$ is a function $\gamma$ that maps $[0, 1]$ to PD matrices, satisfies $\gamma(0) = A$ and $\gamma(1) = B$, and is locally distance minimizing (w.r.t. an appropriate metric). A function $F(M)$ is *geodesically convex* iff the univariate function $F(\gamma(t))$ is convex in $t$ for any PD matrices $A$ and $B$.

In the Euclidean metric, shortest paths are straight lines and such a path is $\gamma(t) = (1-t)A+tB$. In this case, geodesic convexity reduces to classical convexity.

In the standard Riemannian metric over PD matrices, the geodesic path becomes $\gamma(t) = A^{1/2}(A^{-1/2}BA^{-1/2})^t A^{1/2}$. It should be noted that there does not seem to be any global change of variables that would turn geodesically convex functions into Euclidean convex ones; the change of variables is local and varies smoothly over the manifold.

**Operator Scaling.** An example of an optimization problem which is geodesically convex but not convex arises in the problem of *operator scaling* [34, 39]. A tuple of matrices $(A_1, \ldots, A_m)$ defines a positive *operator*[1] $T(X) = \sum_i A_i X A_i^\dagger$, mapping PSD matrices to PSD matrices. The so-called *capacity* of operator $T$ is defined by:

$$\mathsf{cap}(T) \stackrel{\text{def}}{=} \inf_{X \succ 0, \det(X)=1} \det(T(X)) \ .$$

The name "operator scaling" comes from the fact that if the infimum $X^*$ is attainable, then by defining $Y^* = T(X^*)^{-1}$ and re-scaling $\widetilde{A}_i = (Y^*)^{1/2} A_i (X^*)^{1/2}$, we have

$$\textstyle\sum_i \widetilde{A}_i \widetilde{A}_i^\dagger = I \quad \text{and} \quad \sum_i \widetilde{A}_i^\dagger \widetilde{A}_i = I \ .$$

This is also known as saying that the new operator $\widetilde{T}(X) \stackrel{\text{def}}{=} \sum_{i=1}^m \widetilde{A}_i X \widetilde{A}_i^\dagger$ is *doubly stochastic.*

Before our work, the only known algorithmic approach to solve the above capacity optimization problem was by Gurvits [39] in 2004. His algorithm is a natural extension of Sinkhorn's algorithm, which was proposed in 1964 [73] for the simpler task of *matrix scaling.* A complete analysis of Gurvits' algorithm was done in Garg et al. [34]. Unfortunately, Gurvits' algorithm (and Sinkhorn's algorithm too) run in time $\mathsf{poly}(n, \log M, 1/\varepsilon)$, where $M$ denotes the largest magnitude of an entry of $A_i$,[2] and $\varepsilon$ is the desired accuracy. The polynomial dependency on $1/\varepsilon$ is poor and slows down the downstream applications (such as orbit-closure intersection).

**Remark 1.1.** *A special case of operator scaling is the* matrix scaling *problem (cf. [7, 19] and references therein). In matrix scaling, we are given a real matrix with non-negative entries, and asked to re-scale its rows and columns to make it doubly stochastic. In this very special case, one* can *make a change of variables in the appropriate capacity, and make it convex in the Euclidean metric. This affords standard convex optimization techniques, and for this special case, algorithms running in time* $\mathsf{poly}(n, \log M, \log 1/\varepsilon)$ *are known [7, 19, 52, 60].*

It is known that for every positive operator $T$, $\log(\det(T(X)))$ is geodesically convex in $X$ [74]. Also, it is simple to verify that $\log(\det(X))$ is geodesically linear (i.e., both convex and concave)[3]. Hence, if we define the following alternative objective (removing the hard constraint on $\det(X)$)

$$\mathsf{logcap}(X) = \log\det\left(T(X)\right) - \log\det X \tag{1.1}$$

[1] It is also known as a *completely positive* operator.

[2] One can assume $A_i$'s are integral or complex integral without loss of generality.

[3] This should be contrasted with the fact that $\log(\det(X))$ is a concave function in the Euclidean geometry.

then it is geodesically convex over PD matrices $X$. Note that if $\mathsf{cap}(T) > 0$, then $\inf_{X \succ 0} \mathsf{logcap}(X) = \log(\mathsf{cap}(T))$.

Our main result is an algorithm which $\varepsilon$-approximates capacity and runs in time polynomial in $n, m, \log M$ and $\log(1/\varepsilon)$.

**Theorem M1** (informal)**.** *For every $\varepsilon > 0$, there is a deterministic* $\mathsf{poly}(n, m, \log M, \log(1/\varepsilon))$*-time algorithm that finds $X_\varepsilon \succ 0$ satisfying* $\mathsf{logcap}(X_\varepsilon) - \log(\mathsf{cap}(T)) \le \varepsilon$*.*

Because the problem is non-convex, geodesic convexity plays an important role in getting such an algorithm with a polynomial dependency on $\log(1/\varepsilon)$. Our algorithm is a geodesic generalization of the "box-constrained Newton's method" recently introduced in two independent works [7, 19]. In each iteration, our algorithm expands the objective into its second-order Taylor expansion (up to a *geodesic diameter* 1/2), and then solves it via Euclidean convex optimization.

Although we consider a specific application to operator scaling, our algorithm is in fact a general second-order method and applies to any geodesically convex problem (over PD matrices) that satisfies a particular robustness property. This robustness property is much weaker than self-concordance, and was introduced in the Euclidean space by [7, 19]. We believe that our method applies in a similar way to other metrics, and thus may be of much more general applicability.

In contrast, some previous results (e.g. [83, 84]) only analyze *first-order* methods for geodesically convex functions, and thus cannot achieve polynomial dependency on $\log(1/\varepsilon)$ for operator scaling[4]. We hope that more methods from the Euclidean setting would be transported into the geodesic settings and find applications in invariant theory, machine learning, or more broadly in the future.

## 1.2 Invariant theory, orbits and orbit-closures

We start with a short introduction to the basic concepts of invariant theory, focusing on the various notions of *equivalence* under group actions.

Invariant theory [17] is the study of group actions on vector spaces (more generally algebraic varieties) and the functions (usually polynomials) that are left *invariant* under these actions. It is a rich mathematical field in which computational methods are sought and well developed (see [21, 75]). While significant advances have been made in computational problems involving invariant theory, most algorithms still require exponential time (or longer).

Let $G$ be a group which acts *linearly*[5] on a vector space $V$. (In other words, $V$ is a *representation* of $G$.) Invariant theory is nicest when the underlying field is $\mathbb{C}$ and the group $G$ is either finite, the general linear group $\mathsf{GL}_n(\mathbb{C})$, the special linear group $\mathsf{SL}_n(\mathbb{C})$, or a direct product of these groups. Throughout this paper, whenever we say group, we refer to one of these groups because they are general enough to capture most interesting aspects of the theory.

**Invariant Polynomials.** Invariant polynomials are polynomial functions on $V$ left invariant by the action of $G$. Two simplest examples are

- The symmetric group $G = \mathcal{S}_n$ acts on $V = \mathbb{C}^n$ by permuting the coordinates. In this case, the invariant polynomials are *symmetric* polynomials, and are generated by the $n$ elementary symmetric polynomials.
- Group $G = \mathsf{SL}_n(\mathbb{C}) \times \mathsf{SL}_n(\mathbb{C})$ acts on $V = \mathsf{Mat}_n(\mathbb{C}) = \mathbb{C}^{n \times n}$ by a change of bases of the rows and columns, namely left-right multiplication: that is, $(A, B)$ maps $X$ to $AXB^\dagger$. Here, $\det(X)$ is an invariant polynomial and in fact every invariant polynomial must be a univariate polynomial in $\det(X)$.

[4] Capacity is not strongly geodesically convex.

[5] That is $g \cdot (v_1 + v_2) = g \cdot v_1 + g \cdot v_2$ and $g \cdot (cv) = cg \cdot v$.

The above phenomenon that the ring of invariant of polynomials (denoted by $\mathbb{C}[V]^G$) is generated by a finite number of invariant polynomials is not a coincidence. The *finite generation theorem* due to Hilbert [43, 44] states that, for a large class of groups (including the groups mentioned above), the invariant ring must be finitely generated. These two papers of Hilbert are highly influential and laid the foundations of commutative algebra. In particular, "finite basis theorem" and "Nullstellansatz" were proved as "lemmas" on the way towards proving the finite generation theorem!

**Orbits and Orbit-Closures.** The *orbit* of a vector $v \in V$, denoted by $\mathcal{O}_v$, is the set of all vectors obtained by the action of $G$ on $v$. The *orbit-closure* of $v$, denoted by $\overline{\mathcal{O}}_v$, is the closure (under the Euclidean topology[6]) of the orbit $\mathcal{O}_v$. For actions of continuous groups (like $\mathsf{GL}_n(\mathbb{C})$), it is more natural to look at orbit-closures. Call points in the same orbit (or orbit-closure in the continuous setting) *equivalent* under the action of the group. Many fundamental problems in theoretical computer science (and many more across mathematics) can be phrased as questions about such equivalence. Here are some familiar examples:

- Graph isomorphism problem can be phrased as checking if the orbits of two graphs are the same or not, under the action of the symmetric group permuting the vertices.
- Geometric complexity theory (GCT) [64] formulates a variant of $\mathcal{VP}$ vs. $\mathcal{VNP}$ question as checking if the (padded) permanent lies in the orbit-closure of the determinant (of an appropriate size), under the action of the general linear group on polynomials induced by its natural linear action on the variables.
- Border rank (a variant of tensor rank) of a 3-tensor can be formulated as the minimum dimension such that the (padded) tensor lies in the orbit-closure of the unit tensor, under the natural action of $\mathsf{GL}_r(\mathbb{C}) \times \mathsf{GL}_r(\mathbb{C}) \times \mathsf{GL}_r(\mathbb{C})$. In particular, this captures the complexity of matrix multiplication.

**Orbit-closure Intersection.** We study the orbit-closure intersection problem. Given two vectors $v_1, v_2 \in V$, we want to decide whether $\overline{\mathcal{O}}_{v_1} \cap \overline{\mathcal{O}}_{v_2} \neq \varnothing$. By definition, invariant polynomials are constant on the orbits (and thus on orbit-closures as well). Thus, if $\overline{\mathcal{O}}_{v_1} \cap \overline{\mathcal{O}}_{v_2} \neq \varnothing$, then $p(v_1) = p(v_2)$ for all invariant polynomials $p \in \mathbb{C}[V]^G$. A remarkable theorem due to Mumford says that the converse is also true (for a large class of groups including the ones we discussed above).

**Theorem 1.2** ([65]). *Fix an action of a group $G$ on a vector space $V$. Given two vectors $v_1, v_2 \in V$, we have $\overline{\mathcal{O}}_{v_1} \cap \overline{\mathcal{O}}_{v_2} \neq \varnothing$ if and only if $p(v_1) = p(v_2)$ for all $p \in \mathbb{C}[V]^G$.*

The above theorem gives us a way to test if two orbit-closures intersect. However, in most cases, efficient constructions of invariant polynomials (in the sense of succinct descriptions of Mulmuley [63], see also [32]) are not available. In cases where they are available (as we will see is the case for left-right action in Section 1.3), the orbit-closure intersection problem reduces to *polynomial identity testing* that can be solved by randomized poly-time algorithms.

Our optimization approach (see Section 1.2.1) yields *deterministic* poly-time algorithms, and we believe it should work even in settings where efficient constructions of invariants are not available. We describe a general approach next before describing a concrete application for the left-right action in Section 1.3.

[6] It turns out mathematically, it is more natural to look at closure under the Zariski topology. However, for the group actions we study, the Euclidean and Zariski closures match by a theorem due to Mumford [65].

### 1.2.1 Optimization approach to invariant-theoretic problems

We review an optimization approach to invariant-theoretic problems that comes out of the classical works in geometric invariant theory [55, 65]. We start with the *null-cone membership* problem, which is well defined for any group action. A vector $v \in V$ is said to be in the *null cone* if the orbit-closures of $v$ and 0 intersect. Then the null-cone membership problem is to test if a vector $v$ is in the null cone. This is a special case of the orbit-closure intersection problem.

Given a vector $v \in V$, consider the optimization problem which finds *a vector of minimum $\ell_2$-norm* in the orbit-closure of $v$:

$$N(v) = \inf_{g \in G} \|g \cdot v\|_2^2 \tag{1.2}$$

It is easy to see that $v$ is in the null cone iff $N(v) = 0$. For most group actions (think of $G = \mathsf{GL}_n(\mathbb{C})$ for concreteness), the function $f_v(g) = \|g \cdot v\|_2^2$ is not convex in the Euclidean geometry but is geodescially convex (e.g. see [37, 82]). A consequence of geodesic convexity is the so-called Kempf-Ness theorem [55], that states that any critical point (i.e., point with zero gradient) of $f_v(g)$ must be a global minimum. This brings us to *moment maps.*

**Moment map.** Informally, the moment map $\mu_G(v)$ is the gradient of $f_v(g)$ at $g = id$, the identity element of $G$. The Kempf-Ness theorem draws the following beautiful connection between the moment map and $N(v)$. It is a duality theorem which greatly generalizes linear programming duality to a "non-commutative" setting.

**Theorem 1.3** (Kempf and Ness [55]). *Fix an action of group $G$ on a vector space $V$ and let $v \in V$.*

- $N(v) > 0 \iff \exists$ *non-zero* $w \in \overline{\mathcal{O}}_v$ *s.t.* $\mu_G(w) = 0$.
- *The infimum in $N(v)$ is attainable $\iff$ $\mathcal{O}_v$ is closed $\iff$ $\exists$ non-zero $w \in \mathcal{O}_v$ s.t. $\mu_G(w) = 0$.*
- $N(v) > 0 \implies \exists$ *unique non-zero* $w \in \overline{\mathcal{O}}_v$ *s.t.* $\mu_G(w) = 0$. *Uniqueness is upto the action of a maximal compact subgroup $K$ of $G$.*[7]

Theorem 1.3 gives an optimization route to null-cone membership (which was used in [14, 34]): it suffices to find a $w \in \overline{\mathcal{O}}_v$ satisfying $\mu_G(w) = 0$.[8] Of course, one cannot hope to compute $w$ exactly as it may not have finite bit-size. Instead, one can hope that 'computing it approximately' will suffice, but how accurate do we need to approximate this vector? We will shortly return to this. First let us discuss if this optimization approach be extended to orbit-closure intersection[9]. The extension is provided by the following theorem due to Mumford [65]:

**Theorem 1.4** ([65]). *If $\overline{\mathcal{O}}_{v_1} \cap \overline{\mathcal{O}}_{v_2} \neq \varnothing$, then there is a unique closed orbit in $\overline{\mathcal{O}}_{v_1} \cap \overline{\mathcal{O}}_{v_2}$.*

The above theorem essentially follows from Hilbert's Nullstellansatz and the fact that closed orbits are algebraic varieties[10], and hence separated by a polynomial. Theorems 1.3 and 1.4 imply:

**Corollary 1.5.** *Suppose $N(v_1) > 0$ and $N(v_2) > 0$. If $\overline{\mathcal{O}}_{v_1} \cap \overline{\mathcal{O}}_{v_2} \neq \varnothing$, then there is a unique non-zero $w \in \overline{\mathcal{O}}_{v_1} \cap \overline{\mathcal{O}}_{v_2}$ (upto the action of a maximal compact subgroup $K$) s.t. $\mu_G(w) = 0$.*

---

[7] Maximal compact subgroups of the groups we care about are simple to describe. For $\mathsf{GL}_n(\mathbb{C})$, a maximal compact subgroup is the unitary group $\mathsf{U}_n(\mathbb{C})$. For $\mathsf{SL}_n(\mathbb{C})$, it is the special unitary group $\mathsf{SU}_n(\mathbb{C})$.

[8] This yields a "scaling problem" of the variety of "matrix scaling" and "operator scaling", and leads naturally to alternating minimization heuristics for special classes of groups.

[9] One could also consider an optimization problem which tries to minimize the distance between the two orbit-closures, something like $\inf_{g,h \in G} \|g \cdot v_1 - h \cdot v_2\|_2^2$. It is not clear if this optimization problem has nice properties like geodesic convexity.

[10] For the Zariski closure, this statement follows from the definition but it is true for Euclidean closure as well due to Mumford's theorem that Zariski and Euclidean closures match for the groups we are studying.

**Corollary 1.6.** *In other words, orbit-closure intersection reduces to*

- *computing* $w_1 \in \overline{\mathcal{O}}_{v_1}$ *and* $w_2 \in \overline{\mathcal{O}}_{v_2}$ *satisfying* $\mu_G(w_1) = \mu_G(w_2) = 0$ *and*
- *testing if* $w_1$ *and* $w_2$ *are in the same orbit of the action of the maximal compact subgroup* $K$.

Again one cannot hope to compute $w_1$ and $w_2$ exactly as they may not have finite bit-sizes. Instead, one can hope that 'computing them approximately' will suffice, but how accurate do we need to approximate these vectors?

For null-cone membership, in some cases [14, 34, 35, 39], it suffices to calculate $\varepsilon$-accurate vectors in $\mathsf{poly}(n, m, 1/\varepsilon)$ time. For the orbit-closure intersection, we need a faster $\mathsf{poly}(n, m, \log 1/\varepsilon)$-time algorithm because the distance between two non-intersecting orbit-closures could be exponentially small in $n, m, \log(M)$ (see Section 2.4). This is what our algorithm for capacity minimization (see Theorem M1) achieves. We remark that the optimization problem $N(v)$ for the left-right group action (described next), after elementary transformations, translates directly to the capacity optimization problem. The role of $v$ is played by the tuple of matrices $(A_1, \ldots, A_m)$ which define the completely positive operator $T$.

## 1.3 Left-right group action and polynomial identity testing

In this section, we introduce the left-right group action, describe its invariants, and explain how to reduce its orbit-closure intersection to a special case of polynomial identity testing. Finally, we use our operator-scaling algorithm to derandomize this special case of polynomial identity testing.

Left-right action is a generalization of the basic action we saw in Section 1.2. The group $G = \mathsf{SL}_n(\mathbb{C}) \times \mathsf{SL}_n(\mathbb{C})$ acts simultaneously on a *tuple* of matrices by left-right multiplication. That is $(C, D)$ sends $(Z_1, \ldots, Z_m)$ to $(CZ_1D^\dagger, \ldots, CZ_mD^\dagger)$. The following theorem characterizes the invariants for left-right action.

**Theorem 1.7** ([2, 26, 27, 71]). *The invariants for the left-right action are generated by polynomials of the form* $\det\left(\sum_{i=1}^m E_i \otimes Z_i\right)$, *where* $E_i$ *are* $d \times d$ *complex matrices for an arbitrary* $d$.[11]

In remarkable progress recently, Derksen and Makam [24] proved polynomial bounds on the dimension $d$ that one needs to form a generating set (previous bounds were exponential but held for more general group actions [22]). Formally, they proved

**Theorem 1.8** ([24]). *The invariants for the left-right action are generated by polynomials of the form* $\det\left(\sum_{i=1}^m E_i \otimes Z_i\right)$, *where* $E_i$ *are* $d \times d$ *complex matrices for* $d \leq n^5$.

Using Theorem 1.2, this reduces the orbit-closure intersection problem for the left-right action to the following special case of polynomial identity testing (PIT).

**Corollary 1.9.** *The orbit-closures of the two tuples* $(A_1, \ldots, A_m)$ *and* $(B_1, \ldots, B_m)$ *intersect under the left-right action iff* $\det\left(\sum_{i=1}^m Y_i \otimes A_i\right) \equiv \det\left(\sum_{i=1}^m Y_i \otimes B_i\right)$ *for all* $d \leq n^5$. *Here, the matrices* $Y_i$ *are* $d \times d$ *with disjoint sets of variables.*

**Remark 1.10.** *In the above reduction we have a sequence of PIT problems. We can instead reduce it to a single PIT problem by introducing one additional variable. Although it is beyond the scope of this paper, we conjecture that "for all* $d \leq n^5$*" can be replaced with* $d = n^5$. *A similar phenomenon happened in the special case* $B_1 = \cdots = B_m = 0$ *(i.e., the null-cone problem) [24, 48].*

[11] Here the matrices $Z_i$ have entries which are disjoint formal variables.

Corollary 1.9 implies a randomized poly-time algorithm for the orbit-closure problem for the left-right action (randomly picking the entries of the $Y_i$ using the Schwartz-Zippel lemma). Using our algorithm for capacity minimization in Section 1.1 and the invariant-theory framework in Section 1.2.1, we show

**Theorem M2** (informal). *There is a deterministic polynomial-time algorithm for the orbit-closure intersection problem for the left-right action.*

This generalizes the results in [34, 49] where deterministic poly-time algorithms were designed for the null-cone problem. We refer the readers to those papers for applications of the null-cone problem in non-commutative algebra, analysis, and quantum information theory.

Designing a deterministic algorithm for PIT is a major open problem in complexity theory with applications to circuit lower bounds [51]. There has been extensive work on designing deterministic algorithms for identity testing for restricted computational models (e.g. [28, 31, 54, 57, 70]).[12] However, the above results in PIT (corresponding to the null-cone or orbit-closure intersection problems) give rise to very different class of polynomials for which we can *now* solve PIT in deterministic poly-time. This is part of a bigger agenda proposed by Mulmuley [63] to study PIT problems arising in algebraic geometry and invariant theory.

The other novel aspect of the PIT algorithms in [34] and the current paper is that they are based on continuous optimization whereas the original problems are purely algebraic. It is perhaps not surprising that optimization approaches are now coming back to PIT, since many of the fundamental combinatorial optimization problems like bipartite matching, general matching, linear matroid intersection, and linear matroid parity are special cases of PIT [29, 61].

## 1.4 Side result: unitary equivalence testing for the left-right action

When deriving our algorithm for Theorem M2, we in fact need a subroutine for checking if two given tuples are equivalent under the left-right action: given two tuples of matrices $A = (A_1, \ldots, A_m)$ and $B = (B_1, \cdots, B_m)$, check if there exist unitary matrices $U, V$ such that $UA_iV = B_i$ for all $i$.

Recall there is a deterministic polynomial-time algorithm for this problem (for instance combining [18, Theorem 4] and [45, Proposition 15]). There has been a lot of work characterizing the conditions under which two tuples are equivalent up to unitary transformations [33, 50, 72, 80].

However, in this paper, we need an algorithm for an approximate version of the unitary equivalence problem (recall the discussion in Section 1.2.1). We develop a deterministic polynomial-time algorithm for this purpose, where the time complexity has only poly-logarithmic dependency on the approximation parameter $\varepsilon$.

**Theorem M3** (informal). *There is a deterministic* $\mathsf{poly}(n, m, \log M, \log(1/\varepsilon))$*-time algorithm that, given two tuples $A$ and $B$ and $\varepsilon > 0$, outputs*

- *yes if there are unitary matrices $U, V$ s.t. $UAV$ is $\varepsilon$-close to $B$; or*
- *no if for all unitary matrices $U, V$, $UAV$ is $\varepsilon'$-far to $B$, where $\varepsilon' = \varepsilon^{1/\mathsf{poly}(n,m)} M^{\mathsf{poly}(n,m)}$.*

We believe this algorithm may be of independent interests with possibly other applications.

[12] It is perhaps worth pointing out that, the null cone and orbit-closure intersection problems for the *simultaneous conjugation action* can be done in deterministic time using one such computational model — read-once algebraic branching programs [32, 68]. There are also other instances of PIT that can be solved in deterministic poly-time but which do not correspond to any restricted computational models. These include papers in math studying subspaces of singular matrices [9–11, 30, 36] (after all, by Valiant's completeness theorem for determinant [77], PIT is essentially equivalent to testing if a subspace of matrices contains a non-singular matrix), PIT for subspaces of matrices spanned by rank-1 matrices [39, 46, 47] and algorithms for module isomprhism [13, 18].

### 1.5 Open problems

We design an algorithm for operator scaling with time polynomial in input size and $\log(1/\varepsilon)$, and use it to give a deterministic polynomial-time algorithm for the orbit-closure intersection problem for the left-right action. We believe the recent coming together of optimization and invariant theory (from an algorithmic perspective) is a very exciting development and there are many interesting research directions and open problems in this area. We list some of the most interesting ones (from our perspective).

1. In terms of optimization, it is interesting to design efficient algorithms for other classes of geodesically convex functions, especially with time polynomial in $\log(1/\varepsilon)$, even when the function is not strongly convex. Of particular interest is the manifold on PD matrices that is described in Section 1.1. This will directly lead to polynomial-time algorithms for testing null-cone membership for more general actions, e.g. for the natural action of $\mathsf{SL}_n(\mathbb{C}) \times \mathsf{SL}_n(\mathbb{C}) \times \mathsf{SL}_n(\mathbb{C})$ on tensors in $\mathbb{C}^n \otimes \mathbb{C}^n \otimes \mathbb{C}^n$ (see [14] for some partial results).
2. Design black-box PIT algorithms for testing null-cone membership and orbit-closure intersection for the left-right action, even for characteristic 0. Our algorithm is inherently white box.
3. Design efficient deterministic algorithms for the null cone and orbit-closure intersection problems for actions, of $\mathsf{GL}_n(\mathbb{C})$ for concreteness, only assuming polynomial degree bounds on a generating set. The tools in this paper might already be enough to tackle this general problem.

### Independent work

Independent and concurrent to this work, Derksen and Makam [25] have found a different algorithm for testing orbit-closure intersection for the left-right action. Their algorithm is conceptually simpler than ours, and does not use optimization techniques. Their algorithm works over fields of positive characteristic as well, and may be viewed as extending the null-cone membership algorithm in [49].

### Acknowledgements

We would like to thank Peter Bürgisser, Yash Deshpande, Partha Mukhopadhyay, Suvrit Sra, KV Subrahmanyam and Michael Walter for helpful discussions.

This material is based upon work supported by the National Science Foundation under agreement No. CCF-1412958.

## 2 Techniques and proof overview

- Section 2.1 describes a high level plan for our second-order algorithm for geodesically convex optimization. It gives intuitions for Section 4.
- Section 2.2 contains an overview of our optimization algorithm for operator scaling. It gives intuitions for Section 5.
- Section 2.3 contains a proof overview of our diameter bound for the optimal solutions to capacity optimization. It gives intuitions for Section 6.
- Section 2.4 describes our algorithm for the orbit-closure intersection problem for the left-right action. It gives intuitions for Section 7 and Section 9.
- Section 2.5 describes an algorithm for checking approximate unitary equivalence of two tuple of matrices under the left-right action. It gives intuitions for Section 8.

## 2.1 Geodesically convex optimization

In this section, we provide a high-level overview of our general algorithm for minimizing geodesically convex functions over a natural manifold over PD matrices. The algorithm is a geodesic analogue of the box-constrained Newton's method in [7, 19]. The box-constrained Newton's method is related to trust-region methods (see [20] and the references therein). There has been study of Riemannian/geodesic analogues of these trust-region methods [12]. As far as we know, there was no analysis previously that gave a running time polylogarithmically in the error parameter. While we apply our second-order method to a specific metric, the framework is very general and we believe applicable to many other settings.

We say that a function $F$ over PD matrices is g-convex if for every PD matrix $X$ and every Hermitian matrix $\Delta$, $F\left(X^{1/2}e^{s\Delta}X^{1/2}\right)$ is a convex function in $s$. We also assume a robustness condition on the function $F$ which essentially says that the function behaves like a quadratic function in every "small" neighborhood with respect to the metric.

Our algorithm is quite simple. Starting with $X_0 = I$, we update $X_t$ to $X_{t+1}$ by solving a (constrained) Euclidean convex quadratic minimization problem. Define $f^t(\Delta) = F\left(X_t^{1/2}\exp(\Delta)X_t^{1/2}\right)$. Let $q^t$ be the second-order Taylor expansion of $f^t$ around $\Delta = 0$. We have $q^t$ is a convex and quadratic (in the Euclidean space) because $F$ is $g$-convex. Then, we optimize $q^t(\Delta)$ under the convex constraint $\|\Delta\|_2 \leq 1/2$ (i.e., the "box" constraint). Let $\Delta_t$ be the optimal solution, and we update $X_{t+1} = X_t^{1/2}\exp(\Delta_t)X_t^{1/2}$.[13]

We prove this algorithm finds an $\varepsilon$-approximate minimizer of $F(\cdot)$ in $O(R\log(1/\varepsilon))$ iterations. Here, the diameter parameter $R$ is an upper bound on $\log\left(X_t^{-1/2}X^*X_t^{1/2}\right)$, where $X^*$ is some optimal solution for $F$.

Let us give some intuition for the proof of this. We will prove that in each iteration $F(X_t) - F(X^*)$ decreases by a multiplicative factor of roughly $1-\Omega(1/R)$. Denote by $\Delta^* = \log\left(X_t^{-1/2}X^*X_t^{1/2}\right)$, that is, the "direction" from $X_t$ towards $X^*$. Also let $h(s) = f^t(s\Delta^*)$ and $\Delta'_t = \Delta^*/2R$.

We know $h$ is a univariate convex function due to g-convexity of $F$. Therefore,

$$F(X_t) - F(X^*) = h(0) - h(1) \leq 2R\,(h(0) - h(1/2R)) = 2R\left(F(X_t) - f^t\left(\Delta'_t\right)\right) \quad .$$

On the other hand, since $\|\Delta'_t\|_2 \leq 1/2$, we have that $f^t\left(\Delta'_t\right) \approx g^t\left(\Delta'_t\right)$ by the robustness assumption. Therefore, our obtained solution $\Delta_t$ —which minimizes $q^t(\Delta)$ under the convex constraint $\|\Delta\|_2 \leq 1/2$— will be at least no worse than $\Delta'_t$, or in symbols:

$$f^t\left(\Delta'_t\right) \approx g^t\left(\Delta'_t\right) \geq g^t(\Delta_t) \approx f^t(\Delta_t) = F(X_{t+1})$$

Combining the above two inequalities, we have $F(X_t) - F(X_{t+1}) \geq (1 - \Omega(1/R))(F(X_t) - F(X^*))$.

## 2.2 Operator scaling via geodesically convex optimization

Recall that we are given a positive operator $T(X) = \sum_{i=1}^m A_i X A_i^\dagger$, where matrices $A_i$ are $n \times n$ and whose entries are complex numbers with integer coefficients (Gaussian integers).[14] We want to solve the following optimization problem:

$$\mathsf{cap}(T) = \inf_{X \succ 0 \wedge \det(X)=1} \det(T(X))$$

[13] There is a minor difference in the actual algorithm —where we compute $\exp(\Delta_t/e^2)$ instead of $\exp(\Delta_t)$— but we ignore the subtlety here.

[14] If $A_i$ contains rational entries then one can multiply all matrices by the common denominator.

Before going into our algorithm, let us first explain what is known for a commutative special case of the above optimization problem, which is called *matrix scaling*. There one is given a non-negative $n \times n$ matrix $N$ and one wants to solve the following optimization problem [41]:

$$\mathsf{cap}(N) = \inf_{x>0 \,\wedge\, x_1 x_2 \cdots x_n = 1} \prod_{i=1}^{n} (Nx)_i$$

The above program is an instance of geometric programming, so one can formalize it as a convex function and apply the ellipsoid algorithm to solve it to accuracy $\varepsilon$ in time $\mathsf{poly}(n, b, \log(1/\varepsilon))$, where $b$ denotes the bit size of entries in $N$ [52]. In contrast, our operator scaling problem is not convex, and there is no analogue of ellipsoid algorithm for geodesically convex optimization.

Linial et al. [60] presented an algorithm for matrix scaling which also gives a polylogarithmic time dependency in $1/\varepsilon$. Unfortunately, for its natural extensions to operator scaling, we are aware of counter examples (due to matrix non-commutativity) in which their approach fails to generate similar polylogarithmic efficiency.

We apply Section 2.1 to operator scaling. Recall that

$$\mathsf{logcap}(X) = \log\det\left(\textstyle\sum_i A_i X A_i^\dagger\right) - \log\det X$$

is geodesically convex over PD matrices [58, 74]. Unfortunately, in the language of Section 2.1, the diameter parameter $R$ is not polynomially bounded. In particular, the exact minimizer $X^*$ of $\mathsf{logcap}(X)$ may not even be attainable (so can be at infinity). We fix this issues in two steps.

- First, we show (see Section 2.3) that there is an (approximate) minimizer $X^*_\varepsilon$ of $\mathsf{logcap}(X)$ that has a bounded condition number. That is, $\mathsf{logcap}(X^*_\varepsilon) \leq \inf_{X \succ 0} \mathsf{logcap}(X) + \varepsilon$ and $\kappa(X^*_\varepsilon) \stackrel{\text{def}}{=} \lambda_{\max}(X^*_\varepsilon)/\lambda_{\min}(X^*_\varepsilon) \leq \exp(\mathsf{poly}(n, \log M, \log(1/\varepsilon))$ is bounded.
- Second, we add a regularizer $\mathsf{reg}(X) = \mathrm{Tr}X \cdot \mathrm{Tr}X^{-1}$ (which is also g-convex) to the objective. This ensures that when minimizing $F(X) = \mathsf{logcap}(X) + \mu\, \mathsf{reg}(X)$ for some sufficiently small parameter $\mu > 0$, we always have $\kappa(X) \leq \exp(\mathsf{poly}(n, \log M, \log(1/\varepsilon))$.

Finally, since both $\kappa(X^*_\varepsilon)$ and $\kappa(X)$ are bounded, one can show that the diameter parameter $R = O(\log \kappa(X^*_\varepsilon) + \log \kappa(X))$ is also polynomially bounded. We can now apply Section 2.1 directly.

## 2.3 Bounds on eigenvalues of scaling matrices

We want to bound the condition number of a minimizer $X^*$ of the $\mathsf{logcap}(X)$. Note that the infimum of $\inf_{X \succ 0}\{\mathsf{logcap}(X)\}$ may not be attainable, and in such case we want to bound the condition number of some $X^*_\varepsilon$ that satisfies $\mathsf{logcap}(X^*_\varepsilon) - \inf_{X \succ 0}\{\mathsf{logcap}(X)\} \leq \varepsilon$. Let us call such $X^*_\varepsilon$ being $\varepsilon$-minimizers.

We remark that similar bounds for the simpler matrix-scaling case were derived in Kalantari and Khachiyan [52] (for $X^*$) and in Allen-Zhu et al. [7] (for the more general $X^*_\varepsilon$). Unfortunately, these *combinatorial* proofs do not apply to the operator case due to non-commutativity, even when the infimum is attainable.

We take a completely different approach by considering a symmetric formulation of capacity:[15]

$$\widetilde{\mathsf{cap}}(T) = \inf_{X, Y \succ 0, \det(X) = \det(Y) = 1} \mathrm{Tr}\left[X\, T(Y)\right]$$

Optimal solutions for $\widetilde{\mathsf{cap}}(T)$ have direct correspondence to the optimal solutions for $\mathsf{cap}(T)$. The proof considers running gradient flow on the objective $\mathrm{Tr}\,[X\,T(Y)]$. [16] The main trick is to continuously follow the gradient but *normalized* to norm 1. That is

$$\frac{d}{dt}(X_t, Y_t) = \frac{\mathrm{grad}\ \mathrm{Tr}\,[X_t\,T(Y_t)]}{\|\mathrm{grad}\ \mathrm{Tr}\,[X_t\,T(Y_t)]\|_2}$$

where the gradient has to be defined appropriately. Then, we use several known properties of capacity [34, 59] to prove that the gradient flow converges in polynomial time with a linear convergence rate (i.e., error $\varepsilon \propto e^{-O(t)}$ where $t$ is the time). Also since the gradient has norm 1, informally, the log of the condition number of an $\varepsilon$-minimizer shall be bounded by the amount of time that the gradient flow reaches an $\varepsilon$-minimizer. This yields that there exists an $\varepsilon$-minimizer $X^*_\varepsilon$ (one reached by the continuous gradient flow) that has a bounded condition number, that is, $\kappa(X^*_\varepsilon) \stackrel{\mathrm{def}}{=} \lambda_{\max}(X^*_\varepsilon)/\lambda_{\min}(X^*_\varepsilon) \le e^R$ where $R = \mathsf{poly}(n, \log M, \log(1/\varepsilon))$ and $\log M$ is the bit complexity of entries of the matrices $A_i$ defining the operator $T$.

[15] This $\widetilde{\mathsf{cap}}(T)$ is the same as the minimum $\ell_2$-norm optimization, described in Section 1.2.1, for the left-right action.

Note that this is only an existential proof and one cannot algorithmically find an $\varepsilon$-minimizer using this gradient flow. Indeed, if one discretizes the gradient flow, the resulting algorithm will be a first-order method that converges in a number of iterations *polynomially* in $1/\varepsilon$ as opposed to poly-logarithmically (the objective is not strongly geodesically convex). This is why we have to design a separate algorithm (as explained in the next section) to find an $\varepsilon$-minimizer. Note that our proof strategy only yields that there exists an $\varepsilon$-minimizer that has “small” condition number. But as we will describe in the previous section, this will suffice through the use of an appropriate regularizer.

## 2.4 Orbit-closure intersection for left-right action

In this section, we give an overview of our algorithm for orbit-closure intersection. We are given two tuples $A = (A_1, \ldots, A_m)$ and $B = (B_1, \ldots, B_m)$, which we assume integral for simplicity. They are associated with completely positive operators $T_A$ and $T_B$:

$$T_A(X) = \sum_{i=1}^m A_i X A_i^\dagger \quad \text{and} \quad T_B(X) = \sum_{i=1}^m B_i X B_i^\dagger \ .$$

We can assume wlog that both the tuples are not in the null cone since testing null-cone membership for the left-right action is already solved in [34, 49]. This means we can assume $\mathsf{cap}(T_A) > 0$ and $\mathsf{cap}(T_B) > 0$ (as a consequence of the Kempf-Ness theorem, alternatively see [34]).

Recall from Corollary 1.6 that to test orbit-closure intersection for $A$ and $B$, it suffices to

- find tuples $C = (C_1, \ldots, C_m)$ and $D = (D_1, \ldots, D_m)$, in the orbit-closures of $A$ and $B$ respectively, that have moment map 0. For the left-right action, $C$ (or similarly $D$) has moment map 0 if there exists a scalar $\alpha$ s.t. $\sum_{i=1}^m C_i C_i^\dagger = \sum_{i=1}^m C_i^\dagger C_i = \alpha I_n$.
- test whether $C$ and $D$ are equivalent up to left-right multiplications of unitary matrices: that is, whether there exist special unitary matrices $U, V \in \mathsf{SU}_n(\mathbb{C})$ s.t. $U C_i V = D_i$ for all $i \in [m]$.

As argued in Section 1.2.1, we cannot hope for calculating $C$ or $D$ exactly since they do not even have finite bit length. However, we can run our operator scaling algorithm (on the capacity optimization problem) to find tuples $A' = (A'_1, \ldots, A'_m)$ and $B' = (B'_1, \ldots, B'_m)$, in the orbits of $A$ and $B$ respectively, that are exponentially close to $C$ and $D$ respectively. We describe how to do this (this is a standard argument and is explained in Section 6). Suppose $X_\varepsilon$ is s.t. $\log \mathsf{cap}(X_\varepsilon) \le \log \mathsf{cap}(T_A) + \varepsilon$. Then one defines $A'_i = c\,T_A(X_\varepsilon)^{-1/2} A_i X_\varepsilon^{1/2}$, and similarly for $B'$. Here $c$ is a normalization constant to ensure that we remain in the $\mathsf{SL}_n(\mathbb{C}) \times \mathsf{SL}_n(\mathbb{C})$ orbits. Now, if the orbit-closures of $A$ and $B$ intersect, not only $C$ and $D$ are related by unitary matrices, we also know $A'_i \approx_{\delta_1} UB'_iV$ up to some exponentially small error $\delta_1 > 0$. Note that due to our new operator-scaling algorithm, we can make the running time polylogarithmic in $1/\delta_1$.

[16] This is a special case of Kirwan's gradient flow for general group actions [56, 78], and this particular gradient flow and its properties have also been studied in [59].

We will prove (see below) that if the orbit-closures of $A$ and $B$ do not intersect, then the tuples $UA'V$ and $B'$ must be $\delta_2$ (in $\ell_2$ distance) far apart for every pair of unitary matrices $U, V \in \mathsf{U}_n(\mathbb{C})$ (with $\det(UV) \approx 1$). Here $\delta_2$ is some fixed exponentially small parameter, and we shall choose $\delta_1 \ll \delta_2$. In other words, the orbit-closure intersection problem now reduces to checking if there exist unitary matrices s.t. $UA'V$ is close to $B'$. We provide an efficient algorithm for this problem too, and overview of the techniques will be presented in Section 2.5.

**Distance between non-intersecting orbit-closures.** We now explain, how to prove that $UA'V$ and $B'$ must be $\delta_2$-apart if orbit-closures of $A$ and $B$ do not intersect. By Corollary 1.9, there is an invariant polynomial $p$ of degree at most $n^6$ such that $p(A) \neq p(B)$. We can arrange $p$ to have "small" integer coefficients (using the Schwarz-Zippel lemma). Since $p(A) \neq p(B)$ and $A$ and $B$ have integral entries, $p(A)$ and $p(B)$ are both integer valued and must satisfy $|p(A) - p(B)| \geq 1$. Now, since $UA'V$ and $B'$ lie in the orbits of $A$ and $B$ respectively, we have $p(UA'V) = p(A)$ and $p(B') = p(B)$ (by the definition of invariant polynomials), and hence $|p(UA'V) - p(B')| \geq 1$. Since $p$ has polynomial degree and has small integral coefficients, this implies that $UA'V$ and $B'$ have to far apart by a fixed (exponentially small) value $\delta_2$.

We provide a simple example to show that the orbit-closures can be exponentially close. In this example, $m = 1$, so the tuple has only one matrix. Let $A$ be the $(n \times n)$ diagonal matrix whose entries are all 2. Let $B$ be an arbitrary $(n \times n)$ matrix with entries 1's and 2's s.t. $\det(B) = 2^n + 1$. Since the determinants are different, the orbit-closures of $A$ and $B$ do not intersect. The matrix $2I_n$ lies in the orbit of $A$ and $(2^n + 1)^{1/n} I_n$ lies in the orbit of $B$. The $\ell_2$ distance between these is

$$\sqrt{n}\left((2^n+1)^{1/n} - 2\right) = 2\sqrt{n}\left((1 + 1/2^n)^{1/n} - 2\right) \approx \frac{2}{\sqrt{n}2^n}$$

which is exponentially small in the dimension $n$.

**Comparison of null-cone membership with orbit-closure intersection.** We highlight differences of our result from the work of Garg et al. [34] (which solves a simpler null-cone membership problem). Garg et al. [34] used invariant theory and degree bounds to analyze the convergence of Gurvits' algorithm from [39] [17]. For the simpler null cone problem, it sufficed for them to have an algorithm with inverse polynomial dependence on the approximation parameter.[18]In this paper, we need significant more work (on designing operator-scaling algorithms) to achieve a polylogarithmic time dependency on the error as can be seen from the previous example where non-intersecting orbit-closures can be inverse exponentially close (in terms of the input size).

## 2.5 Algorithm for checking unitary equivalence

In Section 2.4, we have essentially reduced the orbit-closure intersection problem to the following unitary equivalence problem. Given two tuples of matrices $A = (A_1, \ldots, A_m)$ and $B = (B_1, \ldots, B_m)$, decide:[19]

- if there exist unitary matrices $U, V \in \mathsf{U}_n(\mathbb{C})$ s.t. the tuples $UAV$ and $B$ are $\varepsilon$ close; or
- for all unitary matrices $U, V \in \mathsf{U}_n(\mathbb{C})$, the tuples $UAV$ and $B$ are $\varepsilon'$ far apart.

[17] Indeed, the most recent version of their paper does not use any degree bounds

[18] The approximation parameter here refers to the $\mathsf{ds}$ notion in Definition 3.7.

[19] Recall that, testing exact equivalence (i.e., for $\varepsilon = 0$) is a much simpler problem.

Here, $\varepsilon \ll \varepsilon'$ and both are exponentially small in the input-size.

What does the left-right action by unitary matrices preserve?

The (real) singular values of individual matrices $A_i$ and $B_i$ are preserved. Therefore, we look for an $i \in [m]$ s.t. the singular values of $A_i$ form at least two distinct clusters. Since singular values in different clusters must be matched differently, we can reduce problem into smaller dimensions each corresponding to one cluster of singular values. However, what if all singular values for $A_i$ are close to each other? This means each $A_i$ must be close to being (a scaling of) a unitary matrix.

Next, let us assume for simplicity that all matrices $A_i$ and $B_i$ are exactly unitary. Since $UA_1V \approx B_1$ if and only if $V \approx A_1^{-1}U^\dagger B_1$, this restricts the search to just $U$ because $V$ can be explicitly written as a function of $U$. Therefore, the new problem we need to solve is the following: does there exist a unitary $U$ s.t. $UA_iA_1^{-1}U^\dagger \approx B_iB_1^{-1}$ for $i \in \{2, \ldots, m\}$.

What does conjugation by a unitary matrix (i.e., left multiplication by $U$ and right by $U^\dagger$) preserve? The eigenvalues! Therefore, similar to the previous step, we can compute the eigenvalues of our new matrices $A_iA_1^{-1}$ and $B_iB_1^{-1}$ for all $i \in \{2, \ldots, m\}$, and look for clusters of eigenvalues to reduce dimensions. If all the eigenvalues are close to each other for every unitary matrices $A_iA_1^{-1}$ and $B_iB_1^{-1}$, then they must both be close to scalings of the identity matrix so all we are left to do is to compare scalars.

Unfortunately, after reducing the dimensions using eigenvalues, we may come back to matrices with different singular values. Therefore, we need to alternatively apply singular-value and eigenvalue decomposition routines, until we are left with identity matrices. It is in fact tricky, but anyways possible, to ensure that the error does not blow up too much in this decomposition process.

## 3 Preliminaries

Given the action of a group $G$ on a vector space $V$, the orbit of a vector $v$, denoted by $\mathcal{O}_v$, is simply $\mathcal{O}_v = \{g \cdot v : g \in G\}$.

Let $\mathbb{Z}[i] = \{a + bi \mid a, b \in \mathbb{Z}\}$ be the set of Gaussian integers and $\mathbb{Q}[i] = \{a + bi \mid a, b \in \mathbb{Q}\}$. Let $\mathsf{Mat}_n(R)$ denote the space of $n \times n$ matrices whose entries are in $R$. We denote by $\mathsf{GL}_n(\mathbb{C})$ the group of $n \times n$ invertible matrices, $\mathsf{U}_n(\mathbb{C})$ the group of $n \times n$ unitary matrices, $\mathsf{SL}_n(\mathbb{C})$ the group of $n \times n$ invertible matrices with determinant 1, and by $\mathsf{SU}_n(\mathbb{C})$ the group of $n \times n$ unitary matrices with determinant 1.

Throughout the rest of this paper, we use bold fonts $\mathbf{X}$ to denote matrices and $\overrightarrow{\mathbf{X}}$ to denote tuples of matrices. We denote by $\|\mathbf{X}\|_2$, $\|\mathbf{X}\|_F$, and $\|\mathbf{X}\|_\infty$ the spectral, Frobenius, and $\ell_\infty$ norm of matrix $\mathbf{X} \in \mathsf{Mat}_n(\mathbb{C})$. (Recall $\|\mathbf{X}\|_\infty = \max_{i,j}\{|\mathbf{X}_{i,j}|\}$.)

Given Hermitian matrices $\mathbf{A}, \mathbf{B} \in \mathsf{Mat}_n(\mathbb{C})$, we use $\mathbf{A} \succeq \mathbf{B}$ to indicate that $\mathbf{B} - \mathbf{A}$ is positive semidefinite (PSD), and $\mathbf{A} \succ \mathbf{B}$ to indicate that $\mathbf{B} - \mathbf{A}$ is positive definite (PD).

For tuples of matrices $\overrightarrow{\mathbf{X}}$, we denote by $\|\overrightarrow{\mathbf{X}}\|_2$ as the usual $\ell_2$ norm when we regard $\overrightarrow{\mathbf{X}}$ as a vector, that is, $\|\overrightarrow{\mathbf{X}}\|_2^2 \stackrel{\text{def}}{=} \sum_{i=1}^m \|\mathbf{X}_i\|_F^2$.

We denote by $\lambda_{\max}(\mathbf{X})$ and $\lambda_{\min}(\mathbf{X})$ the maximum and minimum eigenvalues of an Hermitian matrix $\mathbf{X}$. We denote by $\kappa(\mathbf{X})$ the condition number of a (possibly non-Hermitian) matrix $\mathbf{X}$, which is the ratio between maximum and minimal singular value.

We will also need the following definitions of distance between tuples of matrices:

**Definition 3.1.** *Given two tuples* $\overrightarrow{\mathbf{A}}, \overrightarrow{\mathbf{B}} \in \mathsf{Mat}_n(\mathbb{C})^m$*, we denote by*

$$\Delta_{SU}(\overrightarrow{\mathbf{A}}, \overrightarrow{\mathbf{B}}) \stackrel{\text{def}}{=} \min_{\mathbf{U},\mathbf{V}\in \mathsf{SU}_n(\mathbb{C})} \left\| \mathbf{U}\overrightarrow{\mathbf{A}}\mathbf{V} - \overrightarrow{\mathbf{B}} \right\|_2 \ ;$$

$$\Delta_{U}(\overrightarrow{\mathbf{A}}, \overrightarrow{\mathbf{B}}) \stackrel{\text{def}}{=} \min_{\mathbf{U},\mathbf{V}\in \mathsf{U}_n(\mathbb{C})} \left\| \mathbf{U}\overrightarrow{\mathbf{A}}\mathbf{V} - \overrightarrow{\mathbf{B}} \right\|_2 \ .$$

*the min* $\ell_2$ *distance between* $\overrightarrow{\mathbf{A}}$ *and* $\overrightarrow{\mathbf{B}}$ *up to unitary transformations.*

## 3.1 Capacity fuction

In this section, we introduce various equivalent forms of capacity.

**Definition 3.2.** *A tuple* $\overrightarrow{\mathbf{A}} = (\mathbf{A}_1, \dots, \mathbf{A}_m) \in \mathsf{Mat}_n(\mathbb{C})^m$ *defines an operator* $T_{\overrightarrow{\mathbf{A}}} : M_n(\mathbb{C}) \to M_n(\mathbb{C})$ *given by*

$$T_{\overrightarrow{\mathbf{A}}}(\mathbf{X}) \stackrel{\text{def}}{=} \textstyle\sum_{i=1}^m \mathbf{A}_i \mathbf{X} \mathbf{A}_i^\dagger$$

**Definition 3.3.** *The* capacity *of* $T_{\overrightarrow{\mathbf{A}}}$ *is given by the following optimization problem:*

$$\mathsf{cap}(T_{\overrightarrow{\mathbf{A}}}) \stackrel{\text{def}}{=} \inf_{\mathbf{X}\succ 0} \frac{\det(T_{\overrightarrow{\mathbf{A}}}(\mathbf{X}))}{\det(\mathbf{X})} = \inf_{\mathbf{X}\succ 0, \det(\mathbf{X})=1} \det(T_{\overrightarrow{\mathbf{A}}}(\mathbf{X})) \ .$$

*Slightly abusing notation, when* $\overrightarrow{\mathbf{A}}$ *is fixed in the context, we also define* $\mathsf{cap}(\mathbf{X})$ *as a function over matrices* $\mathbf{X} \in M_n(\mathbb{C})$*:* $\mathsf{cap}(\mathbf{X}) = \frac{\det(T_{\overrightarrow{\mathbf{A}}}(\mathbf{X}))}{\det(\mathbf{X})}$ *, and it satisfies* $\mathsf{cap}(T_{\overrightarrow{\mathbf{A}}}) = \inf_{\mathbf{X}\succ 0} \mathsf{cap}(\mathbf{X})$*.*

For analysis purpose, we also introduce the symmetrized form of capacity:

**Definition 3.4.** *The* symmetric capacity *of* $T_{\overrightarrow{\mathbf{A}}}$ *is given by*

$$\widetilde{\mathsf{cap}}(T_{\overrightarrow{\mathbf{A}}}) \stackrel{\text{def}}{=} \inf_{\mathbf{X},\mathbf{Y}\succ 0, \det(\mathbf{X})=\det(\mathbf{Y})=1} \mathrm{Tr}[\mathbf{X}\, T_{\overrightarrow{\mathbf{A}}}(\mathbf{Y})] \ .$$

**Definition 3.5.** *The minimum norm of* $\overrightarrow{\mathbf{A}}$ *with respect to left-right actions is (compare to (1.2)):*

$$N(\overrightarrow{\mathbf{A}}) = \inf_{\mathbf{B},\mathbf{C}\in \mathsf{SL}_n(\mathbb{C})} \sum_{i=1}^m \|\mathbf{B}\mathbf{A}_i\mathbf{C}\|_F^2$$

The following proposition relates the previous definitions:

**Proposition 3.6.** $N(\overrightarrow{\mathbf{A}}) = \widetilde{\mathsf{cap}}(T_{\overrightarrow{\mathbf{A}}}) = n\, \mathsf{cap}(T_{\overrightarrow{\mathbf{A}}})^{1/n}$*.*

## 3.2 Operator scaling

We now come to the problem of operator scaling. Before that we need the following definitions.

**Definition 3.7 (doubly stochastic).** *Consider an operator* $T_{\overrightarrow{\mathbf{A}}}$ *defined by* $(\mathbf{A}_1, \dots, \mathbf{A}_m)$*. The "distance" of* $T_{\overrightarrow{\mathbf{A}}}$ *to being doubly stochastic is defined as follows:*

$$\mathsf{ds}(T_{\overrightarrow{\mathbf{A}}}) \stackrel{\text{def}}{=} \mathrm{Tr}\left( (\textstyle\sum_{i=1}^m \mathbf{A}_i\mathbf{A}_i^\dagger - \mathbf{I})^2 \right) + \mathrm{Tr}\left( (\textstyle\sum_{i=1}^m \mathbf{A}_i^\dagger\mathbf{A}_i - \mathbf{I})^2 \right) \ .$$

*We say* $T_{\overrightarrow{\mathbf{A}}}$ *is* doubly-stochastic *if* $\mathsf{ds}(T_{\overrightarrow{\mathbf{A}}}) = 0$*.*

The following is a slight variant of the notion of doubly stochastic.

**Definition 3.8** (**doubly balanced**). *The "distance" of $T_{\overrightarrow{\mathbf{A}}}$ to being doubly balanced [59] is defined as follows:*

$$\mathsf{db}(T_{\overrightarrow{\mathbf{A}}}) \stackrel{\text{def}}{=} \mathrm{Tr}\left(\left(\sum_{i=1}^m \mathbf{A}_i\mathbf{A}_i^\dagger - \frac{\mathrm{Tr}(T_{\overrightarrow{\mathbf{A}}}(\mathbf{I}))}{n}\mathbf{I}\right)^2\right) + \mathrm{Tr}\left(\left(\sum_{i=1}^m \mathbf{A}_i^\dagger\mathbf{A}_i - \frac{\mathrm{Tr}(T_{\overrightarrow{\mathbf{A}}}(\mathbf{I}))}{n}\mathbf{I}\right)^2\right) \ .$$

*We say $T_{\overrightarrow{\mathbf{A}}}$ is* doubly-balanced *if $\mathsf{db}(T_{\overrightarrow{\mathbf{A}}}) = 0$.*

We will also need the following definition of scaling.

**Definition 3.9** (**operator scaling**). *Consider an operator $T_{\overrightarrow{\mathbf{A}}}$ defined by $(\mathbf{A}_1, \dots, \mathbf{A}_m)$. An operator $T'$ is called a* scaling of $T_{\overrightarrow{\mathbf{A}}}$ *if there exist invertible matrices $\mathbf{C}, \mathbf{D} \in \mathsf{GL}_n(\mathbb{C})$ s.t.*

$$T'(\mathbf{X}) = \mathbf{C} \cdot T_{\overrightarrow{\mathbf{A}}}(\mathbf{D}^\dagger\mathbf{X}\mathbf{D}) \cdot \mathbf{C}^\dagger = \sum_{i=1}^m \mathbf{C}\mathbf{A}_i\mathbf{D}^\dagger\mathbf{X}\mathbf{D}\mathbf{A}_i^\dagger\mathbf{C}^\dagger \ .$$

*Equivalently, $(\mathbf{A}_1, \dots, \mathbf{A}_m)$ get sent to $(\mathbf{C}\mathbf{A}_1\mathbf{D}^\dagger, \dots, \mathbf{C}\mathbf{A}_m\mathbf{D}^\dagger)$.*

### 3.3 Orbits for left-right action

We recall the following properties of the left-right action.

**Fact 3.10** (left-right action). *Given element $\overrightarrow{\mathbf{A}} \in \mathsf{Mat}_n(\mathbb{C})^m$, then under the left-right action*

- *The* orbit *$\mathcal{O}_{\overrightarrow{\mathbf{A}}} = \{\overrightarrow{\mathbf{B}} = (\mathbf{X}\mathbf{A}_1\mathbf{Y}, \dots, \mathbf{X}\mathbf{B}_m\mathbf{Y}) \mid \mathbf{X}, \mathbf{Y} \in \mathsf{SL}_n(\mathbb{C})\}$.*
- *$\overrightarrow{\mathbf{C}} = (\mathbf{C}_1, \dots, \mathbf{C}_m) \in \overline{\mathcal{O}}_{\overrightarrow{\mathbf{A}}}$ is an* element of minimum norm *(i.e., with the smallest $N(\overrightarrow{\mathbf{C}})$) iff $T_{\overrightarrow{\mathbf{C}}}$ is doubly-balanced. This follows from the Kempf-Ness theorem (see Theorem 1.3).*
- *If $(\mathbf{B}_1, \dots, \mathbf{B}_m), (\mathbf{C}_1, \dots, \mathbf{C}_m)$ are two elements of minimum norm in $\overline{\mathcal{O}}_{\overrightarrow{\mathbf{A}}}$, then there exist unitary matrices $\mathbf{U}, \mathbf{V} \in \mathsf{SU}_n(\mathbb{C})$ such that $(\mathbf{B}_1, \dots, \mathbf{B}_m) = (\mathbf{U}\mathbf{C}_1\mathbf{V}, \dots, \mathbf{U}\mathbf{C}_m\mathbf{V})$.*

### 3.4 Operator scaling with error

We now discuss various forms of approximation for operator scaling.

**Definition 3.11** ($\varepsilon$-**scaling** 1). *Given an operator $T_{\overrightarrow{\mathbf{A}}}$ defined by $(\mathbf{A}_1, \dots, \mathbf{A}_m)$, find a scaling $T'$ of $T_{\overrightarrow{\mathbf{A}}}$ s.t. $T'(\mathbf{I}) = \mathbf{I}$ and $\mathsf{ds}(T') \leq \varepsilon$.*

**Definition 3.12** ($\varepsilon$-**scaling** 2). *Given an operator $T_{\overrightarrow{\mathbf{A}}}$ defined by $(\mathbf{A}_1, \dots, \mathbf{A}_m)$, find a PSD matrix $\mathbf{X}$ s.t. $\log \mathsf{cap}(\mathbf{X}) \leq \log \mathsf{cap}(T_{\overrightarrow{\mathbf{A}}}) + \varepsilon$.*

**Definition 3.13** ($\varepsilon$-**scaling** 3). *Given a tuple of matrices $\overrightarrow{\mathbf{A}} = (\mathbf{A}_1, \dots, \mathbf{A}_m)$, find a tuple $\overrightarrow{\mathbf{A}}'$ in the orbit of $\overrightarrow{\mathbf{A}}$ (w.r.t. the left-right action) s.t. $\Delta_{SU}(\overrightarrow{\mathbf{A}}', \overrightarrow{\mathbf{A}}^*) \leq \varepsilon$. Here $\overrightarrow{\mathbf{A}}^*$ is a tuple of minimum norm in the orbit-closure of $\overrightarrow{\mathbf{A}}$*

We have the following equivalence between the three notions (for different values of $\varepsilon$).

**Theorem 3.14.** *Let operator $T_{\overrightarrow{\mathbf{A}}}$ be defined by $\overrightarrow{\mathbf{A}} = (\mathbf{A}_1, \dots, \mathbf{A}_m) \in \mathsf{Mat}_n(\mathbb{Z}[i])$ where each $\|\mathbf{A}_i\|_\infty \leq M$.*

*(a) ($\varepsilon$-scaling 1 $\Rightarrow$ $\varepsilon'$-scaling 2) Given operator $T'$ that is a scaling of $T_{\overrightarrow{\mathbf{A}}}$ by $(\mathbf{C}, \mathbf{D})$ with $T'(\mathbf{I}) = \mathbf{I}$ and $\mathsf{ds}(T') \leq \varepsilon$, then $\log \mathsf{cap}(\mathbf{D}^\dagger\mathbf{D}) \leq \log \mathsf{cap}(T_{\overrightarrow{\mathbf{A}}}) + \varepsilon'$ for $\varepsilon' = -n \log(1 - \sqrt{n\varepsilon/2})$.*

(b) *($\varepsilon$-scaling 1 $\Leftarrow$ $\varepsilon'$-scaling 2) Given PSD matrix* $\mathbf{X} \in \mathsf{Mat}_n(\mathbf{C})$ *with* $\log \mathsf{cap}(\mathbf{X}) \le \log \mathsf{cap}(T_{\overrightarrow{\mathbf{A}}}) + \varepsilon'$*, we have* $T'(\mathbf{I}) = \mathbf{I}$ *and* $\mathsf{ds}(T') \le 6\varepsilon'$ *where* $T'$ *is a scaling of* $T_{\overrightarrow{\mathbf{A}}}$ *by* $(\mathbf{C}, \mathbf{D}) = \left(\left(T_{\overrightarrow{\mathbf{A}}}(\mathbf{X})\right)^{-1/2}, \mathbf{X}^{1/2}\right)$.

(c) *($\varepsilon$-scaling 1 $\Rightarrow$ $\varepsilon'$-scaling 3) Given operator* $T'$ *that is a scaling of* $T_{\overrightarrow{\mathbf{A}}}$ *by* $(\mathbf{C}, \mathbf{D})$ *with* $T'(\mathbf{I}) = \mathbf{I}$ *and* $\mathsf{ds}(T') \le \varepsilon$*, then letting* $\mathbf{A}'_i = \frac{\mathbf{C}\mathbf{A}_i\mathbf{D}^\dagger}{\det(\mathbf{C})^{1/n}\det(\mathbf{D}^\dagger)^{1/n}}$*, we have* $\overrightarrow{\mathbf{A}}' \in \mathcal{O}_{\overrightarrow{\mathbf{A}}}$ *and* $\Delta_{SU}(\overrightarrow{\mathbf{A}}', \overrightarrow{\mathbf{A}}^*) \le \varepsilon' = \mathsf{poly}(n, m, M, \varepsilon)$*. Here* $\overrightarrow{\mathbf{A}}^*$ *is a tuple of minimum norm in* $\overline{\mathcal{O}}_{\overrightarrow{\mathbf{A}}}$

(d) *($\varepsilon$-scaling 1 $\Leftarrow$ $\varepsilon'$-scaling 3) Given* $\overrightarrow{\mathbf{A}}' \in \mathcal{O}_{\overrightarrow{\mathbf{A}}}$ *that satisfies* $\Delta_{SU}(\overrightarrow{\mathbf{A}}', \overrightarrow{\mathbf{A}}^*) \le \varepsilon'$ *for some minimum norm tuple* $\overrightarrow{\mathbf{A}}^*$ *in* $\overline{\mathcal{O}}_{\overrightarrow{\mathbf{A}}}$*, we have* $T'(\mathbf{I}) = \mathbf{I}$ *and* $\mathsf{ds}(T') \le \varepsilon = \mathsf{poly}(n, m, M, \varepsilon')$ *where* $T'$ *is a scaling of* $T_{\overrightarrow{\mathbf{A}}'}$ *by* $(\mathbf{C}, \mathbf{D}) = \left(T_{\overrightarrow{\mathbf{A}}'}(\mathbf{I})^{-1/2}, \mathbf{I}\right)$.

# 4 Second order method for geodesically convex functions

In this section, we present our algorithm for minimizing a general class of geodesically convex functions over a natural manifold over PD matrices. In the next section, we shall make our algorithm explicit for the operator-scaling problem. (We describe our algorithm for the specific metric over PD matrices, but the reader may observe that it easily generalizes to other manifolds and metrics.)

We start with a few definitions.

**Definition 4.1.** *Let $F$ be a function from $n \times n$ complex PD matrices to reals. We say that*

- *$F$ is* geodesically convex (or g-convex) *if for every PD matrix $\mathbf{X}$ and every Hermitian matrix $\mathbf{\Delta}$, $F\left(\mathbf{X}^{1/2}e^{s\mathbf{\Delta}}\mathbf{X}^{1/2}\right)$ is a convex function in $s \in \mathbb{R}$.*
- *$F$ is* g-second-order robust *if for every PD matrix $\mathbf{X}$ and every Hermitian matrix $\mathbf{\Delta}$ s.t. $\|\mathbf{\Delta}\|_2 \le 1$, $g(s) = F\left(\mathbf{X}^{1/2}e^{s\mathbf{\Delta}}\mathbf{X}^{1/2}\right)$ satisfies $\left|\frac{d^3g}{ds^3}\right| \le 2\frac{d^2g}{ds^2}$.*

We note that the robustness assumption is much weaker than self-concordance, for which interior point methods work (in the Euclidean case).

We propose Algorithm 1 to minimize a function $F(\mathbf{X})$ satisfying the properties above. It initializes itself with $\mathbf{X}_0 = \mathbf{I}$ and is divided into $T$ iterations. In each iteration $t \ge 0$, it

- defines $f^t(\mathbf{\Delta}) \stackrel{\text{def}}{=} F\left(\mathbf{X}_t^{1/2}e^{\mathbf{\Delta}}\mathbf{X}_t^{1/2}\right)$ (see Line 3), and
- minimizes $f^t(\mathbf{\Delta})$ based on its second-order Taylor expansion over $\|\mathbf{\Delta}\|_2 \le 1/2$ (see Line 4).

  **Claim 4.2.** *The objective* $\mathrm{Tr}\left(\nabla f^t(0) \cdot \mathbf{\Delta}\right) + \frac{1}{2e}\mathrm{Tr}\left(\nabla^2 f^t(0) \cdot \mathbf{\Delta} \otimes \mathbf{\Delta}\right)$ *in Line 4 is quadratic and convex in $\mathbf{\Delta}$, and thus minimization problem for $\mathbf{\Delta}_t$ is convex.*

- moves to $\mathbf{X}_{t+1} \leftarrow \mathbf{X}_t^{1/2}e^{\mathbf{\Delta}_t/e^2}\mathbf{X}_t^{1/2}$ where $\mathbf{\Delta}_t$ is the minimizer of $f^t(\mathbf{\Delta})$.

Our main theorem of the section is the following. It says the number of iterations needed is *logarithmical* in the approximation parameter $\varepsilon$, and linearly in $R$, the "diameter" of the problem.

[20] Although it is beyond the scope of this paper, we quickly point out that for practitioners, one can replace the constraint $\|\mathbf{\Delta}\|_2 \le 1/2$ with $\|\mathbf{\Delta}\|_F \le 1/2$. This results in a simpler quadratic minimization problem that can be solved by one-dimensional binary search without invoking general convex optimization solvers. However, it increases the total number of iterations by a factor $\sqrt{n}$.

**Algorithm 1** Second order algorithm for minimizing g-convex functions

**Input:** oracle access to gradients and Hessians of $F$, number of iterations $T$.

**Output:** PD matrix $\mathbf{X}_T \in \mathsf{Mat}_n(\mathbb{C})$.

1: $\mathbf{X}_0 = \mathbf{I}$.
2: **for** $t \leftarrow 0$ **to** $T-1$ **do**
3: define $f^t(\mathbf{\Delta}) \stackrel{\text{def}}{=} F(\mathbf{X}_t^{1/2} e^{\mathbf{\Delta}} \mathbf{X}_t^{1/2})$
4: solve the following convex quadratic minimization problem[20]

$$\mathbf{\Delta}_t = \operatorname*{arg\,min}_{\mathbf{\Delta}\in\mathsf{Mat}_n(\mathbb{C})} \left\{ \mathrm{Tr}\left(\nabla f^t(0)\cdot\mathbf{\Delta}\right) + \frac{1}{2e}\mathrm{Tr}\left(\nabla^2 f^t(0)\cdot\mathbf{\Delta}\otimes\mathbf{\Delta}\right) \,\middle|\, \mathbf{\Delta} \text{ is Hermitian and } \|\mathbf{\Delta}\|_2 \le \frac{1}{2} \right\}$$

$\diamond$ *This is the second-order Taylor expansion of $f^t(\mathbf{\Delta})$ at $\mathbf{\Delta}=0$.*

5: $\mathbf{X}_{t+1} \leftarrow \mathbf{X}_t^{1/2} e^{\mathbf{\Delta}_t/e^2} \mathbf{X}_t^{1/2}$.
6: **end for**
7: **return** $\mathbf{X}_T$.

**Theorem 4.3.** *Suppose we are given a g-convex function $F$ that is g-second-order robust, then Algorithm 1 produces a point $\mathbf{X}_T$ satisfying $F(\mathbf{X}_T) - F(\mathbf{X}^*) \le \varepsilon$ in $T = O\left(R\log\left(\frac{F(\mathbf{I})-F(\mathbf{X}^*)}{\varepsilon}\right)\right)$ iterations. Here*

$$R = \sup_{X:F(\mathbf{X})\le F(\mathbf{I})} \left\| \log\left(\mathbf{X}^{-1/2}\mathbf{X}^*\mathbf{X}^{-1/2}\right)\right\|_2$$

*and $\mathbf{X}^*$ is an arbitrary point (usually thought of as an approximate minimizer of $F$).*

The proof of Theorem 4.3 is based on the recent analysis of Allen-Zhu et al. [7], Cohen et al. [19], but generalized to g-convex functions. See Appendix B.

# 5 Operator scaling: geodesically convex optimization

In this section, we make our Algorithm 1 concrete for the task of operator scaling.

Consider a tuple of matrices $\overrightarrow{\mathbf{A}} = (\mathbf{A}_1, \ldots, \mathbf{A}_m)$ where each $\mathbf{A}_i \in \mathsf{Mat}_n(\mathbb{Z}[i])$ and $\|\mathbf{A}_i\|_\infty \le M$ (i.e., each complex entry of $\mathbf{A}_i$ has modulus at most $M$). In this section, we study the minimization of the following log-capacity function that maps $\mathsf{GL}_n(\mathbb{C})$ to $\mathbb{R}$:

$$\mathsf{logcap}(\mathbf{X}) \stackrel{\text{def}}{=} \log\det\left(\textstyle\sum_{i=1}^m \mathbf{A}_i\mathbf{X}\mathbf{X}^\dagger\mathbf{A}_i^\dagger\right) - \log\det(\mathbf{X}\mathbf{X}^\dagger) \ .$$

As argued in Section 3.4, approximate minimizers of $\mathsf{logcap}(\mathbf{X})$ give us approximate solutions for operator scaling.

**Remark 5.1.** *There is a slight difference between the above definition and Section 4 or Section 1. In this section, $\mathbf{X}\mathbf{X}^\dagger$ corresponds to the PD matrix $\mathbf{X}$ in previous sections. We have adopted this notion to make our notations simpler.*

Unfortunately, although one can show that $\mathsf{logcap}(\mathbf{X})$ is g-convex and g-second-order robust, it does not enjoy a good diameter bound $R$ if we directly apply Theorem 4.3. To fix this issue, we propose to minimize

$$F(\mathbf{X}) \stackrel{\text{def}}{=} \mathsf{logcap}(\mathbf{X}) + \lambda\,\mathsf{reg}(\mathbf{X}) \quad \text{over} \quad \mathbf{X}\in\mathsf{GL}_n(\mathbb{C}) \ , \tag{5.1}$$

where $\mathsf{reg}\colon \mathsf{GL}_n(\mathbb{C}) \to \mathbb{R}_{\ge 0}$ is a regularizer function

$$\mathsf{reg}(\mathbf{X}) \stackrel{\text{def}}{=} \mathrm{Tr}(\mathbf{X}\mathbf{X}^\dagger)\cdot\mathrm{Tr}((\mathbf{X}\mathbf{X}^\dagger)^{-1}) \ ,$$

**Algorithm 2** Operator scaling algorithm

**Input:** $\mathbf{A}_1, \dots, \mathbf{A}_m \in \mathsf{Mat}_n(\mathbb{Z}[i])$ with $\|\mathbf{A}_i\|_\infty \le M$,
$\lambda > 0$ the regularizer weight, and $T$ the number of iterations.

**Output:** matrix $\mathbf{X}_T \in \mathsf{GL}_n(\mathbb{C})$.

1: $\mathbf{X}_0 = \mathbf{I}$.

2: **for** $t \leftarrow 0$ **to** $T-1$ **do**

3: define

$$f^t(\mathbf{\Delta}) \stackrel{\text{def}}{=} F(\mathbf{X}_t e^{\mathbf{\Delta}/2}) = \mathsf{logcap}(\mathbf{X}_t e^{\mathbf{\Delta}/2}) + \lambda\, \mathsf{reg}(\mathbf{X}_t e^{\mathbf{\Delta}/2})$$
$$= \log\det\Big(\textstyle\sum_{i=1}^m \mathbf{A}_i \mathbf{X}_t e^{\mathbf{\Delta}} \mathbf{X}_t^\dagger \mathbf{A}_i^\dagger\Big) - \log\det(\mathbf{X}_t e^{\mathbf{\Delta}} \mathbf{X}_t^\dagger) + \lambda \mathrm{Tr}\Big(\mathbf{X}_t e^{\mathbf{\Delta}} \mathbf{X}_t^\dagger\Big)\, \mathrm{Tr}\Big(\big(\mathbf{X}_t e^{\mathbf{\Delta}} \mathbf{X}_t^\dagger\big)^{-1}\Big)$$

4: solve the following convex quadratic minimization

$$\mathbf{\Delta}_t = \operatorname*{arg\,min}_{\mathbf{\Delta} \in \mathsf{Mat}_n(\mathbb{C})} \left\{ \mathrm{Tr}\left(\nabla f^t(0) \cdot \mathbf{\Delta}\right) + \frac{1}{2e} \mathrm{Tr}\left(\nabla^2 f^t(0) \cdot \mathbf{\Delta} \otimes \mathbf{\Delta}\right) \,\middle|\, \mathbf{\Delta} \text{ is Hermitian and } \|\mathbf{\Delta}\|_2 \le \frac{1}{2} \right\}$$

◇ *see Appendix C.3 for an explicit form*

◇ *This is the second-order Taylor expansion of $f^t(\mathbf{\Delta})$ at $\mathbf{\Delta} = 0$.*

5: $\mathbf{X}_{t+1} \leftarrow \mathbf{X}_t e^{\mathbf{\Delta}_t / e^2}$.

6: rescale $\mathbf{X}_{t+1} \leftarrow \delta \mathbf{X}_{t+1}$ for some $\delta > 0$ so that $\lambda_{\min}(\mathbf{X}_{t+1} \mathbf{X}_{t+1}^\dagger) \in [1, 2]$.

◇ *note that $F(\mathbf{X}_{t+1}) = F(\delta \mathbf{X}_{t+1})$ but rescaling helps improve numerical stability, see Section C.4*

7: **end for**

8: **return** $\mathbf{X}_T$.

and $\lambda > 0$ is some small regularizer weight to be chosen later. One can show that $\mathsf{reg}(\mathbf{X})$ is also g-convex and g-second-order robust. Furthermore, if we minimize $\mathsf{logcap}(\mathbf{X})$ together with $\mathsf{reg}(\mathbf{X})$, then the diameter $R$ in Theorem 4.3 can be well bounded.

**Our Algorithm and Theorem.** We propose Algorithm 2 to minimize $F(\mathbf{X})$. It initializes itself with $\mathbf{X}_0 = \mathbf{I}$ and is divided into $T$ iterations. In each iteration $t \ge 0$, it

- defines $f^t(\mathbf{\Delta}) \stackrel{\text{def}}{=} F(\mathbf{X}_t e^{\mathbf{\Delta}/2})$ (see Line 3), and
- minimizes $f^t(\mathbf{\Delta})$ based on its second-order Taylor expansion over $\|\mathbf{\Delta}\|_2 \le 1/2$ (see Line 4).

**Claim 5.2.** *The objective* $\mathrm{Tr}\left(\nabla f^t(0) \cdot \mathbf{\Delta}\right) + \frac{1}{2e}\mathrm{Tr}\left(\nabla^2 f^t(0) \cdot \mathbf{\Delta} \otimes \mathbf{\Delta}\right)$ *in Line 4 is quadratic and convex in* $\mathbf{\Delta}$*, and thus minimization problem for* $\mathbf{\Delta}_t$ *is convex.*

We give explicit form of this quadratic function in Appendix C.3. The computation of $\mathbf{\Delta}_t$ can be done by standard convex optimization solvers.[21]

We have the following main theorem for Algorithm 2:

**Theorem 5.3.** *Suppose there exists some $\varepsilon > 0$ and*

$$\mathbf{X}_\varepsilon^* \in \mathsf{GL}_n(\mathbb{C}) \quad s.t. \quad \mathsf{logcap}(\mathbf{X}_\varepsilon^*) \le \log \mathsf{cap}(T_{\vec{\mathbf{A}}}) + \varepsilon \ .$$

[21] Although it is beyond the scope of this paper, we quickly point out that for practitioners, one can replace the constraint $\|\mathbf{\Delta}_t\|_2 \le 1/2$ with $\|\mathbf{\Delta}_t\|_F \le 1/2$. This results in a simpler quadratic minimization problem that can be solved by one-dimensional binary search without invoking general convex optimization solvers. However, it increases the total number of iterations by a factor $\sqrt{n}$.

*Then, letting* $\lambda = \frac{\varepsilon}{n^2 \cdot (\kappa(\mathbf{X}^*_\varepsilon))^2}$ *and* $T = O(\mathsf{polylog}(n, m, M, \kappa(\mathbf{X}^*_\varepsilon), \varepsilon^{-1})$*, the above algorithm finds*

$$\mathbf{X} \in \mathsf{GL}_n(\mathbb{C}) \quad s.t. \quad \mathsf{logcap}(\mathbf{X}) \le \log \mathsf{cap}(T_{\overrightarrow{\mathbf{A}}}) + 3\varepsilon$$

*in time complexity* $\mathsf{poly}(n, m, \log M, \log \kappa(\mathbf{X}^*_\varepsilon), \log \frac{1}{\varepsilon})$.

(We prove Theorem 5.3 and carefully deal with numerical errors in Appendix C.)

Note that Theorem 5.3 almost gives us a polynomial-time algorithm for operator scaling. However, we still need to prove that $\kappa(\mathbf{X}^*_\varepsilon)$ is polynomially bounded in the problem size and in $\varepsilon^{-1}$. This is precisely the goal of the next section.[22]

## 6 Operator scaling: final theorem

In this section, we first show bounds on the singular values of the scaling matrices required to make an operator $\varepsilon$-close to doubly stochastic (that is, to make $\mathsf{ds}(T) \le \varepsilon$, see Definition 3.7). Then, we combine these bounds with Theorem 5.3 to state our final theorem for operator scaling.

**Theorem 6.1.** *Let* $\mathbf{A}_1, \dots, \mathbf{A}_m \in \mathsf{Mat}_n(\mathbb{Z}[i])$ *be matrices where* $\|\mathbf{A}_i\|_\infty \le M$ *for each* $i \in [m]$*. Suppose the operator* $T_{\overrightarrow{\mathbf{A}}}$ *defined by* $(\mathbf{A}_1, \dots, \mathbf{A}_m)$ *satisfies* $\mathsf{cap}(T_{\overrightarrow{\mathbf{A}}}) > 0$*. Then, for all* $\varepsilon > 0$*, there exist matrices* $\mathbf{X}, \mathbf{Y} \succ 0$ *such that the operator* $T_{\overrightarrow{\mathbf{B}}}$ *defined by* $(\mathbf{B}_1, \dots, \mathbf{B}_m) = (\mathbf{X}\mathbf{A}_1\mathbf{Y}, \dots, \mathbf{X}\mathbf{A}_m\mathbf{Y})$ *satisfies*

$$\mathsf{ds}(T_{\overrightarrow{\mathbf{B}}}) \le \varepsilon \quad and \quad \|\mathbf{X}\|_2, \|\mathbf{X}^{-1}\|_2, \|\mathbf{Y}\|_2, \|\mathbf{Y}^{-1}\|_2 \le (mnM) \cdot \exp\left(n^{3/2} \log\left(\frac{12mn^4M^2}{\varepsilon}\right)\right) .$$

The proof of Theorem 6.1 (see Appendix D) is based on continuous gradient flow on the function $f(\mathbf{X}, \mathbf{Y}) = \mathrm{Tr}[\mathbf{X}\, T_{\overrightarrow{\mathbf{A}}}(\mathbf{Y})] = \mathrm{Tr}\left(\sum_{i=1}^m \mathbf{X}\mathbf{A}_i\mathbf{Y}\mathbf{A}_i^\dagger\right)$ (subject to the constraints $\det(\mathbf{X}) = \det(\mathbf{Y}) = 1$). We will prove that this gradient flow converges "linearly" and the solution to this gradient flow (after sufficiently long time) give us the required $\mathbf{X}, \mathbf{Y}$ from the theorem.

**Remark 6.2.** *This gradient flow is similar to the one defined by Kirwan [56, 78] for general group actions (here we study only the left-right action). Also, when we were writing this manuscript, we find out that this gradient flow has been independently studied by [59].*

Putting Theorem 6.1 (for bounding $\kappa(\mathbf{X}^*)$) and Theorem 3.14 (for the three equivalent notions of $\varepsilon$-approximation) into Theorem 5.3, we have

**Theorem M1.** *Let* $\mathbf{A}_1, \dots, \mathbf{A}_m \in \mathsf{Mat}_n(\mathbb{Z}[i])$ *be matrices where* $\|\mathbf{A}_i\|_\infty \le M$ *for each* $i \in [m]$*. Suppose the operator* $T_{\overrightarrow{\mathbf{A}}}$ *defined by* $(\mathbf{A}_1, \dots, \mathbf{A}_m)$ *satisfies* $\mathsf{cap}(T_{\overrightarrow{\mathbf{A}}}) > 0$*. Then, for all* $\varepsilon > 0$*, there is an algorithm that runs in time* $\mathsf{poly}(n, m, \log M, \log \varepsilon^{-1})$ *and find*

- $\mathbf{X} \in \mathsf{GL}_n(\mathbb{C})$ *such that* $\log \mathsf{cap}(\mathbf{X}\mathbf{X}^\dagger) \le \log \mathsf{cap}(T_{\overrightarrow{\mathbf{A}}}) + \varepsilon$.
- $\overrightarrow{\mathbf{B}} = (\mathbf{X}\mathbf{A}_1\mathbf{Y}, \dots \mathbf{X}\mathbf{A}_m\mathbf{Y})$ *with* $\mathbf{X}, \mathbf{Y} \in \mathsf{GL}_n(\mathbb{C})$ *such that*
$$T_{\overrightarrow{\mathbf{B}}}(\mathbf{I}) = \mathbf{I} \ and \ \mathsf{ds}(T_{\overrightarrow{\mathbf{B}}}) = \mathsf{db}(T_{\overrightarrow{\mathbf{B}}}) \le \varepsilon.$$
- $\overrightarrow{\mathbf{B}} = (\mathbf{X}\mathbf{A}_1\mathbf{Y}, \dots \mathbf{X}\mathbf{A}_m\mathbf{Y})$ *with* $\mathbf{X}, \mathbf{Y} \in \mathsf{SL}_n(\mathbb{C})$ *such that*
$$\Delta_{SU}(\overrightarrow{\mathbf{B}}, \overrightarrow{\mathbf{A}}^*) \le \varepsilon$$
*where* $\overrightarrow{\mathbf{A}}^*$ *be any tuple of minimum norm in* $\overline{\mathcal{O}}_{\overrightarrow{\mathbf{A}}}$.

[22] Since $\log \mathsf{cap}(T_{\overrightarrow{\mathbf{A}}}) = \inf_{\mathbf{X} \in \mathsf{GL}_n(\mathbb{C})} \mathsf{logcap}(\mathbf{X})$, there may not exist any matrix $\mathbf{X}^*$ satisfying $\log \mathsf{cap}(T_{\overrightarrow{\mathbf{A}}}) = \mathsf{logcap}(\mathbf{X}^*)$. In other words, the condition number $\kappa(\mathbf{X}^*_\varepsilon)$ may tend to infinity as $\varepsilon$ tends to zero.

## 7 Distance between non-intersecting orbit-closures

Recall that the left-right action is the group $G = \mathsf{SL}_n(\mathbb{C}) \times \mathsf{SL}_n(\mathbb{C})$ acting on the vector space $V = \mathsf{Mat}_n(\mathbb{C})^m$ by simultaneous left-right multiplication. We show that under the left-right action, if the closures of two orbits $\overline{\mathcal{O}}_{\vec{\mathbf{A}}}$ and $\overline{\mathcal{O}}_{\vec{\mathbf{B}}}$ do **not** intersect, then any element of bounded norm in $\overline{\mathcal{O}}_{\vec{\mathbf{A}}}$ is *"sufficiently far"* from any element of bounded norm in $\overline{\mathcal{O}}_{\vec{\mathbf{B}}}$, even up to unitary transformations (of determinant essentially 1). The proof is given in Appendix E.

**Lemma 7.1.** *Let $\vec{\mathbf{A}} = (\mathbf{A}_1, \ldots, \mathbf{A}_m)$ and $\vec{\mathbf{B}} = (\mathbf{B}_1, \ldots, \mathbf{B}_m)$ be two tuples in $\mathsf{Mat}_n(\mathbb{Z}[i])^m$ not in the null cone of the left-right action (i.e., $\mathsf{cap}(T_{\vec{\mathbf{A}}}) > 0$ and $\mathsf{cap}(T_{\vec{\mathbf{B}}}) > 0$). Suppose $\|\vec{\mathbf{A}}\|_2, \|\vec{\mathbf{B}}\|_2 < M$ and $\varepsilon = \exp(-n^{20}m \cdot \log(M))$. Let $\vec{\mathbf{A}}' = (\mathbf{A}'_1, \ldots, \mathbf{A}'_m)$ and $\vec{\mathbf{B}}' = (\mathbf{B}'_1, \ldots, \mathbf{B}'_m)$ be elements in $\overline{\mathcal{O}}_{\vec{\mathbf{A}}}$ and $\overline{\mathcal{O}}_{\vec{\mathbf{B}}}$ respectively.*

*(a) If $\overline{\mathcal{O}}_{\vec{\mathbf{A}}} \cap \overline{\mathcal{O}}_{\vec{\mathbf{B}}} = \varnothing$ and $\mathbf{U}, \mathbf{V} \in \mathsf{U}_n(\mathbb{C})$ are such that $|\det(\mathbf{U}\mathbf{V}) - 1| \le \varepsilon$, then*

$$\left\| \mathbf{U}\vec{\mathbf{A}}'\mathbf{V} - \vec{\mathbf{B}}' \right\|_2 \ge \varepsilon.$$

*(b) If $\overline{\mathcal{O}}_{\vec{\mathbf{A}}} \cap \overline{\mathcal{O}}_{\vec{\mathbf{B}}} \ne \varnothing$, then for all $\mathbf{U}, \mathbf{V} \in \mathsf{U}_n(\mathbb{C})$ such that*

$$\left\| \mathbf{U}\vec{\mathbf{A}}'\mathbf{V} - \vec{\mathbf{B}}' \right\|_2 \le \varepsilon,$$

*we must have $|\det(\mathbf{U}\mathbf{V}) - 1| \le \varepsilon^{1/3}$.*

A very useful corollary of this lemma is that, if we find elements $\vec{\mathbf{A}}'$ and $\vec{\mathbf{B}}'$ which are sufficiently close to the elements of minimum norm in $\overline{\mathcal{O}}_{\vec{\mathbf{A}}}$ and $\overline{\mathcal{O}}_{\vec{\mathbf{B}}}$, respectively, to test whether $\overline{\mathcal{O}}_{\vec{\mathbf{A}}} \cap \overline{\mathcal{O}}_{\vec{\mathbf{B}}} = \varnothing$ it is enough to check if there exist unitary matrices $\mathbf{U}, \mathbf{V}$ such that $\left\| \mathbf{U}\vec{\mathbf{A}}'\mathbf{V} - \vec{\mathbf{B}}' \right\|_2$ is sufficiently small. In case we can find such unitary matrices, we only need to check whether $\det(\mathbf{U}\mathbf{V})$ is sufficiently close to 1. If the answer is positive, then Lemma 7.1 above assures that the orbits intersect. Otherwise we are sure that they do not intersect.

## 8 Algorithm for checking unitary equivalence

In this section we consider the following problem that may be of independent interests: given tuples $\vec{\mathbf{A}} = (\mathbf{A}_1, \ldots, \mathbf{A}_m), \vec{\mathbf{B}} = (\mathbf{B}_1, \ldots, \mathbf{B}_m) \in \mathsf{Mat}_n(\mathbb{C})^m$ and values $\varepsilon' \gg \varepsilon > 0$, we want to efficiently decide whether $\Delta_U(\vec{\mathbf{A}}, \vec{\mathbf{B}}) \le \varepsilon$ or $\Delta_U(\vec{\mathbf{A}}, \vec{\mathbf{B}}) \ge \varepsilon'$. Our main theorem is as follows:

**Theorem M3.** *For any $\vec{\mathbf{A}}, \vec{\mathbf{B}} \in \mathsf{Mat}_n(\mathbb{C})^m$ with spectral norm at most $\lambda \ge 2$, and any $\varepsilon > 0$ satisfying $\varepsilon \le \lambda^{-\mathsf{poly}(n,m)}$, there exists some $\varepsilon' = 2^{8n}(6\lambda)^{\frac{1}{n^5}}\varepsilon^{\frac{1}{20mn^{10}}} \gg \varepsilon$ satisfying that*

- *if Algorithm 5 outputs **Yes** with $\mathbf{U}, \mathbf{V} \in \mathsf{U}_n(\mathbb{C})$, then it must satisfy $\|\mathbf{U}\vec{\mathbf{A}}\mathbf{V}^\dagger - \vec{\mathbf{B}}\|_2 \le \varepsilon'$;*
- *if $\Delta_U(\vec{\mathbf{A}}, \vec{\mathbf{B}}) \le \varepsilon$, then Algorithm 5 must output **Yes** with some $\mathbf{U}, \mathbf{V} \in \mathsf{U}_n(\mathbb{C})$.*

*Furthermore, Algorithm 5 runs in deterministic time $\mathsf{poly}(n, m, \log(\lambda/\varepsilon))$.*

The proof of Theorem M3 and the specification of Algorithm 5 can be found in Appendix F.

At a high level, to design Algorithm 5, we first set $\mathbf{A}'_i = \begin{pmatrix} 0 & \mathbf{A}_i \\ 0 & 0 \end{pmatrix}$ and $\mathbf{B}'_i = \begin{pmatrix} 0 & \mathbf{B}_i \\ 0 & 0 \end{pmatrix}$ to be $2n \times 2n$ matrices, and study whether we can find unitary $\mathbf{U} \in \mathsf{U}_{2n}(\mathbb{C})$ such that

$$\left\| \mathbf{U} \overrightarrow{\mathbf{A}}' \mathbf{U}^\dagger - \overrightarrow{\mathbf{B}}' \right\|_2 \le \varepsilon \ .$$

We call this the *simultaneous conjugation problem.*

Obviously, if the sorted list of singular values or eigenvalues of each pair of $\mathbf{A}_i$ and $\mathbf{B}_i$ do not match, then one can declare that the above problem has no solution. Furthermore, intuitively, if there is a large gap to the sorted list of singular values of some $\mathbf{A}_i$ (or of some $\mathbf{B}_i$), then one can find a new basis so that the problem becomes block diagonal, where the two blocks each corresponds to the singular values above/below the gap. This enables us to reduce the dimension of the problem.

Unfortunately, a lot of technical subtlety arises when carefully dealing how error propagates through this decomposition tree. We defer all the technical proofs and the algorithm description to Appendix F.

## 9 Final algorithm for orbit-closure intersection

In this section, we give our deterministic polynomial-time Algorithm 3 for the orbit-closure intersection problem for the left-right action. We are given tuples of matrices $\overrightarrow{\mathbf{A}}$ and $\overrightarrow{\mathbf{B}}$ (with complex integral entries of modulus at most $M$). We first check if either of them is in the null cone using [34, 49]. If both are in the null cone then their orbit-closures intersect; if only one is in the null cone then their orbit-closures do not intersect. So we can assume both tuples are not in the null cone.

For analysis purpose, let $\overrightarrow{\mathbf{A}}^*$ and $\overrightarrow{\mathbf{B}}^*$ be some tuples of minimum $\ell_2$-norm in $\overline{\mathcal{O}}_{\overrightarrow{\mathbf{A}}}$ and $\overline{\mathcal{O}}_{\overrightarrow{\mathbf{B}}}$ respectively. We first invoke the operator-scaling Algorithm 2 (see Theorem M1) to get tuples $\overrightarrow{\mathbf{A}}', \overrightarrow{\mathbf{B}}'$ —in the orbits of $\overrightarrow{\mathbf{A}}$ and $\overrightarrow{\mathbf{B}}$ respectively— which are $\varepsilon/2$-close to $\overrightarrow{\mathbf{A}}^*$ and $\overrightarrow{\mathbf{B}}^*$. In symbols:

$$\Delta_{SU}\left(\overrightarrow{\mathbf{A}}', \overrightarrow{\mathbf{A}}^*\right) \le \varepsilon/2 \quad \text{and} \quad \Delta_{SU}\left(\overrightarrow{\mathbf{B}}', \overrightarrow{\mathbf{B}}^*\right) \le \varepsilon/2$$

where we choose $\varepsilon = \exp(-\mathsf{poly}(n, m, \log M))$. Next, we apply Algorithm 5 (see Theorem M3) to check if $\overrightarrow{\mathbf{A}}'$ and $\overrightarrow{\mathbf{B}}'$ are close up to left-right unitary transformations $\mathbf{U}, \mathbf{V}$. If so, we further check whether $|\det(\mathbf{UV}) - 1|$ is close to zero (see Lemma 7.1).

We have the following final theorem:

**Theorem M2.** *Let $\overrightarrow{\mathbf{A}}, \overrightarrow{\mathbf{B}} \in \mathsf{Mat}_n(\mathbb{Z}[i])^m$ be tuples of matrices where $\|\mathbf{A}_i\|_\infty \le M$ and $\|\mathbf{B}_i\|_\infty \le M$ for every $i \in [m]$. There is a deterministic algorithm (see Algorithm 3) that decides whether $\overline{\mathcal{O}}_{\overrightarrow{\mathbf{A}}} \cap \overline{\mathcal{O}}_{\overrightarrow{\mathbf{B}}} = \varnothing$ or $\overline{\mathcal{O}}_{\overrightarrow{\mathbf{A}}} \cap \overline{\mathcal{O}}_{\overrightarrow{\mathbf{B}}} \ne \varnothing$, in time $\mathsf{poly}(n, m, \log M)$.*

*Proof of Theorem M2.* We only need to prove the correctness.

- If $\overline{\mathcal{O}}_{\overrightarrow{\mathbf{A}}} \cap \overline{\mathcal{O}}_{\overrightarrow{\mathbf{B}}} \ne \varnothing$, we know by Corollary 1.5 that $\overrightarrow{\mathbf{A}}^*$ and $\overrightarrow{\mathbf{B}}^*$ are related by unitary transformations: there exist $\mathbf{U}, \mathbf{V} \in \mathsf{SU}_n(\mathbb{C})$ s.t. $\mathbf{U}\overrightarrow{\mathbf{A}}^*\mathbf{V} = \overrightarrow{\mathbf{B}}^*$. Since $\Delta_{SU}(\overrightarrow{\mathbf{A}}', \overrightarrow{\mathbf{A}}^*) \le \varepsilon/2$ and $\Delta_{SU}(\overrightarrow{\mathbf{B}}', \overrightarrow{\mathbf{B}}^*) \le \varepsilon/2$, by triangle inequality and basic algebra, we have $\Delta_{SU}(\overrightarrow{\mathbf{A}}', \overrightarrow{\mathbf{B}}') \le \varepsilon$.

  Since $\Delta_{SU}(\overrightarrow{\mathbf{A}}', \overrightarrow{\mathbf{B}}') \le \varepsilon$ implies $\Delta_{U}(\overrightarrow{\mathbf{A}}', \overrightarrow{\mathbf{B}}') \le \varepsilon$, we can invoke Theorem M3 which says that Algorithm 5 must output **Yes** with $\mathbf{U}, \mathbf{V} \in \mathsf{U}_n(\mathbb{C})$ satisfying

  $$\left\| \mathbf{U} \overrightarrow{\mathbf{A}}' \mathbf{V} - \overrightarrow{\mathbf{B}}' \right\|_2 \le \varepsilon'.$$

  Invoking Lemma 7.1b, we must have $|\det(\mathbf{UV}) - 1| \le (\varepsilon')^{1/3}$ and therefore our algorithm will correctly output "**Yes**, $\overline{\mathcal{O}}_{\overrightarrow{\mathbf{A}}} \cap \overline{\mathcal{O}}_{\overrightarrow{\mathbf{B}}} \ne \varnothing$."

**Algorithm 3** Algorithm for the orbit-closure intersection problem

**Input:** tuples of matrices $\vec{\mathbf{A}}, \vec{\mathbf{B}}$ in $\mathsf{Mat}_n(\mathbb{Z}[i])^m$ s.t. $\|\mathbf{A}_i\|_\infty, \|\mathbf{B}_i\|_\infty \le M$ for all $i$.
**Output:** **Yes**, meaning $\overline{\mathcal{O}}_{\vec{\mathbf{A}}} \cap \overline{\mathcal{O}}_{\vec{\mathbf{B}}} \neq \varnothing$, or **No**, meaning $\overline{\mathcal{O}}_{\vec{\mathbf{A}}} \cap \overline{\mathcal{O}}_{\vec{\mathbf{B}}} = \varnothing$.

Use the algorithm from [34], decide whether $\vec{\mathbf{A}}$, $\vec{\mathbf{B}}$ are in the null cone.
**if** $\vec{\mathbf{A}}$ and $\vec{\mathbf{B}}$ are both in the null cone **then**
  **return** "**Yes**, $\overline{\mathcal{O}}_{\vec{\mathbf{A}}} \cap \overline{\mathcal{O}}_{\vec{\mathbf{B}}} \neq \varnothing$."
**else if** exactly one of $\vec{\mathbf{A}}$ and $\vec{\mathbf{B}}$ is in the null cone **then**
  **return** "**No**, $\overline{\mathcal{O}}_{\vec{\mathbf{A}}} \cap \overline{\mathcal{O}}_{\vec{\mathbf{B}}} = \varnothing$."
**end if**
use Algorithm 2 to obtain tuples $\vec{\mathbf{A}}' \in \mathcal{O}_{\vec{\mathbf{A}}}$ and $\vec{\mathbf{B}}' \in \mathcal{O}_{\vec{\mathbf{B}}}$ such that ◇ *see Theorem M1*

$$\Delta_{SU}\left(\vec{\mathbf{A}}', \vec{\mathbf{A}}^*\right), \Delta_{SU}\left(\vec{\mathbf{B}}', \vec{\mathbf{B}}^*\right) \le \varepsilon/2, \text{ where } \varepsilon \stackrel{\text{def}}{=} \exp(-\log M \cdot \mathsf{poly}(n,m)).$$

Apply Algorithm 5 with inputs $\vec{\mathbf{A}}, \vec{\mathbf{B}}, \varepsilon$.
**if** Algorithm 5 returns **No then**
  **return** "**No**, $\overline{\mathcal{O}}_{\vec{\mathbf{A}}} \cap \overline{\mathcal{O}}_{\vec{\mathbf{B}}} = \varnothing$"
**else**
  $\mathbf{U}, \mathbf{V} \leftarrow$ the solution that Algorithm 5 returns.
  **if** $|\det(\mathbf{UV}) - 1| \le (\varepsilon')^{1/3}$ **then return Yes else return No**.
 ◇ *parameter $\varepsilon'$ is from Theorem M3*
**end if**

- Suppose $\overline{\mathcal{O}}_{\vec{\mathbf{A}}} \cap \overline{\mathcal{O}}_{\vec{\mathbf{B}}} = \varnothing$.

  If Algorithm 5 returns **No**, then we output "**No**, $\overline{\mathcal{O}}_{\vec{\mathbf{A}}} \cap \overline{\mathcal{O}}_{\vec{\mathbf{B}}} = \varnothing$" as desired.

  Otherwise, let $\mathbf{U}, \mathbf{V} \in \mathsf{U}(n)$ be the output of Algorithm 5 satisfying

$$\left\|\mathbf{U}\vec{\mathbf{A}}'\mathbf{V} - \vec{\mathbf{B}}'\right\|_2 \le \varepsilon' < (\varepsilon')^{1/3}.$$

  Invoking Lemma 7.1a, we know that $|\det(\mathbf{UV}) - 1|$ must be larger than $(\varepsilon')^{1/3}$ and therefore our algorithm will correctly return "**No**, $\overline{\mathcal{O}}_{\vec{\mathbf{A}}} \cap \overline{\mathcal{O}}_{\vec{\mathbf{B}}} = \varnothing$." □

# Appendix

## A Missing Proofs for Section 3

Before that we will need two lemmas that connect the capacity to the distance measure $\mathsf{ds}$ in Definition 3.7. The first one shows that: small distance measure implies "large" capacity.[23]

**Lemma A.1** (Lemma 3.2 in [34], Theorem 3.5.15 in [59]). *For any positive operator $T$,*

$$\mathsf{cap}(T) \ge \left(\frac{\mathrm{Tr}[T(\mathbf{I})]}{n}\right)^n \cdot \left(1 - \sqrt{\frac{n \cdot \mathsf{ds}(T)}{2}}\right)^n \quad \text{and} \quad \widetilde{\mathsf{cap}}(T) \ge \mathrm{Tr}[T(\mathbf{I})] \cdot \left(1 - \sqrt{\frac{n \cdot \mathsf{ds}(T)}{2}}\right)$$

[23] Garg et al. [34] proved it for the special case when $T(\mathbf{I}) = \mathbf{I}$ and Kwok et al. [59] proved it in the general case with better parameters (but with essentially the same proof).

The next lemma shows the other direction: "large" capacity implies small distance measure.[24]

**Lemma A.2** (Lemma 3.5 in [34]). *Let* $T : \mathsf{Mat}_n(\mathbb{C}) \to \mathsf{Mat}_n(\mathbb{C})$ *be any positive operator with* $\mathrm{Tr}[T(\mathbf{I})] = n$. *If* $\mathsf{cap}(T) \geq \exp(-\delta)$ *for some* $\delta \leq 1/6$, *then* $\mathsf{ds}(T) \leq 6\delta$. *Similarly —using Proposition 3.6 to relate* $\mathsf{cap}$ *to* $\widetilde{\mathsf{cap}}$— *if* $\widetilde{\mathsf{cap}}(T) \geq n - \frac{1}{12}$, *then* $\mathsf{ds}(T) \leq 12\,(n - \widetilde{\mathsf{cap}}(T))$.

We will also need the following lower bound on capacity from [34, Theorem 2.18].

**Lemma A.3** ([34]). *Let* $\mathbf{A}_1, \ldots, \mathbf{A}_m \in \mathsf{Mat}_n(\mathbb{Z}[i])$ *be such that* $T_{\overrightarrow{\mathbf{A}}}$ *defined by* $\overrightarrow{\mathbf{A}} = (\mathbf{A}_1, \ldots, \mathbf{A}_m)$ *satisfies* $\mathsf{cap}(T_{\overrightarrow{\mathbf{A}}}) > 0$. *Then* $\mathsf{cap}(T_{\overrightarrow{\mathbf{A}}}) \geq \exp(-2n\log(n))$. *This implies that* $\widetilde{\mathsf{cap}}(T_{\overrightarrow{\mathbf{A}}}) \geq 1/n$.

The following elementary fact discusses the effect of scaling on capacity.

**Fact A.4** ([34, 39]). *Let* $T$ *be the operator defined by* $\mathbf{A}_1, \ldots, \mathbf{A}_m$ *and let* $T_{\mathbf{C},\mathbf{D}}$ *be the operator defined by* $\mathbf{C}\mathbf{A}_1\mathbf{D}^\dagger, \ldots, \mathbf{C}\mathbf{A}_m\mathbf{D}^\dagger$, *where* $\mathbf{C}, \mathbf{D}$ *are invertible matrices. Then*

$$\mathsf{cap}(T_{\mathbf{C},\mathbf{D}}) = |\det(\mathbf{C})|^2 |\det(\mathbf{D})|^2\, \mathsf{cap}(T)$$

The following theorem was proven in [59].[25] It says if $\mathsf{db}(T_{\overrightarrow{\mathbf{B}}})$ is small, then $\overrightarrow{\mathbf{B}}$ is close to the element of minimum norm in the orbit-closure of $\overrightarrow{\mathbf{B}}$. The notion $\Delta_{SU}$ is in Definition 3.1.

**Theorem A.5** (Theorem 3.3.5 in [59]). *Let* $\overrightarrow{\mathbf{A}} = (\mathbf{A}_1, \ldots, \mathbf{A}_m)$ *be a tuple in* $\mathsf{Mat}_n(\mathbb{C})^m$. *Then there exists a tuple,* $\overrightarrow{\mathbf{A}}' = (\mathbf{A}'_1, \ldots, \mathbf{A}'_m)$, *of minimum norm in* $\overline{\mathcal{O}}_{\overrightarrow{\mathbf{A}}}$ *s.t.*

$$\sum_{i=1}^{m} \|\mathbf{A}_i - \mathbf{A}'_i\|_F^2 \leq n^{5/2}\sqrt{\mathsf{db}(T_{\overrightarrow{\mathbf{A}}})}\ .$$

*Therefore, if* $\overrightarrow{\mathbf{A}}' = (\mathbf{A}'_1, \ldots, \mathbf{A}'_m)$ *is an arbitrary tuple of minimum norm in* $\overline{\mathcal{O}}_{\overrightarrow{\mathbf{A}}}$, *then we have:*

$$\Delta_{SU}(\overrightarrow{\mathbf{A}}, \overrightarrow{\mathbf{A}}') \leq n^{5/2}\sqrt{\mathsf{db}(T_{\overrightarrow{\mathbf{A}}})}\ .$$

## A.1 Proof of Theorem 3.14

*Proof of Theorem 3.14.*

- ($\varepsilon$-scaling 1 $\Rightarrow$ $\varepsilon'$-scaling 2)

  By Lemma A.1, $\mathsf{cap}(T') \geq \left(1 - \sqrt{\frac{n\varepsilon}{2}}\right)^n$. Suppose $(\mathbf{C}, \mathbf{D})$ scale $T_{\overrightarrow{\mathbf{A}}}$ to $T'$. Then $T'(\mathbf{I}) = \mathbf{I}$ is equivalent to $T_{\overrightarrow{\mathbf{A}}}\left(\mathbf{D}^\dagger\mathbf{D}\right) = \left(\mathbf{C}^\dagger\mathbf{C}\right)^{-1}$. We have

  $$\left(1 - \sqrt{\frac{n\varepsilon}{2}}\right)^n \leq \mathsf{cap}(T') \stackrel{①}{=} |\det(\mathbf{C})|^2 |\det(\mathbf{D})|^2\, \mathsf{cap}(T_{\overrightarrow{\mathbf{A}}}) = \det\left(\mathbf{C}^\dagger\mathbf{C}\right)\det\left(\mathbf{D}^\dagger\mathbf{D}\right)\mathsf{cap}(T_{\overrightarrow{\mathbf{A}}})\ ,$$

  where equality ① is by Fact A.4. Rearranging and substituting $\mathbf{X} = \mathbf{D}^\dagger\mathbf{D}$ and $\log\mathsf{cap}(\mathbf{X}) = \log\det(T_{\overrightarrow{\mathbf{A}}}(\mathbf{X})) - \log\det(\mathbf{X}) = -\log\det(\mathbf{C}^\dagger\mathbf{C}) - \log\det(\mathbf{D}^\dagger\mathbf{D})$, we get that $\log\mathsf{cap}(\mathbf{X}) \leq \log\mathsf{cap}(T_{\overrightarrow{\mathbf{A}}}) + \varepsilon'$, for $\varepsilon' = -n\log(1 - \sqrt{n\varepsilon/2})$.

[24] This lemma is not stated exactly as follows in [34]; however, it is not hard to derive the following lemma from the proof of [34, Lemma 3.5]).

[25] Kwok et al. [59] proved this theorem by analyzing the limiting points of the gradient flow. Our formulation of their theorem can be obtained by noting that the gradient flow converges to points of minimum norm in the orbit-closure and that all points of minimum norm are related by unitary matrices by the Kempf-Ness Theorem 1.3. We independently discovered the theorem but decide to directly invoke theirs to make our paper shorter.

- ($\varepsilon$-scaling 1 $\Leftarrow$ $\varepsilon'$-scaling 2)

  We know $\mathsf{cap}(\mathbf{X}) \leq \exp(\varepsilon')\,\mathsf{cap}(T_{\overrightarrow{\mathbf{A}}})$. Define an operator $T'$ as follows: $T'(P) = T_{\overrightarrow{\mathbf{A}}}(\mathbf{X})^{-1/2} \cdot T_{\overrightarrow{\mathbf{A}}}(\mathbf{X}^{1/2} P \mathbf{X}^{1/2}) \cdot T_{\overrightarrow{\mathbf{A}}}(\mathbf{X})^{-1/2}$. Then

$$\begin{aligned}\mathsf{cap}(T') &= \det(T_{\overrightarrow{\mathbf{A}}}(\mathbf{X}))^{-1} \det(\mathbf{X})\,\mathsf{cap}(T_{\overrightarrow{\mathbf{A}}}) \\ &= \mathsf{cap}(\mathbf{X})^{-1}\,\mathsf{cap}(T_{\overrightarrow{\mathbf{A}}}) \\ &\geq \exp(\varepsilon')\end{aligned}$$

  Then by Lemma A.2, $\mathsf{ds}(T') \leq 6\varepsilon'$.

- ($\varepsilon$-scaling 1 $\Rightarrow$ $\varepsilon'$-scaling 3)

  We first observe that $\mathsf{cap}(T_{\overrightarrow{\mathbf{A}}}) \leq c_2 \stackrel{\text{def}}{=} \mathsf{poly}(n,m,M)^n$ by the definition of capacity. Since $1 \approx \mathsf{cap}(T') \geq 1/2$, we have that (using Fact A.4)

$$\frac{1}{|\det(\mathbf{C})|^2 |\det(\mathbf{D})|^2} \leq 2c_2$$

  Define

$$\mathbf{A}'_i = \frac{1}{\det(\mathbf{C})^{1/n}\det(\mathbf{D}^\dagger)^{1/n}} \mathbf{C}\mathbf{A}_i\mathbf{D}^\dagger$$

  Then the tuple $\overrightarrow{\mathbf{A}}'$ is in the orbit of $\overrightarrow{\mathbf{A}}$. Then

$$\begin{aligned}\mathsf{db}(T_{\overrightarrow{\mathbf{A}}'}) &= \frac{1}{|\det(\mathbf{C})|^{4/n}|\det(\mathbf{D})|^{4/n}}\mathsf{db}(T') \\ &\leq (2c_2)^{2/n}\mathsf{db}(T') \\ &= (2c_2)^{2/n}\mathsf{ds}(T') \\ &\leq (2c_2)^{2/n} 6\varepsilon\end{aligned}$$

  $\mathsf{db}(T') = \mathsf{ds}(T')$ since $T'(\mathbf{I}) = \mathbf{I}$. Then applying Theorem A.5 completes the proof.

- ($\varepsilon$-scaling 1 $\Leftarrow$ $\varepsilon'$-scaling 3):

  Recall $\mathsf{cap}(T_{\overrightarrow{\mathbf{A}}}) \geq c_1 = \exp(-2n\log(n))$ from Lemma A.3. Let $\overrightarrow{\mathbf{A}}^*$ be a minimum norm tuple in the orbit closure of $\overrightarrow{\mathbf{A}}$. We are given that

$$\Delta_{SU}(\overrightarrow{\mathbf{A}}', \overrightarrow{\mathbf{A}}^*) \leq \varepsilon'$$

  By the Kempf-Ness theorem (Theorem 1.3), we know that $\mathsf{db}(T_{\overrightarrow{\mathbf{A}}^*}) = 0$. We might as well just assume (by acting upon $\overrightarrow{\mathbf{A}}^*$ by appropriate unitaries)

$$\sum_{i=1}^m \|\mathbf{A}'_i - \mathbf{A}^*_i\|_F^2 \leq \varepsilon'$$

  It can then be verified that $\mathsf{db}(T_{\overrightarrow{\mathbf{A}}'}) \leq \delta$ for $\delta = \mathsf{poly}(n,m,M,\varepsilon')$.[26] Define $T''$ as follows:

$$T'' = \frac{n}{\mathrm{Tr}[T_{\overrightarrow{\mathbf{A}}'}(\mathbf{I})]} T_{\overrightarrow{\mathbf{A}}'}$$

[26] Note that we have $\mathsf{poly}(M)$ dependency for the following reason. We have $\|\overrightarrow{\mathbf{A}}^*\|_2 \leq \|\overrightarrow{\mathbf{A}}\|_2$ because $\overrightarrow{\mathbf{A}}^*$ is of the minimum norm. This implies $\|\overrightarrow{\mathbf{A}}^*\|_\infty \leq \mathsf{poly}(n,m,M)$.

Then

$$\mathsf{ds}(T'') = \left(\frac{n}{\mathrm{Tr}[T_{\overrightarrow{\mathbf{A}}'}(\mathbf{I})]}\right)^2 \mathsf{db}(T_{\overrightarrow{\mathbf{A}}'}) \le \left(\frac{n}{\mathrm{Tr}[T_{\overrightarrow{\mathbf{A}}'}(\mathbf{I})]}\right)^2 \delta$$

$\mathrm{Tr}[T_{\overrightarrow{\mathbf{A}}'}(\mathbf{I})]$ can be lower bounded as follows.

$$\mathrm{Tr}[T_{\overrightarrow{\mathbf{A}}'}(\mathbf{I})] \approx \mathrm{Tr}[T_{\overrightarrow{\mathbf{A}}^*}(\mathbf{I})] = N(\overrightarrow{\mathbf{A}}) = n\,\mathsf{cap}(T_{\overrightarrow{\mathbf{A}}})^{1/n} \ge nc_1^{1/n}$$

One does not get $T''(\mathbf{I}) = \mathbf{I}$ but this is easy to achieve by defining

$$T'(\mathbf{P}) = T''(\mathbf{I})^{-1/2} T''(\mathbf{P}) T''(\mathbf{I})^{-1/2}$$

It is not so hard to verify that this operation does not blow up $\mathsf{ds}(T')$ too much. Also note that the $T'$ is a scaling of $T_{\overrightarrow{\mathbf{A}}'}$ by $\left(T_{\overrightarrow{\mathbf{A}}'}(\mathbf{I})^{-1/2}, \mathbf{I}\right)$. □

# B Missing Proofs for Section 4

Before proving Theorem 4.3, we state an elementary proposition that says if a univariate convex function $f(t)$ satisfies $|f'''(t)| \le 2f''(t)$, then it can be well approximated by its second-order Taylor expansion at $|t| \le 1/2$.

**Proposition B.1** (quadratic approximation)**.** *For every $\rho > 0$ and convex function $f : \mathbb{R} \to \mathbb{R}$ satisfying $|f'''(s)| \le \rho f''(s)$ for every $s \in \mathbb{R}$, we have*

$$\forall |s| \le \tfrac{1}{\rho}: \quad f(0) + f'(0)s + \tfrac{1}{2e} f''(0)s^2 \le f(s) \le f(0) + f'(0)s + \tfrac{e}{2} f''(0)s^2$$

*Proof.* We first show $f''(0)e^{-\rho t} \le f''(t) \le f''(0)e^{\rho t}$. To see this, denoting by $h(t) = f''(t)$, we have $|[\log h(t)]'| = \left|\frac{h'(t)}{h(t)}\right| \le \rho$. Therefore,

$$\log h(0) - \rho|t| \le \log h(t) \le \log h(0) + \rho|t|$$

or equivalently

$$h(0)e^{-\rho|t|} \le h(t) \le h(0)e^{\rho|t|} \ .$$

By Taylor expansion, we have that there exists $\xi \in [0,t]$ (or $[t,0]$) such that

$$f(0) + f'(0)t + \frac{e^{-\rho|t|}}{2} f''(0)t^2 \le f(t) = f(0) + f'(0)t + \frac{1}{2} f''(\xi)t^2 \le f(0) + f'(0)t + \frac{e^{\rho|t|}}{2} f''(0)t^2 \ .$$

Using $|t| \le \frac{1}{\rho}$, we complete the proof. □

## B.1 Proof of Theorem 4.3

*Proof of Theorem 4.3.* We prove by induction on $t \ge 0$ the following (which are clearly sufficient):

- $F(\mathbf{X}_{t+1}) \le F(\mathbf{I})$.
- $F(\mathbf{X}_{t+1}) - F(\mathbf{X}^*) \le \left(1 - \frac{1}{2e^2 R}\right)^t (F(\mathbf{I}) - F(\mathbf{X}^*))$.

Suppose the statements hold for all iterations until $t-1$, then we have $F(\mathbf{X}_t) \le F(\mathbf{I})$ and therefore by the assumption in the theorem statement, we have $\left\|\log\left(\mathbf{X}_t^{-1/2}\mathbf{X}^*\mathbf{X}_t^{-1/2}\right)\right\|_2 \le R$.

Suppose $\mathbf{\Delta}^* = \log\left(\mathbf{X}_t^{-1/2}\mathbf{X}^*\mathbf{X}_t^{-1/2}\right)$. Note that $f^t(\mathbf{\Delta}^*) = F(\mathbf{X}^*)$.

Denoting by $\rho = 2R$ and $\mathbf{\Delta}_t' = \frac{\mathbf{\Delta}^*}{\rho}$, we have $\|\mathbf{\Delta}_t'\|_2 \le 1/2$ and hence $\mathbf{\Delta}_t'$ satisfies the constraint $\|\mathbf{\Delta}_t\|_2 \le 1/2$. Now, let us denote by $h(s) \stackrel{\text{def}}{=} f^t(s\mathbf{\Delta}^*)$, which is a univariate function over $s \in \mathbb{R}$.

We know that $h(s)$ is convex because $F$ is g-convex and the convexity of $h(s)$ gives

$$f^t(0) - f^t(\mathbf{\Delta}_t') = h(0) - h\left(\frac{1}{\rho}\right) \geq \frac{1}{\rho}\big(h(0) - h(1)\big) = \frac{1}{\rho}\left(f^t(0) - f^t(\mathbf{\Delta}^*)\right) \tag{B.1}$$

By the g-second-order robustness of $F$ and Proposition B.1, we have that for any $\mathbf{\Delta}$ s.t. $\|\mathbf{\Delta}\|_2 \leq 1/2$,

$$\mathrm{Tr}\left(\nabla f^t(0) \cdot \mathbf{\Delta}\right) + \frac{1}{2e}\mathrm{Tr}\left(\nabla^2 f^t(0) \cdot \mathbf{\Delta} \otimes \mathbf{\Delta}\right) \leq f^t(\mathbf{\Delta}) - f^t(0) \tag{B.2}$$

$$\mathrm{Tr}\left(\nabla f^t(0) \cdot \mathbf{\Delta}\right) + \frac{e}{2}\mathrm{Tr}\left(\nabla^2 f^t(0) \cdot \mathbf{\Delta} \otimes \mathbf{\Delta}\right) \geq f^t(\mathbf{\Delta}) - f^t(0) \tag{B.3}$$

On one hand, we have

$$\begin{aligned}
&\mathrm{Tr}\left(\nabla f^t(0) \cdot \mathbf{\Delta}_t\right) + \frac{1}{2e}\mathrm{Tr}\left(\nabla^2 f^t(0) \cdot \mathbf{\Delta}_t \otimes \mathbf{\Delta}_t\right) \\
&\overset{①}{\leq} \mathrm{Tr}\left(\nabla f^t(0) \cdot \mathbf{\Delta}_t'\right) + \frac{1}{2e}\mathrm{Tr}\left(\nabla^2 f^t(0) \cdot \mathbf{\Delta}_t' \otimes \mathbf{\Delta}_t'\right) \\
&\overset{②}{\leq} -\big(f^t(0) - f^t(\mathbf{\Delta}_t')\big) \overset{③}{\leq} -\frac{1}{\rho}\left(f^t(0) - f^t(\mathbf{\Delta}^*)\right) = -\frac{1}{\rho}\left(F(\mathbf{X}_t) - F(\mathbf{X}^*)\right) \ .
\end{aligned} \tag{B.4}$$

Above, ① is by the optimality of $\mathbf{\Delta}_t$; ② is by (B.2) (and the fact that $\left\|\mathbf{\Delta}_t'\right\|_2 \leq 1/2$); and ③ is by (B.1).

On the other hand, using (B.3) (and the fact that $\|\mathbf{\Delta}_t/e^2\|_2 \leq 1/2$) we have

$$\begin{aligned}
&\left(\mathrm{Tr}\left(\nabla f^t(0) \cdot \mathbf{\Delta}_t\right) + \frac{1}{2e}\mathrm{Tr}\left(\nabla^2 f^t(0) \cdot \mathbf{\Delta}_t \otimes \mathbf{\Delta}_t\right)\right) \\
&= e^2\left(\mathrm{Tr}\left(\nabla f^t(0) \cdot \frac{\mathbf{\Delta}_t}{e^2}\right) + \frac{e}{2}\mathrm{Tr}\left(\nabla^2 f^t(0) \cdot \frac{\mathbf{\Delta}_t}{e^2} \otimes \frac{\mathbf{\Delta}_t}{e^2}\right)\right) \\
&\geq -e^2\left(f^t(0) - f^t\left(\frac{\mathbf{\Delta}_t}{e^2}\right)\right) = -e^2\left(F(\mathbf{X}_t) - F(\mathbf{X}_{t+1})\right) \ .
\end{aligned} \tag{B.5}$$

Since the left hand side of (B.5) is always non-positive by the definition of $\mathbf{\Delta}_t$, we immediately have $F(\mathbf{X}_t) \geq F(\mathbf{X}_{t+1})$ and this proves the first item.

As for the second item, we combine (B.4) and (B.5):

$$\left(F(\mathbf{X}_t) - F(\mathbf{X}_{t+1})\right) \geq \frac{1}{e^2\rho}\left(F(\mathbf{X}_t) - F(\mathbf{X}^*)\right)$$

which after rearranging gives

$$F(\mathbf{X}_{t+1}) - F(\mathbf{X}^*) \leq \left(1 - \frac{1}{2e^2 R}\right)\left(F(\mathbf{X}_t) - F(\mathbf{X}^*)\right) \ . \qquad \square$$

# C Missing Proofs for Section 5

This section is devoted to the full proof of Theorem 5.3. To make our proof self-contained, we do not reply on Section 4.

- In Section C.1, we prove that the function $F(\cdot)$ we introduced in (5.1) is g-convex and g-second-order robust.
- In Section C.2, we prove Theorem 5.3 assuming $\mathbf{\Delta}_t$ and $\mathbf{X}_t$ are calculated exactly.
- In Section C.3, we prove Claim 5.2 which gives an explicit quadratic formula for calculating $\mathbf{\Delta}_t$ in Line 4 of Algorithm 2.
- In Section C.4, we discuss how to implement Algorithm 2 in finite arithmetics.

### C.1 Geodesic properties of our objective

Given any $\mathbf{X} \in \mathsf{GL}_n(\mathbb{C})$ and Hermitian matrix $\mathbf{\Delta} \in \mathsf{Mat}_n(\mathbb{C})$, consider the following univariate function $g\colon \mathbb{R} \to \mathbb{R}$ which captures the behavior of $\mathsf{logcap}(\mathbf{X})$ in the direction of $\mathbf{X}e^{t\mathbf{\Delta}/2}$:

$$g(t) \stackrel{\text{def}}{=} \log\det\left(\textstyle\sum_{i=1}^m \mathbf{A}_i \mathbf{X} e^{t\mathbf{\Delta}} \mathbf{X}^\dagger \mathbf{A}_i^\dagger\right) - \log\det(\mathbf{X}e^{t\mathbf{\Delta}}\mathbf{X}^\dagger) = \mathsf{logcap}(\mathbf{X}e^{t\mathbf{\Delta}/2}) \ .$$

**Lemma C.1.** *We have the following properties of $g$:*

*(a) $g(t)$ is convex over $t \in \mathbb{R}$.*

*(b) If $\|\mathbf{\Delta}\|_2 \le 1$ then we have $\left|\frac{d^3 g}{dt^3}\right| \le 2\frac{d^2 g}{dt^2}$.*

Above, Lemma C.1.a should not be surprising because it is equivalent to the fact $G(\mathbf{M}) = \log\det(\sum_i \mathbf{A}_i \mathbf{M} \mathbf{A}_i^\dagger)$ is geodesically convex over the Riemannian manifold of positive definite matrices [40] (see Section 1.1). In contrast, Lemma C.1.b is a special property that we show for the log capacity function, and is the main property that allows us to build an optimization algorithm with time complexity polynomial in $\log(\varepsilon^{-1})$.

Given any $\mathbf{X} \in \mathsf{GL}_n(\mathbb{C})$ and Hermitian matrix $\mathbf{\Delta} \in \mathsf{Mat}_n(\mathbb{C})$, we also consider the following univariate function $r\colon \mathbb{R} \to \mathbb{R}$ which captures the behavior of $\mathsf{reg}(\mathbf{X})$ in the direction of $\mathbf{X}e^{t\mathbf{\Delta}/2}$:

$$r(t) = \mathrm{Tr}(\mathbf{X}e^{t\mathbf{\Delta}}\mathbf{X}^\dagger) \cdot \mathrm{Tr}((\mathbf{X}e^{t\mathbf{\Delta}}\mathbf{X}^\dagger)^{-1}) = \mathsf{reg}(\mathbf{X}e^{t\mathbf{\Delta}/2}) \ .$$

**Lemma C.2.** *We have the following properties about $r(t)$:*

*(a) $r(t)$ is convex over $t \in \mathbb{R}$.*

*(b) For $\|\mathbf{\Delta}\|_2 \le 1$, $\left|\frac{d^3 r}{dt^3}\right| \le 2\frac{d^2 r}{dt^2}$.*

#### C.1.1 Proof of Lemma C.1

*Proof of Lemma C.1.* The proof is by careful manipulations of matrix algebra. We first show that it suffices to prove the statements for $t = 0$. Indeed, for every $t_0 \in \mathbb{R}$, we have

$$g(t) = \log\det\left(\sum_{i=1}^m \mathbf{A}_i \left(\mathbf{X}e^{t_0\mathbf{\Delta}/2}\right) e^{(t-t_0)\mathbf{\Delta}} \left(e^{t_0\mathbf{\Delta}/2}\mathbf{X}^\dagger\right) \mathbf{A}_i^\dagger\right) - \log\det\left(\left(\mathbf{X}e^{t_0\mathbf{\Delta}/2}\right) e^{(t-t_0)\mathbf{\Delta}} \left(e^{t_0\mathbf{\Delta}/2}\mathbf{X}^\dagger\right)\right) \ .$$

Therefore, if we treat $\left(\mathbf{X}e^{t_0\mathbf{\Delta}/2}\right)$ as the new $\mathbf{X}$, we can replace $t$ with $t - t_0$. Therefore, we only consider $t = 0$ for the remainder of the proof. For notation simplicity, let

$$\mathbf{B}_t = \textstyle\sum_{i=1}^m \mathbf{A}_i \mathbf{X}\big(e^{t\mathbf{\Delta}}\big)\mathbf{X}^\dagger \mathbf{A}_i^\dagger \qquad \mathbf{C}_t = \sum_{i=1}^m \mathbf{A}_i \mathbf{X}\big(\mathbf{\Delta}e^{t\mathbf{\Delta}}\big)\mathbf{X}^\dagger \mathbf{A}_i^\dagger$$
$$\mathbf{D}_t = \textstyle\sum_{i=1}^m \mathbf{A}_i \mathbf{X}\big(\mathbf{\Delta}^2 e^{t\mathbf{\Delta}}\big)\mathbf{X}^\dagger \mathbf{A}_i^\dagger \qquad \mathbf{E}_t = \sum_{i=1}^m \mathbf{A}_i \mathbf{X}\big(\mathbf{\Delta}^3 e^{t\mathbf{\Delta}}\big)\mathbf{X}^\dagger \mathbf{A}_i^\dagger \ .$$

We first calculate the first-order derivative:

$$\frac{dg}{dt} = \frac{d\log\det\mathbf{B}_t}{dt} - \frac{d\log\det(e^{t\mathbf{\Delta}})}{dt} = \mathrm{Tr}\left(\mathbf{B}_t^{-1}\frac{d\mathbf{B}_t}{dt}\right) - \mathrm{Tr}(\mathbf{\Delta}) = \mathrm{Tr}\left(\mathbf{B}_t^{-1}\mathbf{C}_t\right) - \mathrm{Tr}(\mathbf{\Delta}) \tag{C.1}$$

We next calculate the second-order derivative:

$$\begin{aligned}\frac{d^2 g}{dt^2} &= \mathrm{Tr}\left(\mathbf{C}_t \frac{d}{dt}\mathbf{B}_t^{-1}\right) + \mathrm{Tr}\left(\mathbf{B}_t^{-1}\frac{d}{dt}\mathbf{C}_t\right) = -\mathrm{Tr}\left(\mathbf{C}_t\mathbf{B}_t^{-1}\left(\frac{d}{dt}\mathbf{B}_t\right)\mathbf{B}_t^{-1}\right) + \mathrm{Tr}\left(\mathbf{B}_t^{-1}\mathbf{D}_t\right) \\ &= -\mathrm{Tr}(\mathbf{C}_t\mathbf{B}_t^{-1}\mathbf{C}_t\mathbf{B}_t^{-1}) + \mathrm{Tr}\left(\mathbf{B}_t^{-1}\mathbf{D}_t\right) \ .\end{aligned}$$

If we denote by $\widetilde{\mathbf{A}}_i \stackrel{\text{def}}{=} \left(\sum_{i=1}^m \mathbf{A}_i\mathbf{X}\mathbf{X}^\dagger\mathbf{A}_i^\dagger\right)^{-1/2}\mathbf{A}_i\mathbf{X} = \mathbf{B}_0^{-1/2}\mathbf{A}_i\mathbf{X}$, we have $\sum_{i=1}^m \widetilde{\mathbf{A}}_i\widetilde{\mathbf{A}}_i^\dagger = \mathbf{I}$ and

$$
\begin{aligned}
\left.\frac{d^2g}{dt^2}\right|_{t=0} &= -\mathrm{Tr}\left(\left(\sum_{i=1}^m \widetilde{\mathbf{A}}_i\mathbf{\Delta}\widetilde{\mathbf{A}_i}^\dagger\right)^2\right) + \mathrm{Tr}\left(\sum_{i=1}^m \widetilde{\mathbf{A}}_i\mathbf{\Delta}^2\widetilde{\mathbf{A}_i}^\dagger\right)\\
&= -\mathrm{Tr}\left(\left(\sum_{i=1}^m \widetilde{\mathbf{A}}_i\mathbf{\Delta}\widetilde{\mathbf{A}_i}^\dagger\right)^2\right) + \mathrm{Tr}\left(\left(\sum_{i=1}^m \widetilde{\mathbf{A}}_i\widetilde{\mathbf{A}}_i^\dagger\right)\left(\sum_{i=1}^m \widetilde{\mathbf{A}}_i\mathbf{\Delta}^2\widetilde{\mathbf{A}_i}^\dagger\right)\right)\\
&= \frac{1}{2}\sum_{i,j=1}^m \mathrm{Tr}\left(\left(\widetilde{\mathbf{A}}_j^\dagger\widetilde{\mathbf{A}}_i\mathbf{\Delta} - \mathbf{\Delta}\widetilde{\mathbf{A}}_j^\dagger\widetilde{\mathbf{A}}_i\right)\left(\mathbf{\Delta}\widetilde{\mathbf{A}}_i^\dagger\widetilde{\mathbf{A}}_j - \widetilde{\mathbf{A}}_i^\dagger\widetilde{\mathbf{A}}_j\mathbf{\Delta}\right)\right)\\
&= \frac{1}{2}\sum_{i,j=1}^m \mathrm{Tr}\left((\mathbf{P}_{i,j}-\mathbf{Q}_{i,j})(\mathbf{P}_{i,j}-\mathbf{Q}_{i,j})^\dagger\right) \geq 0 \ . \qquad \text{(C.2)}
\end{aligned}
$$

Above, we have denoted by $\mathbf{P}_{i,j} \stackrel{\text{def}}{=} \widetilde{\mathbf{A}}_j^\dagger\widetilde{\mathbf{A}}_i\mathbf{\Delta}$ and $\mathbf{Q}_{i,j} \stackrel{\text{def}}{=} \mathbf{\Delta}\widetilde{\mathbf{A}}_j^\dagger\widetilde{\mathbf{A}}_i$. Since $g''(0) \geq 0$ we have finished the proof of the convexity of $g(\cdot)$. As for the third-order derivative, we have

$$
\begin{aligned}
\left.\frac{d^3g}{dt^3}\right|_{t=0} &= \left.\left(-3\mathrm{Tr}(\mathbf{D}_t\mathbf{B}_t^{-1}\mathbf{C}_t\mathbf{B}_t^{-1}) + 2\mathrm{Tr}(\mathbf{C}_t\mathbf{B}_t^{-1}\mathbf{C}_t\mathbf{B}_t^{-1}\mathbf{C}_t\mathbf{B}_t^{-1}) + \mathrm{Tr}(\mathbf{B}_t^{-1}\mathbf{E}_t)\right)\right|_{t=0}\\
&= 2\mathrm{Tr}\left(\left(\sum_{i=1}^m \widetilde{\mathbf{A}}_i\mathbf{\Delta}\widetilde{\mathbf{A}_i}^\dagger\right)^3\right) - 2\mathrm{Tr}\left(\left(\sum_{i=1}^m \widetilde{\mathbf{A}}_i\mathbf{\Delta}^2\widetilde{\mathbf{A}_i}^\dagger\right)\left(\sum_{i=1}^m \widetilde{\mathbf{A}}_i\mathbf{\Delta}\widetilde{\mathbf{A}_i}^\dagger\right)\right)\\
&\quad + \mathrm{Tr}\left(\sum_{i=1}^m \widetilde{\mathbf{A}}_i\mathbf{\Delta}^3\widetilde{\mathbf{A}_i}^\dagger\right) - \mathrm{Tr}\left(\left(\sum_{i=1}^m \widetilde{\mathbf{A}}_i\mathbf{\Delta}^2\widetilde{\mathbf{A}_i}^\dagger\right)\left(\sum_{i=1}^m \widetilde{\mathbf{A}}_i\mathbf{\Delta}\widetilde{\mathbf{A}_i}^\dagger\right)\right)\\
&= -\sum_{i,j=1}^m \mathrm{Tr}\left(\left(\sum_{i=1}^m \widetilde{\mathbf{A}}_i\mathbf{\Delta}\widetilde{\mathbf{A}_i}^\dagger\right)(\mathbf{P}_{i,j}-\mathbf{Q}_{i,j})(\mathbf{P}_{i,j}-\mathbf{Q}_{i,j})^\dagger\right)\\
&\quad + \mathrm{Tr}\left(\sum_{i=1}^m \widetilde{\mathbf{A}}_i\mathbf{\Delta}^3\widetilde{\mathbf{A}_i}^\dagger\right) - \mathrm{Tr}\left(\left(\sum_{i=1}^m \widetilde{\mathbf{A}}_i\mathbf{\Delta}^2\widetilde{\mathbf{A}_i}^\dagger\right)\left(\sum_{i=1}^m \widetilde{\mathbf{A}}_i\mathbf{\Delta}\widetilde{\mathbf{A}_i}^\dagger\right)\right)\\
&= \sum_{i,j=1}^m \mathrm{Tr}\left((\mathbf{P}_{i,j}-\mathbf{Q}_{i,j})(\mathbf{P}_{i,j}-\mathbf{Q}_{i,j})^\dagger\left(\mathbf{\Delta} - \left(\sum_{i=1}^m \widetilde{\mathbf{A}}_i\mathbf{\Delta}\widetilde{\mathbf{A}_i}^\dagger\right)\right)\right) \ . \qquad \text{(C.3)}
\end{aligned}
$$

Finally, we know that as long as $\|\mathbf{\Delta}\|_2 \leq 1$, we have

$$
-2\mathbf{I} \preceq \mathbf{\Delta} - \mathbf{I} \preceq \mathbf{\Delta} - \left(\sum_{i=1}^m \widetilde{\mathbf{A}}_i\mathbf{\Delta}\widetilde{\mathbf{A}_i}^\dagger\right) \preceq \mathbf{\Delta} + \mathbf{I} \preceq 2\mathbf{I}
$$

and therefore combining (C.2) and (C.3) , we conclude that

$$
\left|\left.\frac{d^3g}{dt^3}\right|_{t=0}\right| \leq 2\left.\frac{d^2g}{dt^2}\right|_{t=0} \ . \qquad \square
$$

### C.1.2 Proof of Lemma C.2

*Proof of Lemma C.2.* Similar to the proof of Lemma C.1, we only need to prove the lemma at $t = 0$. For notational simplicity, let us denote by

$$
x_k = \mathrm{Tr}(\mathbf{X}\mathbf{\Delta}^k\mathbf{X}^\dagger) \quad \text{and} \quad y_k = \mathrm{Tr}(\mathbf{X}^{-1}(-\mathbf{\Delta})^k(\mathbf{X}^{-1})^\dagger) = \mathrm{Tr}((-\mathbf{\Delta})^k(\mathbf{X}\mathbf{X}^\dagger)^{-1}) \ ,
$$

We first calculate the first-order derivative:

$$\frac{dr}{dt} = \mathrm{Tr}(\mathbf{X}\boldsymbol{\Delta}e^{t\boldsymbol{\Delta}}\mathbf{X}^\dagger)\cdot\mathrm{Tr}((\mathbf{X}e^{t\boldsymbol{\Delta}}\mathbf{X}^\dagger)^{-1}) - \mathrm{Tr}(\mathbf{X}e^{t\boldsymbol{\Delta}}\mathbf{X}^\dagger)\cdot\mathrm{Tr}((\mathbf{X}e^{t\boldsymbol{\Delta}}\mathbf{X}^\dagger)^{-2}(\mathbf{X}\boldsymbol{\Delta}e^{t\boldsymbol{\Delta}}\mathbf{X}^\dagger)) \ .$$

Using the fact that $\mathbf{X}^\dagger(\mathbf{X}\mathbf{X}^\dagger)^{-2}\mathbf{X} = (\mathbf{X}^{-1})^\dagger\mathbf{X}^{-1}$, we immediately have

$$\frac{dr}{dt}\Big|_{t=0} = x_1y_0 + x_0y_1 \ . \tag{C.4}$$

Following similar arguments, we can also show

$$\frac{d^2r}{dt^2}\bigg|_{t=0} = x_2y_0 + 2x_1y_1 + x_0y_2 \quad \text{and} \quad \frac{d^3r}{dt^3}\bigg|_{t=0} = x_3y_0 + 3x_2y_1 + 3x_1y_2 + x_0y_3 \ . \tag{C.5}$$

Now, without loss of generality (by unitary transformation), let us assume $\boldsymbol{\Delta} = \mathsf{diag}(\sigma_1, \cdots \sigma_n)$ is diagonal. Also, let us denote by $h_1, \dots, h_n \in \mathbb{R}_{\geq 0}$ the diagonal entries of $\mathbf{X}^\dagger\mathbf{X}$, and by $h'_1, \dots, h'_n \in \mathbb{R}_{\geq 0}$ the diagonal entries of $(\mathbf{X}\mathbf{X}^\dagger)^{-1}$. Then, for the second-order derivative we have

$$\begin{aligned}
\frac{d^2r}{dt^2}\bigg|_{t=0} &= \mathrm{Tr}(\boldsymbol{\Delta}^2(\mathbf{X}^\dagger\mathbf{X}))\cdot\mathrm{Tr}((\mathbf{X}\mathbf{X}^\dagger)^{-1}) + \mathrm{Tr}(\boldsymbol{\Delta}^2(\mathbf{X}\mathbf{X}^\dagger)^{-1})\cdot\mathrm{Tr}(\mathbf{X}^\dagger\mathbf{X}) \\
&\quad - 2\mathrm{Tr}(\boldsymbol{\Delta}(\mathbf{X}^\dagger\mathbf{X}))\cdot\mathrm{Tr}(\boldsymbol{\Delta}(\mathbf{X}\mathbf{X}^\dagger)^{-1}) \\
&= \left(\sum_{i=1}^n \sigma_i^2 h_i\right)\left(\sum_{i=1}^n h'_i\right) + \left(\sum_{i=1}^n \sigma_i^2 h'_i\right)\left(\sum_{i=1}^n h_i\right) - 2\left(\sum_{i=1}^n \sigma_i h_i\right)\left(\sum_{i=1}^n \sigma_i h'_i\right) \\
&= \sum_{i,j=1}^n h_i h'_j(\sigma_i - \sigma_j)^2 \geq 0
\end{aligned} \tag{C.6}$$

This proves the convexity of $r(\cdot)$. As for the third-order derivative, we have

$$\begin{aligned}
\frac{d^3r}{dt^3}\bigg|_{t=0} &= \left(\sum_{i=1}^n \sigma_i^3 h_i\right)\left(\sum_{i=1}^n h'_i\right) + 3\left(\sum_{i=1}^n \sigma_i^2 h'_i\right)\left(\sum_{i=1}^n \sigma_i h_i\right) \\
&\quad - \left(\sum_{i=1}^n \sigma_i^3 h'_i\right)\left(\sum_{i=1}^n h_i\right) - 3\left(\sum_{i=1}^n \sigma_i^2 h_i\right)\left(\sum_{i=1}^n \sigma_i h'_i\right) \\
&= \sum_{i,j=1}^n h_i h'_j(\sigma_i^3 + 3\sigma_i\sigma_j^2 - 3\sigma_i^2\sigma_j - \sigma_j^3) = \sum_{i,j=1}^n h_i h'_j(\sigma_i - \sigma_j)^2(\sigma_i - \sigma_j) \ .
\end{aligned}$$

Therefore, as long as $\|\boldsymbol{\Delta}\|_2 \leq 1$, we have $|\sigma_i - \sigma_j| \leq 2$ so

$$\left|\frac{d^3r}{dt^3}\Big|_{t=0}\right| \leq 2\sum_{i,j=1}^n h_i h'_j(\sigma_i - \sigma_j)^2 = 2\frac{d^2r}{dt^2}\bigg|_{t=0} \ . \qquad \square$$

## C.2 Convergence analysis

Using the regularizer, we argue that as long as $F(\mathbf{X})$ is not much larger than the initial point $F(\mathbf{X}_0) = F(\mathbf{I})$ (which is a very mild assumption), the condition number of $\mathbf{X}$ must be polynomially bounded.

**Claim C.3.** *If $\mathbf{X}$ satisfies $F(\mathbf{X}) \leq F(\mathbf{I}) + \mathsf{poly}(n, m, M)$ and $\inf_{\mathbf{X}\in\mathsf{GL}_n(\mathbb{C})} \mathsf{logcap}(\mathbf{X}) > -\infty$, then*

$$\kappa(\mathbf{X}) \leq \kappa_0 \quad \textit{and} \quad \mathsf{reg}(\mathbf{X}) \leq \kappa_0$$

*where $\kappa_0 \stackrel{\text{def}}{=} \mathsf{poly}(n, m, M, \kappa(\mathbf{X}^*_\varepsilon), \varepsilon^{-1})$.*

Since both $\mathsf{logcap}(\mathbf{X})$ and $\mathsf{reg}(\mathbf{X})$ satisfy the property that the second-order derivative is bounded by the third-order one (see Lemma C.1.b and C.2.b), we can use Proposition B.1 to argue that in each iteration $t \geq 0$ of Algorithm 2, it suffices for us to consider the second-order Taylor approximation of $F(\mathbf{X}_t e^{\mathbf{\Delta}/2})$ up to radius $\|\mathbf{\Delta}\|_2 \leq \frac{1}{2}$. This allows us to decrease the objective, and can be summarized as the following lemma:

**Lemma C.4.** *If in each iteration $t \geq 0$, $\mathbf{\Delta}_t$ and $\mathbf{X}_{t+1}$ are calculated exactly, then*

- $F(\mathbf{X}_{t+1}) \leq F(\mathbf{X}_t)$*; and*
- $F(\mathbf{X}_{t+1}) - F(\mathbf{X}^*_\varepsilon) \leq \left(1 - \frac{1}{8e^2 \log \kappa_0}\right)(F(\mathbf{X}_t) - F(\mathbf{X}^*_\varepsilon))$.

Having set up all the technical claims, let us first provide a proof of our main theorem.

*Proof of Theorem 5.3.* We have

$$\mathsf{reg}(\mathbf{X}^*_\varepsilon) = \mathrm{Tr}(\mathbf{X}^*_\varepsilon(\mathbf{X}^*_\varepsilon)^\dagger) \cdot \mathrm{Tr}\big((\mathbf{X}^*_\varepsilon(\mathbf{X}^*_\varepsilon)^\dagger)^{-1}\big) \leq n^2 \frac{\lambda_{\max}(\mathbf{X}^*_\varepsilon(\mathbf{X}^*_\varepsilon)^\dagger)}{\lambda_{\min}(\mathbf{X}^*_\varepsilon(\mathbf{X}^*_\varepsilon)^\dagger)} = n^2(\kappa(\mathbf{X}^*_\varepsilon))^2$$

and therefore $F(\mathbf{X}^*_\varepsilon) \leq \log \mathsf{cap}(T_{\vec{\mathbf{A}}}) + \varepsilon + \lambda\, \mathsf{reg}(\mathbf{X}^*_\varepsilon) \leq \log \mathsf{cap}(T_{\vec{\mathbf{A}}}) + 2\varepsilon$.

Now, if in each iteration $\mathbf{\Delta}_t$ and $\mathbf{X}_{t+1}$ are computed exactly, then Lemma C.4 implies that in $T = O(\log \kappa_0 \cdot \log(nmM\varepsilon^{-1}))$ iterations, we have

$$F(\mathbf{X}_T) - F(\mathbf{X}^*_\varepsilon) \leq \left(1 - \frac{1}{8e^2 \log \kappa_0}\right)^T (F(\mathbf{I}) - F(\mathbf{X}^*_\varepsilon)) \leq \varepsilon$$

and thus

$$\mathsf{logcap}(\mathbf{X}_T) \leq F(\mathbf{X}_T) \leq F(\mathbf{X}^*_\varepsilon) + \varepsilon \leq \log \mathsf{cap}(T_{\vec{\mathbf{A}}}) + 3\varepsilon \ .$$

In Section C.4, we shall argue that even if matrices $\mathbf{\Delta}_t$ and $\mathbf{X}_t$ are calculated to bit complexity $\mathsf{poly}(n, m, \log M, \log \kappa(\mathbf{X}^*_\varepsilon), \log \frac{1}{\varepsilon})$ in each iteration $t$, we can still satisfy $F(\mathbf{X}_T) - F(\mathbf{X}^*_\varepsilon) \leq \varepsilon$. □

### C.2.1 Proof of Claim C.3

*Proof of Claim C.3.* We know

$$\mathsf{logcap}(\mathbf{I}) = \log\det(\sum_{i=1}^m \mathbf{A}_i \mathbf{A}_i^\dagger) \leq \mathsf{poly}(n, m, M)$$

as well as

$$-\,\mathsf{logcap}(\mathbf{X}) \leq -\inf_{\mathbf{X} \in \mathsf{GL}_n(\mathbb{C})} \mathsf{logcap}(\mathbf{X}) \leq \mathsf{poly}(n)$$

from Lemma A.3.Therefore, the inequality $F(\mathbf{X}) \leq F(\mathbf{I}) + \mathsf{poly}(n, m, M)$ implies:

$$\lambda\, \mathsf{reg}(\mathbf{X}) \leq \lambda\, \mathsf{reg}(\mathbf{I}) + \mathsf{logcap}(\mathbf{I}) - \mathsf{logcap}(\mathbf{X}) + \mathsf{poly}(n, m, M) \leq \lambda\, \mathsf{reg}(\mathbf{I}) + \mathsf{poly}(n, m, M)$$

Therefore, $\mathsf{reg}(\mathbf{X}) \leq \frac{1}{\lambda}\mathsf{poly}(n, m, M) + \mathsf{reg}(\mathbf{I}) \leq \mathsf{poly}(n, m, M, \kappa(\mathbf{X}^*_\varepsilon), \varepsilon^{-1})$.

By $(\kappa(\mathbf{X}))^2 = \big(\|\mathbf{X}\|_2^2 \|\mathbf{X}^{-1}\|_2^2\big) \leq \mathsf{reg}(\mathbf{X})$ we complete the proof. □

### C.2.2 Proof of Lemma C.4

*Proof of Lemma C.4.* We prove by induction on $t \geq 0$. Suppose the statements hold for all iterations until $t-1$, then we have $F(\mathbf{X}_t) \leq F(\mathbf{I})$ and therefore by Claim C.3, we have $\kappa(\mathbf{X}_t) \leq \kappa_0$. Also, by the definition of $\kappa_0$, we have $\kappa(\mathbf{X}^*_\varepsilon) \leq \kappa_0$.

Let Hermitian matrix $\mathbf{\Delta}_a \stackrel{\text{def}}{=} \log\big(\mathbf{X}_t^{-1}\mathbf{X}_\varepsilon^*(\mathbf{X}_\varepsilon^*)^\dagger(\mathbf{X}_t^{-1})^\dagger\big)$. Then we have:

$$\begin{aligned}\lambda_{\max}(\mathbf{\Delta}_a) - \lambda_{\min}(\mathbf{\Delta}_a) &= \log\kappa\left(\mathbf{X}_t^{-1}\mathbf{X}_\varepsilon^*(\mathbf{X}_\varepsilon^*)^\dagger\mathbf{X}_t^{-\dagger}\right) \\ &\le \log\kappa\left(\mathbf{X}_t\mathbf{X}_t^\dagger\right) + \log\kappa\left(\mathbf{X}_\varepsilon^*(\mathbf{X}_\varepsilon^*)^\dagger\right) \le 4\log\kappa_0 \ .\end{aligned}$$

Therefore, if we define $\mathbf{\Delta}_b = \mathbf{\Delta}_a - \lambda_{\min}(\mathbf{\Delta}_a)\mathbf{I}$, then we have: $\|\mathbf{\Delta}_b\|_2 \le 4\log\kappa_0$. Moreover, by definition,

$$\begin{aligned}f^t(\mathbf{\Delta}_b) &= \log\det\left(\sum_{i=1}^m \mathbf{A}_i\mathbf{X}_t e^{\mathbf{\Delta}_b}\mathbf{X}_t^\dagger\mathbf{A}_i^\dagger\right) - \log\det(\mathbf{X}_t e^{\mathbf{\Delta}_b}\mathbf{X}_t^\dagger) + \lambda\,\mathsf{reg}(\mathbf{X}_t e^{\mathbf{\Delta}_b/2}) \\ &= \log\det\left(\sum_{i=1}^m \mathbf{A}_i\frac{\mathbf{X}_\varepsilon^*(\mathbf{X}_\varepsilon^*)^\dagger}{e^{\lambda_{\min}(\mathbf{\Delta}_a)}}\mathbf{A}_i^\dagger\right) - \log\det\left(\frac{\mathbf{X}_\varepsilon^*(\mathbf{X}_\varepsilon^*)^\dagger}{e^{\lambda_{\min}(\mathbf{\Delta}_a)}}\right) \\ &\quad + \lambda\left(\mathrm{Tr}\left(\frac{\mathbf{X}_\varepsilon^*(\mathbf{X}_\varepsilon^*)^\dagger}{e^{\lambda_{\min}(\mathbf{\Delta}_a)}}\right)\cdot\mathrm{Tr}\left(\left(\frac{\mathbf{X}_\varepsilon^*(\mathbf{X}_\varepsilon^*)^\dagger}{e^{\lambda_{\min}(\mathbf{\Delta}_a)}}\right)^{-1}\right)\right) \\ &= \mathsf{logcap}(\mathbf{X}_\varepsilon^*) + \lambda\,\mathsf{reg}(\mathbf{X}_\varepsilon^*) = F(\mathbf{X}_\varepsilon^*)\end{aligned}$$

Now, let us denote by $h(s) \stackrel{\text{def}}{=} f^t(s\mathbf{\Delta}_b)$, which is a univariate function over $s \in \mathbb{R}$. We know that $h(s)$ is convex by Lemma C.1.a and C.2.a. Denoting by $\rho = 8\log\kappa_0 \ge 1$ and $\mathbf{\Delta}_t^* = \frac{\mathbf{\Delta}_b}{\rho}$, we have $\|\mathbf{\Delta}_t^*\|_2 \le 1/2$ and the convexity of $h(s)$ gives

$$f^t(0) - f^t(\mathbf{\Delta}_t^*) = h(0) - h\Big(\frac{1}{\rho}\Big) \ge \frac{1}{\rho}\big(h(0) - h(1)\big) = \frac{1}{\rho}\left(f^t(0) - f^t(\mathbf{\Delta}_b)\right) \tag{C.7}$$

We also know that $|h'''(s)| \le 2h''(s)$ by Lemma C.1.b and C.2.b, and thus by Proposition B.1, we know for every $\mathbf{\Delta}$ satisfying $\|\mathbf{\Delta}\|_2 \le \frac{1}{2}$, we have:

$$\mathrm{Tr}\left(\nabla f^t(0)\cdot\mathbf{\Delta}\right) + \frac{1}{2e}\mathrm{Tr}\left(\nabla^2 f^t(0)\cdot\mathbf{\Delta}\otimes\mathbf{\Delta}\right) \le f^t(\mathbf{\Delta}) - f^t(0) \tag{C.8}$$

$$\mathrm{Tr}\left(\nabla f^t(0)\cdot\mathbf{\Delta}\right) + \frac{e}{2}\mathrm{Tr}\left(\nabla^2 f^t(0)\cdot\mathbf{\Delta}\otimes\mathbf{\Delta}\right) \ge f^t(\mathbf{\Delta}) - f^t(0) \tag{C.9}$$

On one hand, using (C.8) we have

$$\begin{aligned}&\mathrm{Tr}\left(\nabla f^t(0)\cdot\mathbf{\Delta}_t\right) + \frac{1}{2e}\mathrm{Tr}\left(\nabla^2 f^t(0)\cdot\mathbf{\Delta}_t\otimes\mathbf{\Delta}_t\right) \\ &\overset{①}{\le} \mathrm{Tr}\left(\nabla f^t(0)\cdot\mathbf{\Delta}_t^*\right) + \frac{1}{2e}\mathrm{Tr}\left(\nabla^2 f^t(0)\cdot\mathbf{\Delta}_t^*\otimes\mathbf{\Delta}_t^*\right) \\ &\overset{②}{\le} -\big(f^t(0) - f^t(\mathbf{\Delta}_t^*)\big) \overset{③}{\le} -\frac{1}{\rho}\left(f^t(0) - f^t(\mathbf{\Delta}_b)\right) = -\frac{1}{\rho}\left(F(\mathbf{X}_t) - F(\mathbf{X}_\varepsilon^*)\right) \ .\end{aligned} \tag{C.10}$$

Above, ① is by the optimality of $\mathbf{\Delta}_t$; ② is by (C.8); and ③ is by (C.7).

On the other hand, using (C.9) we have

$$\begin{aligned}&\left(\mathrm{Tr}\left(\nabla f^t(0)\cdot\mathbf{\Delta}_t\right) + \frac{1}{2e}\mathrm{Tr}\left(\nabla^2 f^t(0)\cdot\mathbf{\Delta}_t\otimes\mathbf{\Delta}_t\right)\right) \\ &= e^2\left(\mathrm{Tr}\left(\nabla f^t(0)\cdot\frac{\mathbf{\Delta}_t}{e^2}\right) + \frac{e}{2}\mathrm{Tr}\left(\nabla^2 f^t(0)\cdot\frac{\mathbf{\Delta}_t}{e^2}\otimes\frac{\mathbf{\Delta}_t}{e^2}\right)\right) \\ &\ge -e^2\left(f^t(0) - f^t\left(\frac{\mathbf{\Delta}_t}{e^2}\right)\right) = -e^2\left(F(\mathbf{X}_t) - F(\mathbf{X}_{t+1})\right) \ .\end{aligned} \tag{C.11}$$

Since the left hand side of (C.11) is always non-positive by the definition of $\mathbf{\Delta}_t$, we immediately have $F(\mathbf{X}_t) \ge F(\mathbf{X}_{t+1})$ and this proves the first item.

As for the second item, we combine (C.10) and (C.11):

$$(F(\mathbf{X}_t) - F(\mathbf{X}_{t+1})) \geq \frac{1}{e^2 \rho} (F(\mathbf{X}_t) - F(\mathbf{X}^*_\varepsilon))$$

which after rearranging gives

$$F(\mathbf{X}_{t+1}) - F(\mathbf{X}^*_\varepsilon) \leq \left(1 - \frac{1}{8e^2 \log \kappa_0}\right) (F(\mathbf{X}_t) - F(\mathbf{X}^*_\varepsilon)) \ . \qquad \square$$

## C.3 Proof of Claim 5.2

*Proof of Claim 5.2.* Note that this should follow from the geodesic convexity of $F(X)$. We have actually explicitly calculated in the proofs of Lemma C.1 and C.2 (see Eqs. (C.1), (C.2), (C.4), (C.5)) and (C.6)) that for any Hermitian matrix $\mathbf{\Delta}$,

$$\begin{aligned}
\mathrm{Tr}\left(\nabla f^t(0) \cdot \mathbf{\Delta}\right) &= \mathrm{Tr}\left(\sum_{i=1}^m \mathbf{\Delta} \widetilde{\mathbf{A}}_i^\dagger \widetilde{\mathbf{A}}_i\right) - \mathrm{Tr}(\mathbf{\Delta}) \\
&\quad + \lambda \left(\mathrm{Tr}(\mathbf{\Delta} \mathbf{X}_t^\dagger \mathbf{X}_t) \cdot \mathrm{Tr}((\mathbf{X}_t \mathbf{X}_t^\dagger)^{-1}) - \mathrm{Tr}(\mathbf{X}_t \mathbf{X}_t^\dagger) \cdot \mathrm{Tr}(\mathbf{\Delta}(\mathbf{X}_t \mathbf{X}_t^\dagger)^{-1})\right) \\
\mathrm{Tr}\left(\nabla^2 f^t(0) \cdot \mathbf{\Delta} \otimes \mathbf{\Delta}\right) &= \frac{1}{2} \sum_{i,j=1}^m \mathrm{Tr}\left(\left(\widetilde{\mathbf{A}}_j^\dagger \widetilde{\mathbf{A}}_i \mathbf{\Delta} - \mathbf{\Delta} \widetilde{\mathbf{A}}_j^\dagger \widetilde{\mathbf{A}}_i\right)\left(\mathbf{\Delta} \widetilde{\mathbf{A}}_i^\dagger \widetilde{\mathbf{A}}_j - \widetilde{\mathbf{A}}_i^\dagger \widetilde{\mathbf{A}}_j \mathbf{\Delta}\right)\right) \\
&\quad + \lambda \Big(\mathrm{Tr}(\mathbf{\Delta}^2 \mathbf{X}_t^\dagger \mathbf{X}_t) \cdot \mathrm{Tr}((\mathbf{X}_t \mathbf{X}_t^\dagger)^{-1}) - 2\mathrm{Tr}(\mathbf{\Delta} \mathbf{X}_t \mathbf{X}_t^\dagger) \cdot \mathrm{Tr}(\mathbf{\Delta}(\mathbf{X}_t \mathbf{X}_t^\dagger)^{-1}) \\
&\quad + \mathrm{Tr}(\mathbf{X}_t^\dagger \mathbf{X}_t) \cdot \mathrm{Tr}(\mathbf{\Delta}^2 (\mathbf{X}_t \mathbf{X}_t^\dagger)^{-1})\Big) \in \mathbb{R}_{\geq 0} \qquad \text{(C.12)}
\end{aligned}$$

where $\widetilde{\mathbf{A}}_i \stackrel{\text{def}}{=} \left(\sum_{i=1}^m \mathbf{A}_i \mathbf{X}_t \mathbf{X}_t^\dagger \mathbf{A}_i^\dagger\right)^{-1/2} \mathbf{A}_i \mathbf{X}_t$. Therefore, the objective

$$\mathrm{Tr}\left(\nabla f^t(0) \cdot \mathbf{\Delta}\right) + \frac{1}{2e} \mathrm{Tr}\left(\nabla^2 f^t(0) \cdot \mathbf{\Delta} \otimes \mathbf{\Delta}\right)$$

is in fact quadratic and convex in $\mathbf{\Delta}$. (That is, if one writes the real and imaginary parts of the entries $\mathbf{\Delta}_{i,j}$ for $i \geq j$ into a real vector $\delta$, then the objective can be written as $\delta^\top \mathbf{M} \delta + b^\top \delta$ for some real PSD matrix $\mathbf{M} \succeq 0$ and real vector $b$.) Also, the constraint $\|\mathbf{\Delta}\|_2 \leq \frac{1}{2}$ is also convex in $\mathbf{\Delta}$. Hence the minimization problem is convex. $\square$

## C.4 Implementation in Finite Arithmetics

To implement Algorithm 2 in finite arithmetics, we maintain $\mathbf{X}_t$ so that

- $\lambda_{\min}(\mathbf{X}_t \mathbf{X}_t^\dagger) \in [1, 2]$,[27] and;
- the (real and imaginary) entries of $\mathbf{X}_t$ are integral multiples of $\xi \stackrel{\text{def}}{=} \frac{1}{\mathsf{poly}(n,m,M,\kappa(\mathbf{X}^*_\varepsilon),\varepsilon^{-1})}$.

Using the fact that $\kappa(\mathbf{X}_t) \leq \kappa_0 = \mathsf{poly}(n, m, M, \kappa(\mathbf{X}^*_\varepsilon), \varepsilon^{-1})$ from Claim C.3, we immediately conclude that the entries of $\mathbf{X}_t$ will not exceed $\mathsf{poly}(\kappa_0)$ so we can store them in word size $O(\log \kappa_0)$.

Next, when calculating the minimizer $\mathbf{\Delta}_t$, we wish to calculate the entries of $\mathbf{\Delta}_t$ up to additive accuracy $\xi$. This can be done in time $\mathsf{poly}(n, m, M, \log \kappa(\mathbf{X}^*_\varepsilon), \log \varepsilon^{-1})$ because as explicitly calculated in (C.12), the quadratic function in $\mathbf{\Delta}_t$ has all the coefficients encodable in bit complexity $O(\log \kappa_0)$. Let the resulting Hermitian matrix be $\widehat{\mathbf{\Delta}}_t$ and we have $\|\widehat{\mathbf{\Delta}}_t - \mathbf{\Delta}_t\| \leq n\xi$.

[27] Recall that we can always scale re-scale $\mathbf{X}$ by a positive constant factor without affecting the value $F(\mathbf{X})$.

Finally, we wish to calculate $\mathbf{X}_{t+1} = \delta \mathbf{X}_t e^{\mathbf{\Delta}_t/e^2}$ where $\delta > 0$ is some scaling factor which ensures $\lambda_{\min}(\mathbf{X}_{t+1}\mathbf{X}_{t+1}^\dagger) \in [1, 2]$. (Recall that $F(\mathbf{X})$ is invariant up to scaling.) From Lemma C.4 we have

$$F(\mathbf{X}_{t+1}) \leq F(\mathbf{X}_t) \quad \text{and} \quad F(\mathbf{X}_{t+1}) - F(\mathbf{X}_\varepsilon^*) \leq \left(1 - \frac{1}{8e^2 \log \kappa_0}\right)(F(\mathbf{X}_t) - F(\mathbf{X}_\varepsilon^*)) \ .$$

To calculate $\mathbf{X}_{t+1}$ in finite arithmetics, we first find some value $\delta \in [1/10, 10]$ so that

$$\lambda_{\min}(\overline{\mathbf{X}}_{t+1}\overline{\mathbf{X}}_{t+1}^\dagger) \in [1.2, 1.8] \quad \text{where} \quad \overline{\mathbf{X}}_{t+1} = \delta \mathbf{X}_t e^{\widehat{\mathbf{\Delta}}_t/e^2} \ .$$

This is always possible because $\lambda_{\min}(\mathbf{X}_t\mathbf{X}_t^\dagger) \in [1, 2]$ and $\|\widehat{\mathbf{\Delta}}_t\|_2 \leq 1/2 + n\xi < 1$, so it suffices to consider $\delta$ in the range $[1/10, 10]$ and up to constant accuracy (say, 0.1).

Next, after $\delta$ is explicitly calculated, we wish to compute $\overline{\mathbf{X}}_{t+1} = \delta \mathbf{X}_t e^{\widehat{\mathbf{\Delta}}_t/e^2}$ but due to numerical error, in polynomial time we can only obtain some

$$\widehat{\mathbf{X}}_{t+1} = \overline{\mathbf{X}}_{t+1} + \mathbf{C} \quad \text{where} \quad \|\mathbf{C}\| \leq n\xi \ .$$

Using Claim C.5 (see below) we have

$$(1 + 3n\xi)\overline{\mathbf{X}}_{t+1}\overline{\mathbf{X}}_{t+1}^\dagger \preceq \widehat{\mathbf{X}}_{t+1}\widehat{\mathbf{X}}_{t+1}^\dagger \preceq (1 - 3n\xi)\overline{\mathbf{X}}_{t+1}\overline{\mathbf{X}}_{t+1}^\dagger \ .$$

In addition, since $\mathbf{X}_{t+1} = \delta \mathbf{X} e^{\mathbf{\Delta}_t/e^2}$ and $\|\mathbf{\Delta}_t - \widehat{\mathbf{\Delta}}_t\|_2 \leq n\xi$, we have

$$e^{2n\xi}\mathbf{X}_{t+1}\mathbf{X}_{t+1}^\dagger \preceq \overline{\mathbf{X}}_{t+1}\overline{\mathbf{X}}_{t+1}^\dagger \preceq e^{-2n\xi}\mathbf{X}_{t+1}\mathbf{X}_{t+1}^\dagger \ .$$

Putting them together, we have

$$(1 + 10n\xi)\mathbf{X}_{t+1}\mathbf{X}_{t+1}^\dagger \preceq \widehat{\mathbf{X}}_{t+1}\widehat{\mathbf{X}}_{t+1}^\dagger \preceq (1 - 10n\xi)\mathbf{X}_{t+1}\mathbf{X}_{t+1}^\dagger \ .$$

This implies two things:

- $|\,\mathsf{logcap}(\mathbf{X}_{t+1}) - \mathsf{logcap}(\widehat{\mathbf{X}}_{t+1})| \leq O(n^2\xi)$

  This is because if $0 \preceq \mathbf{A} \preceq (1+\delta)\mathbf{B}$ then $\log\det \mathbf{A} \leq (1+\delta)^n \log\det \mathbf{B}$.
- $|\,\mathsf{reg}(\mathbf{X}_{t+1}) - \mathsf{reg}(\widehat{\mathbf{X}}_{t+1})| \leq O(n\xi\kappa_0)$

  This is because if $0 \preceq (1-\delta)\mathbf{B} \preceq \mathbf{A} \preceq (1+\delta)\mathbf{B}$ then $\mathrm{Tr}(\mathbf{A}) \cdot \mathrm{Tr}(\mathbf{A}^{-1}) \leq (1+\delta)\mathrm{Tr}(\mathbf{B}) \cdot \frac{1}{1-\delta}\mathrm{Tr}(\mathbf{B}^{-1})$, as well as $\mathsf{reg}(\mathbf{X}_{t+1}) \leq \kappa_0$ from Claim C.3.

In sum, we conclude that

$$F(\widehat{\mathbf{X}}_{t+1}) \leq F(\mathbf{X}_{t+1}) + O(n^2 + \lambda n \kappa_0) \cdot \xi$$

which is a small additive error on top of the function value $F(\cdot)$. Since $\xi$ is sufficiently small and the total number of iteration $T$ is no more than $T = O(\log \kappa_0 \cdot \log(nmM\varepsilon^{-1}))$ (see the proof of Theorem 5.3), this error is negligible.

**Claim C.5.** *For any $\mathbf{X} \in \mathsf{GL}_n(\mathbb{C})$ with $\lambda_{\min}(\mathbf{X}\mathbf{X}^\dagger) \geq 1$ and $\mathbf{C} \in \mathsf{Mat}_n(\mathbb{C})$, if $\|\mathbf{C}\|_2 \leq c$, then*

$$(1 + 2c + c^2)\mathbf{X}\mathbf{X}^\dagger \succeq (\mathbf{X} + \mathbf{C})(\mathbf{X} + \mathbf{C})^\dagger \succeq (1 - 2c)\mathbf{X}\mathbf{X}^\dagger \ .$$

*Proof.* Since

$$(\mathbf{X} + \mathbf{C}/c)(\mathbf{X} + \mathbf{C}/c)^\dagger \succeq 0 \quad \text{and} \quad (\mathbf{X} - \mathbf{C}/c)(\mathbf{X} - \mathbf{C}/c)^\dagger \succeq 0 \ ,$$

expanding them out we have

$$c\mathbf{X}\mathbf{X}^\dagger + \mathbf{C}\mathbf{C}^\dagger/c \succeq \mathbf{C}\mathbf{X}^\dagger + \mathbf{X}\mathbf{C}^\dagger \succeq -c\mathbf{X}\mathbf{X}^\dagger - \mathbf{C}\mathbf{C}^\dagger/c \ .$$

Using the fact that $0 \preceq \mathbf{C}\mathbf{C}^\dagger \preceq c^2\mathbf{I}$ and $\mathbf{X}\mathbf{X}^\dagger \succeq \mathbf{I}$, the above inequality chain implies

$$(1 + 2c + c^2)\mathbf{X}\mathbf{X}^\dagger \succeq \mathbf{X}\mathbf{X}^\dagger + \mathbf{C}\mathbf{X}^\dagger + \mathbf{X}\mathbf{C}^\dagger + \mathbf{C}\mathbf{C}^\dagger = (\mathbf{X} + \mathbf{C})(\mathbf{X} + \mathbf{C})^\dagger \succeq (1 - 2c)\mathbf{X}\mathbf{X}^\dagger \ . \quad \square$$

# D Missing Proofs for Section 6

We need the following elementary lemma about differential equations concerning real matrices. The first two points can be found for instance in [59, Lemma 3.4.4].

**Proposition D.1.** *Consider two differential equations on complex $n \times n$ matrices:*

$$\mathbf{E}^{(0)} = \mathbf{I}_n,\ \frac{d}{dt}\mathbf{E}^{(t)} = \mathbf{C}^{(t)}\mathbf{E}^{(t)} \quad \textit{and} \quad \mathbf{F}^{(0)} = \mathbf{I}_n,\ \frac{d}{dt}\mathbf{F}^{(t)} = \mathbf{F}^{(t)}\mathbf{D}^{(t)}.$$

*Above, $\mathbf{C}^{(t)}$ and $\mathbf{D}^{(t)}$ are Hermitian $n \times n$ matrices satisfying $\mathrm{Tr}\left(\mathbf{C}^{(t)}\right) = \mathrm{Tr}\left(\mathbf{D}^{(t)}\right) = 0$ and $\left\|\mathbf{C}^{(t)}\right\|_F, \left\|\mathbf{D}^{(t)}\right\|_F \leq 1$. Also, $\mathbf{C}^{(t)}$ and $\mathbf{D}^{(t)}$ are Lipschitz continuous in $t \geq 0$. Then:*

*(a)* [59] *there is a unique solution to the above differential equations.*

*(b)* [59] $\det\left(\mathbf{E}^{(t)}\right) = \det\left(\mathbf{F}^{(t)}\right) = 1$ *for all $t \geq 0$.*

*(c)* $\frac{d}{dt}\mathbf{E}^{(t)\dagger}\mathbf{E}^{(t)} = 2\mathbf{E}^{(t)\dagger}\mathbf{C}^{(t)}\mathbf{E}^{(t)}$ *and* $\frac{d}{dt}\mathbf{F}^{(t)}\mathbf{F}^{(t)\dagger} = 2\mathbf{F}^{(t)}\mathbf{D}^{(t)}\mathbf{F}^{(t)\dagger}$.

*(d)* $\mathrm{Tr}\big[\mathbf{E}^{(t)\dagger}\mathbf{E}^{(t)}\big], \mathrm{Tr}\big[\big(\mathbf{E}^{(t)\dagger}\mathbf{E}^{(t)}\big)^{-1}\big], \mathrm{Tr}\big[\mathbf{F}^{(t)}\mathbf{F}^{(t)\dagger}\big], \mathrm{Tr}\big[\big(\mathbf{F}^{(t)}\mathbf{F}^{(t)\dagger}\big)^{-1}\big] \leq n \cdot e^{2t}$.

*Proof of Proposition D.1.* Point (c) follows from elementary calculations. By symmetry, we will only prove point (d) for $\mathbf{E}^{(t)}$. Using point (c),

$$\begin{aligned}\frac{d}{dt}\mathrm{Tr}\left[\mathbf{E}^{(t)\dagger}\mathbf{E}^{(t)}\right] = 2\mathrm{Tr}\left[\mathbf{E}^{(t)\dagger}\mathbf{C}^{(t)}\mathbf{E}^{(t)}\right] = 2\mathrm{Tr}\left[\mathbf{E}^{(t)}\mathbf{E}^{(t)\dagger}\mathbf{C}^{(t)}\right] &\leq 2\mathrm{Tr}\left[\mathbf{E}^{(t)}\mathbf{E}^{(t)\dagger}\right] \\ &= 2\mathrm{Tr}\left[\mathbf{E}^{(t)\dagger}\mathbf{E}^{(t)}\right]\end{aligned}$$

The inequality follows from the fact that the spectral norm of $\mathbf{C}$ satisfies $\|\mathbf{C}\|_2 \leq \|\mathbf{C}\|_F \leq 1$. Now, the above inequality along with the initial condition at $t = 0$ implies $\mathrm{Tr}\left[\mathbf{E}^{(t)\dagger}\mathbf{E}^{(t)}\right] \leq n \cdot e^{2t}$.

For the inverse, we use $\frac{d}{dt}\left(\mathbf{G}^{(t)}\right)^{-1} = -\left(\mathbf{G}^{(t)}\right)^{-1}\left(\frac{d}{dt}\mathbf{G}^{(t)}\right)\left(\mathbf{G}^{(t)}\right)^{-1}$ from matrix calculus:

$$\begin{aligned}\frac{d}{dt}\mathrm{Tr}\left[\left(\mathbf{E}^{(t)\dagger}\mathbf{E}^{(t)}\right)^{-1}\right] &= -2\mathrm{Tr}\left[\left(\mathbf{E}^{(t)\dagger}\mathbf{E}^{(t)}\right)^{-1}\mathbf{E}^{(t)\dagger}\mathbf{C}^{(t)}\mathbf{E}^{(t)}\left(\mathbf{E}^{(t)\dagger}\mathbf{E}^{(t)}\right)^{-1}\right] \\ &= -2\mathrm{Tr}\left[\left(\mathbf{E}^{(t)\dagger}\right)^{-1}\left(\mathbf{E}^{(t)}\right)^{-1}\mathbf{C}^{(t)}\right] \leq 2\mathrm{Tr}\left[\left(\mathbf{E}^{(t)\dagger}\right)^{-1}\left(\mathbf{E}^{(t)}\right)^{-1}\right] \\ &= 2\mathrm{Tr}\left[\left(\mathbf{E}^{(t)\dagger}\mathbf{E}^{(t)}\right)^{-1}\right] \quad .\end{aligned}$$

Hence, we also have $\mathrm{Tr}\left[\left(\mathbf{E}^{(t)\dagger}\mathbf{E}^{(t)}\right)^{-1}\right] \leq n \cdot e^{2t}$. ☐

## D.1 Proof of Theorem 6.1

*Proof of Theorem 6.1.* Consider the following two differential equations:

$$\mathbf{E}^{(0)} = \mathbf{I}_n,\ \frac{d}{dt}\mathbf{E}^{(t)} = \mathbf{C}^{(t)}\mathbf{E}^{(t)} \quad \text{and} \quad \mathbf{F}^{(0)} = \mathbf{I}_n,\ \frac{d}{dt}\mathbf{F}^{(t)} = \mathbf{F}^{(t)}\mathbf{D}^{(t)}.$$

where

$$\mathbf{C}^{(t)} \stackrel{\text{def}}{=} -\frac{1}{\ell^{(t)}}\left(\sum_{i=1}^{m}\mathbf{A}_i^{(t)}\mathbf{A}_i^{(t)\dagger} - \frac{s^{(t)}}{n}\mathbf{I}_n\right) \quad \text{and} \quad \mathbf{D}^{(t)} \stackrel{\text{def}}{=} -\frac{1}{\ell^{(t)}}\left(\sum_{i=1}^{m}\mathbf{A}_i^{(t)\dagger}\mathbf{A}_i^{(t)} - \frac{s^{(t)}}{n}\mathbf{I}_n\right)$$

Above, we have denoted by $\mathbf{A}_i^{(t)} \stackrel{\text{def}}{=} \mathbf{E}^{(t)}\mathbf{A}_i\mathbf{F}^{(t)}$, $s^{(t)} \stackrel{\text{def}}{=} \sum_{i=1}^m \left\|\mathbf{A}_i^{(t)}\right\|_F^2$, and

$$\ell^{(t)} \stackrel{\text{def}}{=} \sqrt{\mathrm{Tr}\left[\left(\sum_{i=1}^m \mathbf{A}_i^{(t)}\mathbf{A}_i^{(t)\dagger} - \frac{s^{(t)}}{n}\mathbf{I}_n\right)^2\right] + \mathrm{Tr}\left[\left(\sum_{i=1}^m \mathbf{A}_i^{(t)\dagger}\mathbf{A}_i^{(t)} - \frac{s^{(t)}}{n}\mathbf{I}_n\right)^2\right]} \ .$$

We now analyze the rate of decay of $s^{(t)}$:

$$
\begin{aligned}
\frac{d}{dt}s^{(t)} &= \frac{d}{dt}\mathrm{Tr}\left[\sum_{i=1}^{m}\mathbf{A}_i^{(t)}\mathbf{A}_i^{(t)\dagger}\right] \\
&= \frac{d}{dt}\mathrm{Tr}\left[\sum_{i=1}^{m}\mathbf{E}^{(t)}\mathbf{A}_i\mathbf{F}^{(t)}\mathbf{F}^{(t)\dagger}\mathbf{A}_i^{\dagger}\mathbf{E}^{(t)\dagger}\right] \\
&= \frac{d}{dt}\mathrm{Tr}\left[\sum_{i=1}^{m}\mathbf{E}^{(t)\dagger}\mathbf{E}^{(t)}\mathbf{A}_i\mathbf{F}^{(t)}\mathbf{F}^{(t)\dagger}\mathbf{A}_i^{\dagger}\right] \\
&\overset{①}{=} 2\mathrm{Tr}\left[\sum_{i=1}^{m}\mathbf{E}^{(t)\dagger}\mathbf{C}^{(t)}\mathbf{E}^{(t)}\mathbf{A}_i\mathbf{F}^{(t)}\mathbf{F}^{(t)\dagger}\mathbf{A}_i^{\dagger}\right] + 2\mathrm{Tr}\left[\sum_{i=1}^{m}\mathbf{E}^{(t)\dagger}\mathbf{E}^{(t)}\mathbf{A}_i\mathbf{F}^{(t)}\mathbf{D}^{(t)}\mathbf{F}^{(t)\dagger}\mathbf{A}_i^{\dagger}\right] \\
&= 2\mathrm{Tr}\left[\mathbf{C}^{(t)}\sum_{i=1}^{m}\mathbf{A}_i^{(t)}\mathbf{A}_i^{(t)\dagger}\right] + 2\mathrm{Tr}\left[\sum_{i=1}^{m}\mathbf{A}_i^{(t)\dagger}\mathbf{A}_i^{(t)}\mathbf{D}^{(t)}\right] \\
&\overset{②}{=} 2\mathrm{Tr}\left[\mathbf{C}^{(t)}\left(\sum_{i=1}^{m}\mathbf{A}_i^{(t)}\mathbf{A}_i^{(t)\dagger} - \frac{s^{(t)}}{n}\mathbf{I}_n\right)\right] + 2\mathrm{Tr}\left[\left(\sum_{i=1}^{m}\mathbf{A}_i^{(t)\dagger}\mathbf{A}_i^{(t)} - \frac{s^{(t)}}{n}\mathbf{I}_n\right)\mathbf{D}^{(t)}\right] \\
&= -2\ell^{(t)}
\end{aligned}
$$

Above, equality ① follows from Proposition D.1.c, and equality ② follows from the fact that $\mathrm{Tr}\left[\mathbf{C}^{(t)}\right] = \mathrm{Tr}\left[\mathbf{D}^{(t)}\right] = 0$.

Let operators $T^{(t)}$ and $T_{\mathbf{A}^{(t)}}$ respectively be defined by

$$
\left(\sqrt{\frac{n}{s^{(t)}}}\mathbf{A}_1^{(t)}, \dots, \sqrt{\frac{n}{s^{(t)}}}\mathbf{A}_m^{(t)}\right) \quad \text{and} \quad \left(\mathbf{A}_1^{(t)}, \dots, \mathbf{A}_m^{(t)}\right) .
$$

Recall the $\widetilde{\mathsf{cap}}(T_{\vec{\mathbf{A}}}) \overset{\text{def}}{=} \inf_{\mathbf{X},\mathbf{Y}\succ 0, \det(\mathbf{X})=\det(\mathbf{Y})=1}\mathrm{Tr}[\mathbf{X}\, T_{\vec{\mathbf{A}}}(\mathbf{Y})]$ from Definition 3.4 and we apply Lemma A.1 to infer that

$$
n - \widetilde{\mathsf{cap}}\left(T^{(t)}\right) = \mathrm{Tr}\left[T^{(t)}(\mathbf{I})\right] - \widetilde{\mathsf{cap}}\left(T^{(t)}\right) \le \frac{n^{3/2}}{\sqrt{2}}\sqrt{\mathsf{ds}\left(T^{(t)}\right)}
$$

Multiplying both sides of the inequality by $\frac{s^{(t)}}{n}$ and noting that $\frac{s^{(t)}}{n}\widetilde{\mathsf{cap}}\left(T^{(t)}\right) = \widetilde{\mathsf{cap}}\left(T_{\vec{\mathbf{A}}}\right)$ and also that $\frac{s^{(t)}}{n}\sqrt{\mathsf{ds}\left(T^{(t)}\right)} = \ell^{(t)}$, we obtain that

$$
s^{(t)} - \widetilde{\mathsf{cap}}\left(T_{\mathbf{A}^{(t)}}\right) \le \frac{n^{3/2}}{\sqrt{2}}\ell^{(t)}
$$

Now using Proposition D.1.b, we have $\det\left(\mathbf{E}^{(t)}\right) = \det\left(\mathbf{F}^{(t)}\right) = 1$ and therefore $\widetilde{\mathsf{cap}}\left(T_{\mathbf{A}^{(t)}}\right) = \widetilde{\mathsf{cap}}(T_{\vec{\mathbf{A}}})$ owing to the definition of $\widetilde{\mathsf{cap}}$ (see Definition 3.4). One can then apply the above inequality to get

$$
s^{(t)} - \widetilde{\mathsf{cap}}(T_{\vec{\mathbf{A}}}) = s^{(t)} - \widetilde{\mathsf{cap}}\left(T_{\mathbf{A}^{(t)}}\right) \le \frac{n^{3/2}}{\sqrt{2}}\ell^{(t)} = -\frac{n^{3/2}}{\sqrt{8}}\frac{d}{dt}\left(s^{(t)} - \widetilde{\mathsf{cap}}(T_{\vec{\mathbf{A}}})\right)
$$

Rearranging both sides we have

$$
\frac{d}{dt}\left(s^{(t)} - \widetilde{\mathsf{cap}}(T_{\vec{\mathbf{A}}})\right) \le -\frac{\sqrt{8}}{n^{3/2}}\left(s^{(t)} - \widetilde{\mathsf{cap}}(T_{\vec{\mathbf{A}}})\right)
$$

Since initially it satisfies $s^{(0)} \leq mn^2M^2$, we have

$$s^{(t)} - \widetilde{\mathsf{cap}}(T_{\overrightarrow{\mathbf{A}}}) \leq s^{(0)} \cdot \exp\left(-\frac{\sqrt{8}t}{n^{3/2}}\right) \leq mn^2M^2 \cdot \exp\left(-\frac{\sqrt{8}t}{n^{3/2}}\right)$$

Now, choose $k = n^{3/2}\log\left(\frac{12mn^4M^2}{\varepsilon}\right)$ and it satisfies $s^{(k)} - \widetilde{\mathsf{cap}}(T_{\overrightarrow{\mathbf{A}}}) \leq \frac{\varepsilon}{12n^2}$. Therefore,

$$n - \widetilde{\mathsf{cap}}\left(T^{(k)}\right) = \frac{n}{s^{(k)}}\left(s^{(k)} - \widetilde{\mathsf{cap}}(T_{\overrightarrow{\mathbf{A}}})\right) \leq \frac{\varepsilon}{12ns^{(k)}} \leq \frac{\varepsilon}{12}$$

By Lemma A.3 we have $s^{(k)} \geq 1/n$. Applying Lemma A.2 we conclude that

$$\mathsf{ds}\left(T^{(k)}\right) \leq \varepsilon \ .$$

Now let us define $\mathbf{X} = \left(\frac{n}{s^{(k)}}\right)^{1/2}\left(\mathbf{E}^{(k)\dagger}\mathbf{E}^{(k)}\right)^{1/2}$ and $\mathbf{Y} = \left(\mathbf{F}^{(k)}\mathbf{F}^{(k)\dagger}\right)^{1/2}$. Let $\mathbf{B}_i = \mathbf{X}\mathbf{A}_i\mathbf{Y}$ and let $T_{\overrightarrow{\mathbf{B}}}$ be the operator defined by $(\mathbf{B}_1, \dots, \mathbf{B}_m)$. Then one can verify that for any integer $r \geq 1$ (we only need it for $r = 1, 2$),

$$\mathrm{Tr}\left[(\textstyle\sum_{i=1}^m \mathbf{B}_i\mathbf{B}_i^\dagger - \mathbf{I})^r\right] = \mathrm{Tr}\left[(\textstyle\sum_{i=1}^m \frac{n}{s^{(t)}}\mathbf{A}_i^{(t)}(\mathbf{A}_i^{(t)})^\dagger - \mathbf{I})^r\right]$$

and similarly for $\mathrm{Tr}\left[(\sum_{i=1}^m \mathbf{B}_i^\dagger\mathbf{B}_i - \mathbf{I})^r\right]$. Therefore, we have

$$\mathsf{ds}\left(T_{\overrightarrow{\mathbf{B}}}\right) = \mathsf{ds}\left(T^{(k)}\right) \leq \varepsilon$$

What is left is to analyze the bounds on the eigenvalues of $\mathbf{X}$ and $\mathbf{Y}$. For this we use Proposition D.1.d which tells us

$$\mathrm{Tr}\left[\mathbf{E}^{(k)\dagger}\mathbf{E}^{(k)}\right], \mathrm{Tr}\left[\left(\mathbf{E}^{(k)\dagger}\mathbf{E}^{(k)}\right)^{-1}\right], \mathrm{Tr}\left[\mathbf{F}^{(k)}\mathbf{F}^{(k)\dagger}\right], \mathrm{Tr}\left[\left(\mathbf{F}^{(k)}\mathbf{F}^{(k)\dagger}\right)^{-1}\right] \leq n \cdot e^{2k} \ .$$

Along with the bounds $1/n \leq s^{(k)} \leq s^{(0)} \leq mn^2M^2$, this completes the proof. □

# E Missing Proofs for Section 7

Before proceeding to the proofs, we list below some properties about the left-right action to make our notions more concrete. Some of them follow from the general definitions, but some (such as the description of elements of minimum norm) require the use of the Kempf-Ness theorem as discussed in Section 1.2.1.

**Lemma E.1** (Corollary 1.9). *The orbit-closures of the two tuples $(\mathbf{A}_1, \dots, \mathbf{A}_m)$ and $(\mathbf{B}_1, \dots, \mathbf{B}_m)$ intersect under the left-right action iff $\det\left(\sum_{i=1}^m \mathbf{Y}_i \otimes \mathbf{A}_i\right) \equiv \det\left(\sum_{i=1}^m \mathbf{Y}_i \otimes \mathbf{B}_i\right)$ for all $d \leq n^5$. Here, the matrices $\mathbf{Y}_i$ are $d \times d$ with disjoint sets of variables.*

**Proposition E.2** (Proposition 1.12 from [23]). *If $(\mathbf{A}_1, \dots, \mathbf{A}_m) \in \mathsf{Mat}_n(\mathbb{C})^m$ is not in the null cone of the left-right action, then for any $d \geq n^5$ we have that $\det\left(\sum_{i=1}^m \mathbf{Y}_i \otimes \mathbf{A}_i\right) \not\equiv 0$, where matrices $\mathbf{Y}_i$ are $d \times d$ generic matrices on disjoint sets of variables.*

## E.1 Auxiliary lemma for polynomials

We will need the fact that (nonzero) polynomials with small degree cannot vanish on all points with non-negative integer coordinates bounded by the individual degrees. This follows from Alon's combinatorial nullstellensatz [8, Theorem 1.2].

**Lemma E.3** ([8]). *If $p(x_1, \ldots, x_n) \in \mathbb{C}[x_1, \ldots, x_n]$ is a (nonzero) polynomial where the individual degree of the variable $x_i$ is at most $d_i$, then there exists $(a_1, \ldots, a_n) \in \mathbb{Z}^n_{\geq 0}$ such that $a_i \leq d_i$, for which $p(a_1, \ldots, a_n) \neq 0$.*

Let $\mathbf{a}, \mathbf{b} \in \mathbb{R}^r$ be vectors of real entries. We say that $\mathbf{a} \leq \mathbf{b}$ iff $a_i \leq b_i$ for all $i \in [r]$, and $\mathbf{a} < \mathbf{b}$ iff $\mathbf{a} \leq \mathbf{b}$ and $\mathbf{a} \neq \mathbf{b}$. For any $\mathbf{a}, \mathbf{b} \in \mathbb{N}^r$, we write $\binom{\mathbf{a}}{\mathbf{b}} = \prod_{i=1}^{r} \binom{a_i}{b_i}$.

**Proposition E.4.** *Let $p(\mathbf{x}) \in \mathbb{C}[\mathbf{x}]$ be a homogeneous polynomial of degree $D$ on $r$ variables $\mathbf{x} = (x_1, \ldots, x_r)$, such that the norm of each of its coefficients is upper bounded by a positive integer $M$. If $\mathbf{a}, \mathbf{c} \in \mathbb{C}^r$ are vectors such that $\|\mathbf{a} - \mathbf{c}\|_\infty \leq \beta \in (0, 1)$ and $\|\mathbf{a}\|_\infty \leq \alpha$, where $\alpha \geq 1$, then the following inequality holds:*

$$|p(\mathbf{c}) - p(\mathbf{a})| \leq \beta \cdot \exp(Dr \log(M\alpha))$$

*Proof.* Since $p(\mathbf{x})$ is a homogeneous polynomial of degree $D$, we can write it as

$$p(\mathbf{x}) = \sum_{\mathbf{e} \in \{0,1,\ldots,D\}^r \wedge \|\mathbf{e}\|_1 = D} p_{\mathbf{e}} \cdot \mathbf{x}^{\mathbf{e}} \ ,$$

where each $p_{\mathbf{e}} \in \mathbb{C}$ has absolute value bounded by $M$ and $\mathbf{x}^{\mathbf{e}}$ goes through all monomials of degree exactly $D$. Letting $\mathbf{b} = \mathbf{c} - \mathbf{a}$, we have that $\mathbf{c} = \mathbf{a} + \mathbf{b}$ and thus $p(\mathbf{c}) - p(\mathbf{a}) = p(\mathbf{a} + \mathbf{b}) - p(\mathbf{a})$.

Note that $(\mathbf{a} + \mathbf{b})^{\mathbf{e}} = \sum_{\mathbf{d} \leq \mathbf{e}} \binom{\mathbf{e}}{\mathbf{d}} \mathbf{a}^{\mathbf{d}} \cdot \mathbf{b}^{\mathbf{e} - \mathbf{d}}$. Thus, we have that

$$\begin{aligned}
|(\mathbf{a} + \mathbf{b})^{\mathbf{e}} - \mathbf{a}^{\mathbf{e}}| &= \left| \sum_{\mathbf{d} < \mathbf{e}} \binom{\mathbf{e}}{\mathbf{d}} \mathbf{a}^{\mathbf{d}} \cdot \mathbf{b}^{\mathbf{e} - \mathbf{d}} \right| \leq \sum_{\mathbf{d} < \mathbf{e}} \binom{\mathbf{e}}{\mathbf{d}} \left| \mathbf{a}^{\mathbf{d}} \right| \cdot \beta \\
&\leq \beta \cdot \sum_{\mathbf{d} \leq \mathbf{e}} \binom{\mathbf{e}}{\mathbf{d}} \left| \mathbf{a}^{\mathbf{d}} \right| \leq \beta \cdot \sum_{\mathbf{d} \leq \mathbf{e}} \binom{\mathbf{e}}{\mathbf{d}} \alpha^{\|\mathbf{e}\|_1} \\
&= \beta \cdot \alpha^D \cdot \sum_{\mathbf{d} \leq \mathbf{e}} \binom{\mathbf{e}}{\mathbf{d}} = \beta \cdot \alpha^D \cdot 2^{\|\mathbf{e}\|_1} = \beta \cdot (2\alpha)^D
\end{aligned}$$

Since $p(\mathbf{c}) - p(\mathbf{a}) = \sum_{\mathbf{e} \in Mon(D)} p_{\mathbf{e}} \cdot [(\mathbf{a} + \mathbf{b})^{\mathbf{e}} - \mathbf{a}^{\mathbf{e}}]$, using the inequality above and the triangle inequality, we have:

$$\begin{aligned}
|p(\mathbf{c}) - p(\mathbf{a})| &\leq \sum_{\mathbf{e} \in Mon(D)} |p_{\mathbf{e}}| \cdot |(\mathbf{a} + \mathbf{b})^{\mathbf{e}} - \mathbf{a}^{\mathbf{e}}| \\
&\leq \sum_{\mathbf{e} \in Mon(D)} M \cdot \beta \cdot (2\alpha)^D \leq \beta \cdot \binom{D + r - 1}{D} \cdot M \cdot (2\alpha)^D
\end{aligned}$$

□

## E.2 Proof of Lemma 7.1

We now prove a lemma which says that if the orbit-closures of $\overline{\mathcal{O}}_{\vec{\mathbf{A}}}, \overline{\mathcal{O}}_{\vec{\mathbf{B}}}$ do not intersect, then there exists an invariant polynomial with small coefficients that distinguishes them.

**Lemma E.5.** *Let $\mathbf{A} = (\mathbf{A}_1, \ldots, \mathbf{A}_m)$ and $\mathbf{B} = (\mathbf{B}_1, \ldots, \mathbf{B}_m)$ be two points in $\mathsf{Mat}_n(\mathbb{C})^m$. If $\overline{\mathcal{O}}_{\vec{\mathbf{A}}} \cap \overline{\mathcal{O}}_{\vec{\mathbf{B}}} = \varnothing$ then there exists some integer $d \in [n^5]$ and $(\mathbf{Z}_1, \ldots, \mathbf{Z}_m) \in \mathsf{Mat}_d(\mathbb{Z})^m$ with $\|\mathbf{Z}_i\|_\infty \leq n$ for which*

$$\det\left(\textstyle\sum_{i=1}^m \mathbf{Z}_i \otimes \mathbf{A}_i\right) \neq \det\left(\textstyle\sum_{i=1}^m \mathbf{Z}_i \otimes \mathbf{B}_i\right).$$

*In particular, if* $\mathbf{A}, \mathbf{B} \in \mathsf{Mat}_n(\mathbb{Z}[i])^m$ *are integral we have that*

$$|\det(\sum_{i=1}^m \mathbf{Z}_i \otimes \mathbf{A}_i) - \det(\sum_{i=1}^m \mathbf{Z}_i \otimes \mathbf{B}_i)| \geq 1.$$

*Proof of Lemma E.5.* By Lemma E.1, we know that there exists a $d \in [n^5]$ s.t.

$$\det\left(\sum_{i=1}^m \mathbf{Y}_i \otimes \mathbf{A}_i\right) \not\equiv \det\left(\sum_{i=1}^m \mathbf{Y}_i \otimes \mathbf{B}_i\right).$$

as *polynomials.* Since the degree of each variable in the polynomials $\det(\sum_{i=1}^m \mathbf{Y}_i \otimes \mathbf{A}_i)$ and $\det(\sum_{i=1}^m \mathbf{Z}_i \otimes \mathbf{B}_i)$ is upper bounded by $n$, Lemma E.3 tells us that there exist $(\mathbf{Z}_1, \ldots, \mathbf{Z}_m) \in \mathsf{Mat}_d(\mathbb{Z})^m$ such that $\|\mathbf{Z}_i\|_\infty \leq n$ and

$$\det\left(\sum_{i=1}^m \mathbf{Z}_i \otimes \mathbf{A}_i\right) \neq \det\left(\sum_{i=1}^m \mathbf{Z}_i \otimes \mathbf{B}_i\right).$$

Since both sides are complex integral which are distinct, we have that their difference has modulus at least 1. ☐

We are now ready to prove Lemma 7.1, which we restate here for convenience:

**Lemma 7.1.** *Let* $\overrightarrow{\mathbf{A}} = (\mathbf{A}_1, \ldots, \mathbf{A}_m)$ *and* $\overrightarrow{\mathbf{B}} = (\mathbf{B}_1, \ldots, \mathbf{B}_m)$ *be two tuples in* $\mathsf{Mat}_n(\mathbb{Z}[i])^m$ *not in the null cone of the left-right action (i.e.,* $\mathsf{cap}(T_{\overrightarrow{\mathbf{A}}}) > 0$ *and* $\mathsf{cap}(T_{\overrightarrow{\mathbf{B}}}) > 0$*). Suppose* $\|\overrightarrow{\mathbf{A}}\|_2, \|\overrightarrow{\mathbf{B}}\|_2 < M$ *and* $\varepsilon = \exp(-n^{20}m \cdot \log(M))$*. Let* $\overrightarrow{\mathbf{A}}' = (\mathbf{A}'_1, \ldots, \mathbf{A}'_m)$ *and* $\overrightarrow{\mathbf{B}}' = (\mathbf{B}'_1, \ldots, \mathbf{B}'_m)$ *be elements in* $\overline{\mathcal{O}}_{\overrightarrow{\mathbf{A}}}$ *and* $\overline{\mathcal{O}}_{\overrightarrow{\mathbf{B}}}$ *respectively.*

*(a) If* $\overline{\mathcal{O}}_{\overrightarrow{\mathbf{A}}} \cap \overline{\mathcal{O}}_{\overrightarrow{\mathbf{B}}} = \varnothing$ *and* $\mathbf{U}, \mathbf{V} \in \mathsf{U}_n(\mathbb{C})$ *are such that* $|\det(\mathbf{UV}) - 1| \leq \varepsilon$*, then*

$$\left\|\mathbf{U}\overrightarrow{\mathbf{A}}'\mathbf{V} - \overrightarrow{\mathbf{B}}'\right\|_2 \geq \varepsilon.$$

*(b) If* $\overline{\mathcal{O}}_{\overrightarrow{\mathbf{A}}} \cap \overline{\mathcal{O}}_{\overrightarrow{\mathbf{B}}} \neq \varnothing$*, then for all* $\mathbf{U}, \mathbf{V} \in \mathsf{U}_n(\mathbb{C})$ *such that*

$$\left\|\mathbf{U}\overrightarrow{\mathbf{A}}'\mathbf{V} - \overrightarrow{\mathbf{B}}'\right\|_2 \leq \varepsilon,$$

*we must have* $|\det(\mathbf{UV}) - 1| \leq \varepsilon^{1/3}$*.*

*Proof of Lemma 7.1a.* If $\overline{\mathcal{O}}_{\overrightarrow{\mathbf{A}}} \cap \overline{\mathcal{O}}_{\overrightarrow{\mathbf{B}}} = \varnothing$, Lemma E.5 tells us that there exist $d \leq n^5$ and $(\mathbf{Z}_1, \ldots, \mathbf{Z}_m) \in \mathsf{Mat}_d(\mathbb{Z})^m$ with $\|\mathbf{Y}_i\|_\infty \leq n$ such that

$$\left|\det\left(\sum_{i=1}^m \mathbf{Z}_i \otimes \mathbf{A}_i\right) - \det\left(\sum_{i=1}^m \mathbf{Z}_i \otimes \mathbf{B}_i\right)\right| \geq 1. \tag{E.1}$$

Let $(\mathbf{X}_1, \ldots, \mathbf{X}_m)$ be a tuple of $m$ generic $n \times n$ matrices (i.e. entries are disjoint formal variables), $(\mathbf{Z}_1, \ldots, \mathbf{Z}_m)$ be the tuple above, and let polynomial $p(\mathbf{X}_1, \ldots, \mathbf{X}_m) \overset{\text{def}}{=} \det(\sum_{i=1}^m \mathbf{Z}_i \otimes \mathbf{X}_i)$. We have that $\deg(p) = nd$, the number of variables of $p$ is $n^2 m$ and each coefficient of $p$ is upper bounded by $\exp(3nd \log(nd))$. This bound on the coefficients holds because $p$ is the linear projection of a determinant of dimension $nd$, with linear forms whose coefficients are bounded by $n$.

We prove by way of contradiction. If $\mathbf{U}, \mathbf{V} \in \mathsf{U}_n(\mathbb{C})$ were such that $|\det(\mathbf{UV}) - 1| \leq \varepsilon$ and $\left\|\mathbf{U}\overrightarrow{\mathbf{A}}'\mathbf{V} - \overrightarrow{\mathbf{B}}'\right\|_2 \leq \varepsilon$, Proposition E.4 would imply that

$$|p(\mathbf{U}\overrightarrow{\mathbf{A}}'\mathbf{V}) - p(\overrightarrow{\mathbf{B}}')| \leq \varepsilon \cdot \exp(nd \cdot n^2 m \cdot 3nd \log(ndM)) = \varepsilon \cdot \exp(n^{15} m \log(M)) < 1/n.$$

However, as $\overrightarrow{\mathbf{A}}'$ and $\overrightarrow{\mathbf{B}}'$ are in $\overline{\mathcal{O}}_{\overrightarrow{\mathbf{A}}}$ and $\overline{\mathcal{O}}_{\overrightarrow{\mathbf{B}}}$, respectively, we have that $p(\overrightarrow{\mathbf{A}}') = p(\overrightarrow{\mathbf{A}})$ and that $p(\overrightarrow{\mathbf{B}}') = p(\overrightarrow{\mathbf{B}})$. Moreover, given the definition of $p$, we have $p(\mathbf{U}\overrightarrow{\mathbf{A}}'\mathbf{V}) = \det(\mathbf{UV})^d \cdot p(\overrightarrow{\mathbf{A}}')$. Thus:

$$\begin{aligned}1 \leq |p(\overrightarrow{\mathbf{A}}) - p(\overrightarrow{\mathbf{B}})| &= |p(\overrightarrow{\mathbf{A}}') - p(\overrightarrow{\mathbf{B}}')| \\ &\leq |p(\overrightarrow{\mathbf{A}}') - p(\mathbf{U}\overrightarrow{\mathbf{A}}'\mathbf{V})| + |p(\mathbf{U}\overrightarrow{\mathbf{A}}'\mathbf{V}) - p(\overrightarrow{\mathbf{B}}')| \\ &\leq |\det(\mathbf{UV})^d - 1| \cdot |p(\overrightarrow{\mathbf{A}}')| + 1/n \leq \varepsilon \cdot M \cdot \exp(n^{15}m) + 1/n < 1\end{aligned}$$

which is a contradiction. This proves Lemma 7.1a. □

*Proof of Lemma 7.1b.* Let $\mathbf{U}, \mathbf{V} \in \mathsf{U}_n(\mathbb{C})$ be a pair satisfying $\left\|\mathbf{U}\overrightarrow{\mathbf{A}}'\mathbf{V} - \overrightarrow{\mathbf{B}}'\right\|_2 \leq \varepsilon$. We need to prove that $|\det(\mathbf{UV}) - 1| \leq \varepsilon$.

Let $d$ be any integer $\geq n^5$. Since $\overrightarrow{\mathbf{B}}$ is not in the null cone, Proposition E.2 and Lemma E.1 imply that $\det(\sum_{i=1}^m \mathbf{B}_i \otimes \mathbf{X}_i) \not\equiv 0$, where $\mathbf{X}_i$ are $d \times d$ generic matrices on disjoint sets of variables. Moreover by Lemma E.1, we have $\det(\sum_{i=1}^m \mathbf{A}_i \otimes \mathbf{X}_i) \equiv \det(\sum_{i=1}^m \mathbf{B}_i \otimes \mathbf{X}_i)$.

By Lemma E.3, there are $\overrightarrow{\mathbf{Z}} \in \mathsf{Mat}_d(\mathbb{Z})^m$ with $\|\overrightarrow{\mathbf{Z}}\|_\infty \leq n$ such that $\det(\sum_{i=1}^m \mathbf{B}_i \otimes \mathbf{Z}_i) \neq 0$. Since $\|\overrightarrow{\mathbf{B}}' - \mathbf{U}\overrightarrow{\mathbf{A}}'\mathbf{V}\|_2 \leq \varepsilon$, Proposition E.4 implies

$$\begin{aligned}\varepsilon^{1/2} \geq \varepsilon \cdot \exp(n^{15} m \log(M)) &\geq \left|\det(\sum_{i=1}^m \mathbf{B}'_i \otimes \mathbf{Z}_i) - \det(\sum_{i=1}^m (\mathbf{U}\mathbf{A}'_i\mathbf{V}) \otimes \mathbf{Z}_i)\right| \\ &= \left|\det(\sum_{i=1}^m \mathbf{A}_i \otimes \mathbf{Z}_i) - \det(\mathbf{UV})^d \cdot \det(\sum_{i=1}^m \mathbf{A}_i \otimes \mathbf{Z}_i)\right| \\ &= |1 - \det(\mathbf{UV})^d| \cdot \left|\det(\sum_{i=1}^m \mathbf{A}_i \otimes \mathbf{Z}_i)\right| \geq |1 - \det(\mathbf{UV})^d|\end{aligned}$$

where the last inequality is true since $\det(\sum_{i=1}^m \mathbf{A}_i \otimes \mathbf{Z}_i) \in \mathbb{Z}[i]$, as $\overrightarrow{\mathbf{Z}} \in \mathsf{Mat}_d(\mathbb{Z})^m$ and $\overline{\mathcal{O}}_{\overrightarrow{\mathbf{A}}}$ contains a Gaussian integral point. Thus, for $d$ equal to $n^5$ or $n^5 + 1$, we have that

$$\varepsilon^{1/2} \geq |1 - \det(\mathbf{UV})^d|.$$

This implies, in particular, that $|\det(\mathbf{UV}) - 1| \leq \varepsilon^{1/3}$. □

# F Missing Proofs for Section 8

This section is served for proving Theorem M3 and constructing our Theorem M3 to check whether two given tuples of matrices are close under (simultaneous) unitary transformation. Recall from Section 8 that, to check (simultaneous) unitary equivalence, it suffices to study

**Problem F.1** (simultaneous conjugation)**.** *Given $\overrightarrow{\mathbf{A}}, \overrightarrow{\mathbf{B}} \in \mathsf{Mat}_n(\mathbb{C})^m$, where the Frobenius norm of each $\mathbf{A}_i, \mathbf{B}_i$ is upper bounded by $\lambda$ and $\varepsilon > 0$, find a unitary matrix $\mathbf{U} \in \mathsf{U}_n(\mathbb{C})$ such that*

$$\left\|\mathbf{U}\overrightarrow{\mathbf{A}}\mathbf{U}^\dagger - \overrightarrow{\mathbf{B}}\right\|_2 \leq \varepsilon \ .$$

*We denote an instance of this problem by the tuple $(\overrightarrow{\mathbf{A}}, \overrightarrow{\mathbf{B}}, \varepsilon, n)$.*

We generalize the above problem into a block-diagonal form:

**Problem F.2** (block-diagonal simultaneous conjugation). *Given* $\overrightarrow{\mathbf{A}}, \overrightarrow{\mathbf{B}} \in \mathsf{Mat}_n(\mathbb{C})^m$, $\varepsilon > 0$, $\lambda > 0$ *such that each* $\mathbf{A}_i, \mathbf{B}_i$ *has Frobenius norm* $\leq \lambda$, *and positive integers* $r_1, \ldots, r_p$ *such that* $r_1 + \cdots + r_p = n$, *find unitary matrices* $\mathbf{U}_i \in \mathsf{U}_{r_i}(\mathbb{C})$ *for which* $\mathbf{U} = \mathit{diag}(\mathbf{U}_1, \ldots, \mathbf{U}_p) \in \mathsf{U}_n(\mathbb{C})$ *satisfies*

$$\left\| \mathbf{U}\overrightarrow{\mathbf{A}}\mathbf{U}^\dagger - \overrightarrow{\mathbf{B}} \right\|_2 \leq \varepsilon.$$

*We denote an instance of this problem by the tuple* $(\overrightarrow{\mathbf{A}}, \overrightarrow{\mathbf{B}}, \varepsilon, \mathbf{r})$, *where* $\mathbf{r} = (r_1, \ldots, r_p)$.

Due to the block diagonal structure of the problem above, we can partition the tuples $\overrightarrow{\mathbf{A}}, \overrightarrow{\mathbf{B}}$ into blocks $\{\mathbf{A}_{i,j,k}, \mathbf{B}_{i,j,k} \mid i \in [m], j, k \in [p]\}$ where $\mathbf{A}_{i,j,k}, \mathbf{B}_{i,j,k} \in \mathbb{C}^{r_j \times r_k}$. Then, the problem is to find unitary matrices $\mathbf{U}_1, \cdots, \mathbf{U}_p$ such that

$$\forall i \in [m],\ j, k \in [p] : \|\mathbf{U}_j \mathbf{A}_{i,j,k} \mathbf{U}_k^\dagger - \mathbf{B}_{i,j,k}\|_2 \leq \varepsilon\ .$$

**Main Idea.** At a high level, our algorithm will recursively decompose any instance of Problem (F.2) into small instances of Problem (F.1) where both the singular value gap and the eigenvalue gap of each matrix $\mathbf{A}_i, \mathbf{B}_i$ is small. We call such small instances *"near identity,"* because each $\mathbf{A}_i$ and $\mathbf{B}_i$ must be close to being a complex scaler multiple of the identity matrix (see Lemma F.6). Since checking simultaneous conjugation for "near-identity" cases is trivial, we can piece together the solutions of those small problems back into a solution of Problem (F.2).

**Roadmap.**

- In Section F.1, we give matrix lemmas that we be served for the purpose of decomposing matrices into smaller pieces.
- In Section F.2, we state lemmas which reduce instances of Problem (F.2) into smaller pieces.
- In Section F.3, we state our algorithms: Algorithm 4 for decomposing an instance of Problem (F.2) recursively into "near identity" instances, and Algorithm 5 for checking unitary equivalence.
- In Section F.4, we prove the correctness of the algorithms, and thus prove Theorem M3.

## F.1 Matrix Lemmas

- In Section F.1.1, we show that if a matrix $\mathbf{A}$ is close to matrix $\mathbf{B}$ under unitary transformation (i.e., $\mathbf{UAV} \approx \mathbf{B}$), and if $\mathbf{A}$ has a gap in its *singular values*, then we can find a suitable basis where $\mathbf{A}$ and $\mathbf{B}$ are close to being block diagonal. (See Lemma F.4)
- In Section F.1.2, we show that if a matrix $\mathbf{A}$ is close to matrix $\mathbf{B}$ under unitary conjugation (i.e., $\mathbf{UAU}^\dagger \approx \mathbf{B}$), and if $\mathbf{A}$ has a gap in its *eigenvalues*, then we can find a suitable basis where $\mathbf{A}$ and $\mathbf{B}$ are close to being block diagonal. (See Lemma F.5)
- In Section F.1.3, we show that if a matrix $\mathbf{A}$ has its singular values being close to each other, and eigenvalues also being close to each other, then it must be close to a scalar of the identity matrix. (See Lemma F.6.)

### F.1.1 Wedin using singular values

We first recall a lemma (known as Wedin's theorem) that tells us that we can divide matrices into smaller blocks whenever there is a gap in their (sorted) singular values:

**Lemma F.3** (Wedin [79]). *Let* $n \geq d$ *be integers,* $\varepsilon > 0$ *a positive real, and*

$$\mathbf{\Sigma} = \begin{pmatrix} \mathit{diag}(\sigma_1, \cdots, \sigma_d) \\ 0 \end{pmatrix}, \mathbf{\Sigma}' = \begin{pmatrix} \mathit{diag}(\sigma'_1, \cdots, \sigma'_d) \\ 0 \end{pmatrix}$$

*be two matrices in* $\mathbb{R}^{n\times d}$ *with non-negative, non-increasing diagonal entries. If there exists* $i \in [d-1]$ *for which* $\sigma_i - \sigma_{i+1} > \varepsilon$*, then for any* $\mathbf{U} \in \mathsf{U}_n(\mathbb{C}), \mathbf{V} \in \mathsf{U}_d(\mathbb{C})$ *unitary matrices such that*

$$\|\mathbf{U\Sigma V} - \mathbf{\Sigma}'\|_2 \le \varepsilon$$

*we must have*

$$\mathbf{U} = \begin{pmatrix} \mathbf{U}_{1,1} & \mathbf{U}_{1,2} \\ \mathbf{U}_{2,1} & \mathbf{U}_{2,2} \end{pmatrix} \quad \text{and} \quad \mathbf{V} = \begin{pmatrix} \mathbf{V}_{1,1} & \mathbf{V}_{1,2} \\ \mathbf{V}_{2,1} & \mathbf{V}_{2,2} \end{pmatrix}$$

*with* $\mathbf{U}_{1,1}, \mathbf{V}_{1,1} \in \mathsf{Mat}_i(\mathbb{C})$*, such that:*

$$\|\mathbf{U}_{1,2}\|_2, \quad \|\mathbf{U}_{2,1}\|_2, \quad \|\mathbf{V}_{1,2}\|_2, \quad \|\mathbf{V}_{2,1}\|_2 \quad \le \quad \frac{\varepsilon}{\sigma_i - \sigma_{i+1} - \varepsilon}.$$

Lemma F.3 has the following corollary. If two matrices $\mathbf{A}$ and $\mathbf{B}$ (of the same dimension) are close up to left-right unitary transformation, and if $\mathbf{A}$ has a gap in its (sorted) singular values, then we can find a left-right uniform transformation to send $\mathbf{A}$ and $\mathbf{B}$ into $\mathbf{A}' = \begin{pmatrix} \mathbf{A}'_{11} & \mathbf{A}'_{12} \\ \mathbf{A}'_{21} & \mathbf{A}'_{22} \end{pmatrix}$ and $\mathbf{B}' = \begin{pmatrix} \mathbf{B}'_{11} & \mathbf{B}'_{12} \\ \mathbf{B}'_{21} & \mathbf{B}'_{22} \end{pmatrix}$, such that the corresponding pairs of matrices $(\mathbf{A}_{i,j}, \mathbf{B}_{ij})$ are also close up to unitary transformation.

We summarize this corollary formally as below:

**Lemma F.4** (corollary 1 of Lemma F.3). *Let* $\mathbf{A}, \mathbf{B} \in \mathbb{C}^{n\times d}$ *be two matrices with* $n \ge d$*, having spectral norm at most* $\lambda$*, and* $\varepsilon > 0$*. For every* $\delta > \varepsilon$*, we can decide in time* $\mathsf{poly}\left(n, \log\frac{\lambda}{\varepsilon}\right)$ *which of the following holds:*

1. *All singular values of* $\mathbf{A}$ *(*$n$ *of them, including zeros if* $n > d$*) are within pairwise distance* $n\delta$*; and the same holds for* $\mathbf{B}$*.*
2. *There exists a positive integer* $r \in [n-1]$ *and unitary matrices* $\mathbf{U}_L, \mathbf{U}_R \in \mathsf{U}_n(\mathbb{C})$ *such that the following holds:*

$$\text{for any } \mathbf{U} \in \mathsf{U}_n(\mathbb{C}),\ \mathbf{V} \in \mathsf{U}_d(\mathbb{C}) \text{ with} \quad \|\mathbf{UAV} - \mathbf{B}\|_2 \le \varepsilon$$

$$\implies \text{there exists } \mathbf{U}_1 \in \mathsf{U}_r(\mathbb{C}), \mathbf{U}_2 \in \mathsf{U}_{n-r}(\mathbb{C}) \text{ such that} \quad \|\mathbf{U}_L \mathit{diag}(\mathbf{U}_1, \mathbf{U}_2)\mathbf{U}_R - \mathbf{U}\|_2 \le \frac{\varepsilon}{\delta - \varepsilon}.$$

   *Moreover, we can compute* $r$ *and matrices* $\mathbf{U}_L, \mathbf{U}_R$ *in time* $\mathsf{poly}\left(n, \log\frac{\lambda}{\varepsilon}\right)$ *via SVD decomposition on* $\mathbf{A}, \mathbf{B}$*. Here,* $\mathit{diag}(\mathbf{U}_1, \mathbf{U}_2) = \begin{pmatrix} \mathbf{U}_1 & 0 \\ 0 & \mathbf{U}_2 \end{pmatrix}$*.*

*Proof sketch.* Let $\mathbf{A}' = \begin{pmatrix} \mathbf{A} & 0 \end{pmatrix}$ and $\mathbf{B}' = \begin{pmatrix} \mathbf{B} & 0 \end{pmatrix}$ be $n \times n$ matrices, $\mathbf{U}' = \mathbf{U}$ and $\mathbf{V}' = \begin{pmatrix} \mathbf{V} & 0 \\ 0 & \mathbf{I}_{n-d} \end{pmatrix}$. Then we have $\|\mathbf{UAV} - \mathbf{B}\|_2 \le \varepsilon \implies \|\mathbf{U}'\mathbf{A}'\mathbf{V}' - \mathbf{B}'\|_2 \le \varepsilon$. We can compute the SVD decomposition of $\mathbf{A}', \mathbf{B}'$ as $\mathbf{A}' = \mathbf{U}_1 \mathbf{\Sigma} \mathbf{V}_1$, and $\mathbf{B}' = \mathbf{U}_2 \mathbf{\Sigma}' \mathbf{V}_2$, and apply Lemma F.3 on $\mathbf{U}'' = \mathbf{U}_2^\dagger \mathbf{U}' \mathbf{U}_1$, $\mathbf{V}'' = \mathbf{V}_1 \mathbf{V}' \mathbf{V}_2^\dagger$, $\mathbf{\Sigma}$, $\mathbf{\Sigma}'$. □

To sum up, if the singular values of $\mathbf{A}$ (or $\mathbf{B}$) are sufficiently different, and if there exists unitary matrices $\mathbf{U}, \mathbf{V}$ such that $\|\mathbf{UAV} - \mathbf{B}\|_2$ is sufficiently small, then we can apply Lemma F.4 to reduce the task of finding $\mathbf{U}$ to the smaller-sized tasks of finding $\mathbf{U}_1$ and $\mathbf{U}_2$. This reduces the dimension of the problem.

Unfortunately, if the (real) singular values of $\mathbf{A}$ (and $\mathbf{B}$) are all close to each other, we cannot further reduce the dimension using Lemma F.4. In such a case, $\mathbf{A}$ and $\mathbf{B}$ are both (close to being) unitary matrices, but their (complex) eigenvalues may still be very different. Therefore, finding $\mathbf{U}$ is still a non-trivial task, as we shall see in the next subsection.

### F.1.2 Wedin using eigenvalues

The next lemma allows us to reduce the dimension if the eigenvalues of $\mathbf{A}$ (or $\mathbf{B}$) are far apart.

**Lemma F.5** (corollary 2 of Lemma F.3). *Let $\mathbf{A}, \mathbf{B} \in \mathsf{Mat}_n(\mathbb{C})$ with spectral norm at most $\lambda$ and for $\varepsilon > 0$. Suppose*

$$\text{there exists } \mathbf{U} \in \mathsf{U}_n(\mathbb{C}) \text{ such that } \quad \left\|\mathbf{U}\mathbf{A}\mathbf{U}^\dagger - \mathbf{B}\right\|_2 \le \varepsilon \ .$$

*Then, for every $\delta > \varepsilon$, we can decide in time $\mathsf{poly}\left(n, \log\lambda, \log\frac{1}{\varepsilon}\right)$ which of the following holds:*

1. *All the eigenvalues of $\mathbf{A}$ (and of $\mathbf{B}$) are within pairwise distance $n\delta$ (in complex modulus).*
2. *There exists $r \in [n-1]$ and unitary matrices $\mathbf{U}_L, \mathbf{U}_R \in \mathsf{U}_n(\mathbb{C})$ and $\mathbf{U}_1 \in \mathsf{U}_r(\mathbb{C}), \mathbf{U}_2 \in \mathsf{U}_{n-r}(\mathbb{C})$ such that*
$$\|\mathbf{U}_L \mathit{diag}(\mathbf{U}_1, \mathbf{U}_2)\mathbf{U}_R - \mathbf{U}\|_2 \le \frac{\varepsilon}{\delta - \varepsilon}.$$
*Moreover, we can find the integer $r$ and the matrices $\mathbf{U}_L, \mathbf{U}_R \in \mathsf{U}_n(\mathbb{C})$ in time $\mathsf{poly}\left(n, \log\lambda, \log\frac{1}{\varepsilon}\right)$.*

*Proof of Lemma F.5.* Suppose there exist eigenvalues $\lambda_1, \lambda_2$ of $\mathbf{B}$ with (unit) eigenvectors $u_1, u_2$ such that $|\lambda_1 - \lambda_2| \ge n\delta$. Let $\mathbf{A}_1 = \mathbf{A} - \lambda_1\mathbf{I}$, $\mathbf{B}_1 = \mathbf{B} - \lambda_1\mathbf{I}$, and let $\sigma_1 \ge \sigma_2 \ge \cdots \ge \sigma_n$ be the singular values of $\mathbf{B}_1$. Then the following holds:

1. $\left\|\mathbf{U}\mathbf{A}_1\mathbf{U}^\dagger - \mathbf{B}_1\right\|_2 = \left\|\mathbf{U}\mathbf{A}\mathbf{U}^\dagger - \mathbf{B}\right\|_2 \le \varepsilon$.
2. $\sigma_n = 0$, since $\mathbf{B}_1 u_1 = 0$.
3. $\sigma_1 \ge |\lambda_1 - \lambda_2| \ge n\delta$, since $\|\mathbf{B}_1 u_2\|_2 = \|(\lambda_2 - \lambda_1)u_2\|_2 = |\lambda_1 - \lambda_2|$.

Thus, we can apply Lemma F.4 to $\mathbf{A}_1$ and $\mathbf{B}_1$ to complete the proof.[28] □

To sum up, if $\mathbf{A}$ and $\mathbf{B}$ are square matrices with $\|\mathbf{U}\mathbf{A}\mathbf{U}^\dagger - \mathbf{B}\|_2$ being sufficiently small for some unknown unitary matrix $\mathbf{U}$, then we can use Lemma F.5 to reduce the task of finding $\mathbf{U}$ to smaller-sized problems, as long as the eigenvalues of $\mathbf{A}$ (or $\mathbf{B}$) are sufficiently different.

We emphasize that even if a matrix has identical eigenvalues, it may not be a multiple of the identity matrix. For instance, matrix $\begin{pmatrix} 1 & 1 \\ 0 & 1 \end{pmatrix}$ has a unique eigenvalue 1 (of multiplicity 2) but is not identity. In this case, however, the matrix has two distinct singular values.

### F.1.3 Stopping criterion

We can alternatively apply Lemma F.4 and Lemma F.5 until *both singular and eigenvalues* of a matrix $\mathbf{A}$ (and $\mathbf{B}$) become approximately equal. In such a case, the matrix $\mathbf{A}$ (and $\mathbf{B}$) must be a scalar multiple of the identity, see the lemma below:

**Lemma F.6** (identity matrix). *Given matrix $\mathbf{A} \in \mathsf{Mat}_n(\mathbb{C})$ and $\delta > 0$ such that, all the singular values of $\mathbf{A}$ are within distance $\delta$ and all eigenvalues of $\mathbf{A}$ are within distance $\delta$, then*

$$\exists c \in \mathbb{C}\colon \quad \|\mathbf{A} - c\mathbf{I}\|_2 \le 11n\delta.$$

*Proof of Lemma F.6.* We can assume that $\|\mathbf{A}\|_2 > 0$ as otherwise we can set $c = 0$ and are done.

Since all the singular values of $\mathbf{A}$ are within distance $\delta$, we know that there exists a unitary matrix $\mathbf{Q}$ and real $\gamma > 0$ such that $\|\mathbf{A} - \gamma\mathbf{Q}\|_2 \le \delta$.

[28] This approach is fundamentally different from taking $\mathbf{A}\mathbf{A}^\dagger$ first and then subtracting $\lambda\mathbf{I}$, since $\mathbf{A}_1\mathbf{A}_1^\dagger = (\mathbf{A} - \lambda_1\mathbf{I})(\mathbf{A} - \lambda_1\mathbf{I})^\dagger$ and not of the form $\mathbf{A}\mathbf{A}^\dagger - \lambda\mathbf{I}$.

If the eigenvalues of $\mathbf{Q}$ are within distance $\le \frac{10n\delta}{\gamma}$, there exists some complex $\xi \in \mathbb{C}$ so that $\|\xi\mathbf{I} - \mathbf{Q}\| \le \frac{10n\delta}{\gamma}$. In this case, we will be done (by choosing $c = \gamma\xi$) because

$$\|\mathbf{A} - \gamma\xi\mathbf{I}\|_2 \le \|\mathbf{A} - \gamma\mathbf{Q}\|_2 + \|\gamma\mathbf{Q} - \gamma\xi\mathbf{I}\|_2 = \|\mathbf{A} - \gamma\mathbf{Q}\|_2 + \gamma\|\mathbf{Q} - \xi\mathbf{I}\|_2 \le 11n\delta \ .$$

Therefore, let us suppose, by way of contradiction, that there exist eigenvalues $\lambda_1, \lambda_2$ of $\mathbf{Q}$ such that $\gamma \cdot |\lambda_1 - \lambda_2| \ge 10n\delta$. Writing $\mathbf{A} = \gamma\mathbf{Q} + \mathbf{E}$ with $\|\mathbf{E}\|_2 \le \delta$, we can consider the following matrix-valued function $\mathbf{A}(t)$ defined as:

$$\mathbf{A}(t) = \gamma\mathbf{Q} + t\mathbf{E}$$

Applying Theorem F.7, we get continuous functions $\mu_1(t), \cdots, \mu_n(t)$ representing the eigenvalues of $\mathbf{A}(t)$. Since all the eigenvalues of $\mathbf{A} = \mathbf{A}(1)$ are all within distance $\delta$, we know that there exists $x \in \mathbb{C}$ such that $|\mu_j(1) - x| \le \delta$ for all $j = 1, 2, \ldots, n$.

Since $|\lambda_1 - \lambda_2| \ge \frac{10n\delta}{\gamma}$, we cannot both have $|\gamma\lambda_1 - x| < 5n\delta$ and $|\gamma\lambda_2 - x| < 5n\delta$. Therefore, without loss of generality we assume $|\gamma\lambda_1 - x| \ge 5n\delta$. Also without loss of generality, let $\mu_1(0) = \gamma\lambda_1$ so $|\mu_1(1) - \mu_1(0)| \ge 4n\delta$.

Therefore, there exists $t \in [0, 1]$ such that for the eigenvalues $\lambda_1, \ldots, \lambda_n$ of $\mathbf{Q}$, we have: $\forall i \in [n], |\gamma\lambda_i - \mu_1(t)| \ge 2\delta$. (This is so because if we plot $\mu_1 \colon [0, 1] \to \mathbb{C}$ on the complex plane, it is a curve connecting two points of distance $|\mu_1(0) - \mu_1(1)| \ge 4n\delta$. On the other hand, each $j \in [n]$ defines an open ball $\{x \in \mathbb{C} \colon |x - \gamma\lambda_j| < 2\delta\}$ on this plane. Such $n$ balls of diameter $4\delta$ cannot cover the entire path, so there must exist $t \in [0, 1]$ with $|\mu_1(t) - \gamma\lambda_j| \ge 2\delta$ for all $j \in [n]$.)

Thus, the following holds for any unit vector $v \in \mathbb{C}^n$:

$$\begin{aligned}
\|(\mu_1(t)\mathbf{I} - \mathbf{A}(t))v\|_2 &\ge \|(\mu_1(t)\mathbf{I} - \gamma\mathbf{Q} - t\mathbf{E})v\|_2 \\
&\ge \|(\mu_1(t)\mathbf{I} - \gamma\mathbf{Q})v\|_2 - \|\mathbf{E}v\|_2 && \text{(triangle inequality and } t \le 1) \\
&\ge \min_{i \in [d]} |\mu_1(t) - \gamma\lambda_i| - \delta > 0 && (\mathbf{Q} \text{ is a normal and } \|E\|_2 \le \delta)
\end{aligned}$$

The above inequality contradicts the fact that $\lambda_1(t)$ is an eigenvalue of $\mathbf{A}(t)$. This completes the proof. □

**Theorem F.7** (continuity of eigenvalues [53], Theorem 5.2). *For every integer $n > 0$, if $\mathbf{A}(t) \colon [0, 1] \to \mathsf{Mat}_n(\mathbb{C})$ is a continuous function, then there exist $n$ continuous functions $\lambda_1(t), \cdots, \lambda_n(t) \colon [0, 1] \to \mathbb{C}$ such that the eigenvalues of $\mathbf{A}(t)$ are equal to the set $\{\lambda_1(t), \cdots \lambda_n(t)\}$ for all $t \in [0, 1]$.*

## F.2 Decomposition Lemmas

We first introduce some notations.

**Definition F.8.** *Given an instance $(\overrightarrow{\mathbf{A}}, \overrightarrow{\mathbf{B}}, \varepsilon, \mathbf{r})$ of Problem (F.2), where $\mathbf{r} = (r_1, \ldots, r_p)$, and given $S \subset [p]$, let $(\overrightarrow{\mathbf{A}}^S, \overrightarrow{\mathbf{B}}^S, \lambda, \varepsilon, \mathbf{r}^S)$ be defined as follows: $\overrightarrow{\mathbf{A}}^S$ is a tuple of matrices where $\mathbf{A}_i^S$ is the block matrix given by $(\mathbf{A}_{ijk})$ where $j, k \in S$ (analogous definition for $\overrightarrow{\mathbf{B}}^S$), $\mathbf{r}^S = (r_i)_{i \in S}$.*

**Definition F.9** (complexity potential). *Given an instance $(\overrightarrow{\mathbf{A}}, \overrightarrow{\mathbf{B}}, \varepsilon, \mathbf{r})$ where $\overrightarrow{\mathbf{A}}, \overrightarrow{\mathbf{B}} \in \mathsf{Mat}_n(\mathbb{C})^m$, define its complexity potential function $P((\overrightarrow{\mathbf{A}}, \overrightarrow{\mathbf{B}}, \varepsilon, \mathbf{r})) \stackrel{\text{def}}{=} m \cdot n^2 + n^3$.*

As we shall see in our lemmas in this subsection, when decomposing an instance of Problem (F.2) into smaller pieces, the total complexity potential never increases.

**Definition F.10** (spectral indicator graph). *Let $(\overrightarrow{\mathbf{A}}, \overrightarrow{\mathbf{B}}, \varepsilon, \mathbf{r})$ be an instance of Problem (F.2), where $\mathbf{r} = (r_1, \ldots, r_p)$. For any $\delta > 0$, denote by $G_\delta(V, E)$ a graph with $V = [p]$, and $\{j, k\} \in E$ if there is* *one index $i \in [m]$ such that at least one of $\mathbf{A}_{i,j,k}$, $\mathbf{B}_{i,j,k}$, $\mathbf{A}_{i,k,j}$ or $\mathbf{B}_{i,k,j}$ has spectral norm at least $(n+1)\delta$.*

There are four main ways in which we can reduce the dimension of an instance from Problem (F.2). The first, and easiest to do, is when the spectral indicator graph is disconnected.

**Lemma F.11** (when graph is disconnected)**.** *Let $(\overrightarrow{\mathbf{A}}, \overrightarrow{\mathbf{B}}, \varepsilon, \mathbf{r})$, where $\mathbf{r} = (r_1, \ldots, r_p)$ and $p > 1$. Let $\delta > 0$. If $G_\delta$ is disconnected, $S$ is a connected component of $G_\delta$ and $T = [p] \setminus S$, then problems $(\overrightarrow{\mathbf{A}}^S, \overrightarrow{\mathbf{B}}^S, \varepsilon, \mathbf{r}^S)$ and $(\overrightarrow{\mathbf{A}}^T, \overrightarrow{\mathbf{B}}^T, \varepsilon, \mathbf{r}^T)$ satisfy the following conditions*

1. *If $\mathsf{diag}(\mathbf{U}_i)_{i \in [p]}$ is a solution to $(\overrightarrow{\mathbf{A}}, \overrightarrow{\mathbf{B}}, \varepsilon, \mathbf{r})$, then we have $\mathsf{diag}(\mathbf{U}_i)_{i \in S}$ is a solution to $(\overrightarrow{\mathbf{A}}^S, \overrightarrow{\mathbf{B}}^S, \varepsilon, \mathbf{r}^S)$ and $\mathsf{diag}(\mathbf{U}_j)_{j \in T}$ is a solution to $(\overrightarrow{\mathbf{A}}^T, \overrightarrow{\mathbf{B}}^T, \varepsilon, \mathbf{r}^T)$.*
2. *If $\tau > 0$ and $\mathsf{diag}(\mathbf{U}_i)_{i \in S}$ and $\mathsf{diag}(\mathbf{U}_j)_{j \in T}$ are respectively solutions to $(\overrightarrow{\mathbf{A}}^S, \overrightarrow{\mathbf{B}}^S, \tau, \mathbf{r}^S)$ and $(\overrightarrow{\mathbf{A}}^T, \overrightarrow{\mathbf{B}}^T, \tau, \mathbf{r}^T)$, then $\mathsf{diag}(\mathbf{U}_i)_{i \in [p]}$ is a solution to $(\overrightarrow{\mathbf{A}}, \overrightarrow{\mathbf{B}}, 2\tau + (n+1)^3\delta, \mathbf{r})$.*
3. *$P((\overrightarrow{\mathbf{A}}, \overrightarrow{\mathbf{B}}, \varepsilon, \mathbf{r})) > P((\overrightarrow{\mathbf{A}}^S, \overrightarrow{\mathbf{B}}^S, \varepsilon, \mathbf{r}^S)) + P((\overrightarrow{\mathbf{A}}^T, \overrightarrow{\mathbf{B}}^T, \varepsilon, \mathbf{r}^T))$*

In the second case, if there is a sufficiently large singular value gap in some block of the matrix (with respect to an edge of $G_\delta$), we can use Lemma F.4 to reduce the problem dimension. Formally,

**Lemma F.12** (large singular value gap)**.** *Let $(\overrightarrow{\mathbf{A}}, \overrightarrow{\mathbf{B}}, \varepsilon, \mathbf{r})$, where $\mathbf{r} = (r_1, \ldots, r_p)$, $\lambda > 0$ is an upper bound on the Frobenius norm of each $\mathbf{A}_i, \mathbf{B}_i$ and $\delta > 2\varepsilon$. If $G_\delta$ is connected and for an edge $\{j,k\}$ of $G_\delta$ the singular value gap of one of $\mathbf{A}_{i,j,k}, \mathbf{B}_{i,j,k}, \mathbf{A}_{i,k,j}, \mathbf{B}_{i,k,j}$ is $\geq \delta$, we can, in time $\mathsf{poly}\left(m, n, \log \frac{\lambda}{\varepsilon}\right)$, find $\overrightarrow{\mathbf{A}}'$, $\overrightarrow{\mathbf{B}}'$ and $\mathbf{s}$, where $\mathbf{s} = (s_1, \ldots, s_{p+1})$, such that the following hold:*

1. *The Frobenius norm of each block $\mathbf{A}'_{i,j,k}$ and $\mathbf{B}'_{i,j,k}$ is at most $\lambda$*
2. *If there is a solution to $(\overrightarrow{\mathbf{A}}, \overrightarrow{\mathbf{B}}, \varepsilon, \mathbf{r})$ then there exists a solution to $(\overrightarrow{\mathbf{A}}', \overrightarrow{\mathbf{B}}', (1 + 6\lambda/\delta)\varepsilon, \mathbf{s})$.*
3. *If $\tau > 0$, any solution to $(\overrightarrow{\mathbf{A}}', \overrightarrow{\mathbf{B}}', \tau, \mathbf{s})$ can be converted into a solution to $(\overrightarrow{\mathbf{A}}, \overrightarrow{\mathbf{B}}, \tau, \mathbf{r})$ in time $\mathsf{poly}\left(n, m, \log \frac{\lambda}{\tau}\right)$.*
4. *$P((\overrightarrow{\mathbf{A}}, \overrightarrow{\mathbf{B}}, \varepsilon, \mathbf{r})) = P((\overrightarrow{\mathbf{A}}', \overrightarrow{\mathbf{B}}', (1 + 6\lambda/\delta)\varepsilon, \mathbf{s}))$*

After the first two cases, we are *only left* to analyze the case (1) when the spectral indicator graph $G_\delta$ is connected and (2) the singular value gap of $\mathbf{A}_{i,j,k}, \mathbf{B}_{i,j,k}$ for all edges $\{j,k\}$ of graph $G_\delta$ and all $i \in [m]$ are $\leq \delta$. This means, in particular,

- each such matrix $\mathbf{A}_{i,j,k}$ or $\mathbf{B}_{i,j,k}$ must be invertible (because they have spectral norm $\geq (n+1)\delta$ but singular value gap $\leq \delta$) and thus must be a square matrix, meaning $r_j = r_k$. By the connectedness of the graph, this implies $r_1 = r_2 = \cdots = r_p = r$.
- each such matrix $\mathbf{A}_{i,j,k}$ or $\mathbf{B}_{i,j,k}$ must be close to being a scalar times a unitary matrix.

In the third case, we reduce the dimension of such instance whenever $p > 1$:

**Lemma F.13** (near unitary)**.** *Let $(\overrightarrow{\mathbf{A}}, \overrightarrow{\mathbf{B}}, \varepsilon, \mathbf{r})$, where $\mathbf{r} = (r_1, \ldots, r_p)$ with $p > 1$ and each $r_i = r$. Let $\delta > \varepsilon$ be a positive real number and $\lambda > 1$ be an upper bound on the Frobenius norm of all blocks $\mathbf{A}_{i,j,k}, \mathbf{B}_{i,j,k}$. If $G_\delta$ is connected and all blocks $\mathbf{A}_{i,j,k}, \mathbf{B}_{i,j,k}$ have singular value gap $\leq \delta$, then we can find $\overrightarrow{\mathbf{A}}'$, $\overrightarrow{\mathbf{B}}' \in \mathsf{Mat}_r(\mathbf{C})^{mp^2+p}$ in deterministic time $\mathsf{poly}(m, n, \log(\lambda/\varepsilon))$ such that the following hold:*

1. *the spectral norms of $\mathbf{A}'_\ell, \mathbf{B}'_\ell$ are bounded by $\lambda$, for $\ell \in [mp^2 + p]$.*

2. *If* $\mathsf{diag}(\mathbf{U}_i)_{i\in[p]}$ *is a solution to* $(\overrightarrow{\mathbf{A}}, \overrightarrow{\mathbf{B}}, \varepsilon, \mathbf{r})$*, then* $\mathbf{U}_1$ *is a solution to* $\left(\overrightarrow{\mathbf{A}}', \overrightarrow{\mathbf{B}}', 3(2\lambda/\delta)^{4p}\varepsilon, r\right)$.
3. *Any solution to* $\left(\overrightarrow{\mathbf{A}}', \overrightarrow{\mathbf{B}}', \tau, r\right)$ *can be converted into a solution to* $\left(\overrightarrow{\mathbf{A}}, \overrightarrow{\mathbf{B}}, 4(\lambda/\delta)^{4p+1}\tau, \mathbf{r}\right)$ *in deterministic time* $\mathsf{poly}(n, m, \log(\lambda/\tau))$.
4. $P((\overrightarrow{\mathbf{A}}, \overrightarrow{\mathbf{B}}, \varepsilon, \mathbf{r})) > P((\overrightarrow{\mathbf{A}}', \overrightarrow{\mathbf{B}}', 3(2\lambda/\delta)^{4p}\varepsilon, r))$

In the fourth case, we reduce the dimension of whenever $p = 1$ and there is some sufficiently large eigenvalue gap. We apply Lemma F.5 to further reduce the problem dimension.

**Lemma F.14** (large eigenvalue gap)**.** *Let* $(\overrightarrow{\mathbf{A}}, \overrightarrow{\mathbf{B}}, \varepsilon, r)$*,* $\delta > \varepsilon$ *be a positive real, and* $\lambda > 1$ *be an upper bound on the spectral norm of all matrices* $\mathbf{A}_i, \mathbf{B}_i$*. Assume that each* $\mathbf{A}_i, \mathbf{B}_i$ *has singular value gap* $\leq \delta$*.*

*If there is* $i \in [m]$ *such that* $\mathbf{A}_i$ *or* $\mathbf{B}_i$ *has two eigenvalues which are* $(r\delta)$*-far apart, we can, in time* $\mathsf{poly}\left(m, n, \log \frac{\lambda}{\varepsilon}\right)$*, find* $\overrightarrow{\mathbf{A}}', \overrightarrow{\mathbf{B}}' \in \mathsf{Mat}_r(\mathbb{C})^m$ *and* $\mathbf{r} = (r_1, r_2)$ *satisfying* $r = r_1 + r_2$ *and:*

1. *The spectral norms of each block* $\mathbf{A}'_{i,j,k}$ *and* $\mathbf{B}'_{i,j,k}$ *are at most* $\lambda$
2. *If there is a solution to* $(\overrightarrow{\mathbf{A}}, \overrightarrow{\mathbf{B}}, \varepsilon, r)$ *then there exists a solution to* $(\overrightarrow{\mathbf{A}}', \overrightarrow{\mathbf{B}}', (1 + 2\lambda/\delta)\varepsilon, \mathbf{r})$.
3. *If* $\tau > 0$*, any solution to* $(\overrightarrow{\mathbf{A}}', \overrightarrow{\mathbf{B}}', \tau, \mathbf{r})$ *can be converted, in time* $\mathsf{poly}\left(n, m, \log \frac{\lambda}{\tau}\right)$*, into a solution to* $(\overrightarrow{\mathbf{A}}, \overrightarrow{\mathbf{B}}, \tau, r)$.
4. $P((\overrightarrow{\mathbf{A}}, \overrightarrow{\mathbf{B}}, \varepsilon, \mathbf{r})) = P((\overrightarrow{\mathbf{A}}', \overrightarrow{\mathbf{B}}', (1 + 2\lambda/\delta)\varepsilon, \mathbf{s}))$

Finally, if we cannot apply any of the four reduction lemmas above, it must be the case that $p = 1$ and our problem is of the form $(\overrightarrow{\mathbf{A}}, \overrightarrow{\mathbf{B}}, \varepsilon, r)$ where each matrix $\mathbf{A}_i, \mathbf{B}_i$ has singular value gap $\leq \delta$ and eigenvalue gap $\leq \delta$. In this case, Lemma F.6 applies and therefore we have that each $\mathbf{A}_i$ and $\mathbf{B}_i$ is close (up to scalar) to the identity matrix.

### F.2.1 Proof of Lemma F.11

*Proof of Lemma F.11.* We can compute the spectral norm of each block $\mathbf{A}_{i,j,k}$ and $\mathbf{B}_{i,j,k}$, and therefore construct the graph $G_\delta$, in deterministic $\mathsf{poly}(n, m, \log(\lambda/\varepsilon))$ time. If $G_\delta$ is disconnected, let $S$ be a disconnected component of $G_\delta$ and $T = [p] \setminus S$. In this case, after rearranging the blocks in the matrices in $\overrightarrow{\mathbf{A}}, \overrightarrow{\mathbf{B}}$, we can write

$$\overrightarrow{\mathbf{A}} = \begin{pmatrix} \overrightarrow{\mathbf{A}}^S & 0 \\ 0 & \overrightarrow{\mathbf{A}}^T \end{pmatrix} + \overrightarrow{\mathbf{E}} \quad \text{and} \quad \overrightarrow{\mathbf{B}} = \begin{pmatrix} \overrightarrow{\mathbf{B}}^S & 0 \\ 0 & \overrightarrow{\mathbf{B}}^T \end{pmatrix} + \overrightarrow{\mathbf{F}},$$

where $\left\|\overrightarrow{\mathbf{E}}\right\|_2, \left\|\overrightarrow{\mathbf{F}}\right\|_2 \leq (n+1)^2\delta$, as each nonzero block $\{j, k\}$ in $\overrightarrow{\mathbf{E}}, \overrightarrow{\mathbf{F}}$ corresponds to a block where $\{j, k\}$ is not an edge of $G_\delta$, and by definition this implies that the spectral norm of these blocks is smaller than $(n + 1)\delta$.

From this construction one can easily see that the three properties are satisfied. □

### F.2.2 Proof of Lemma F.12

*Proof of Lemma F.12.* We can compute the SVD of each block $\mathbf{A}_{i,j,k}$ and $\mathbf{B}_{i,j,k}$, and therefore compute the largest singular value gap of each block in deterministic $\mathsf{poly}(n, \log(\lambda/\varepsilon))$ time.

Since our hypothesis is that there exist an edge of $G_\delta$ with singular value gap $\geq \delta$, this means (by the definition of $G_\delta$) some block must simultaneously have spectral norm $\geq (n + 1)\delta$ and singular value gap $\geq \delta$. Let $\{\mathbf{A}_{i,j,k}, \mathbf{B}_{i,j,k}, \mathbf{A}_{i,k,j}, \mathbf{B}_{i,k,j}\}$ be such a block and w.l.o.g. assume that $\mathbf{A}_{i,j,k}$ has spectral norm $\geq (n+1)\delta$ and singular value gap $\geq \delta$.

By Lemma F.4, we can find unitary matrices $\mathbf{U}_L, \mathbf{U}_R$ and positive integers $r'_j, r''_j$ satisfying $r'_j + r''_j = r_j$ such that there exist unitary matrices $\mathbf{U}'_j, \mathbf{U}''_j$ with dimensions $r'_j, r''_j$ and

$$\|\mathbf{U}_L \mathsf{diag}(\mathbf{U}'_j, \mathbf{U}''_j)\mathbf{U}_R - \mathbf{U}_j\|_2 \leq \frac{\varepsilon}{\delta - \varepsilon}.$$

Thus, the following holds for every $k \in [p]$, $k \neq j$, $i \in [m]$:

$$\left\|\mathbf{U}_L \mathsf{diag}(\mathbf{U}'_j, \mathbf{U}''_j)\mathbf{U}_R \mathbf{A}_{i,j,k}\mathbf{U}_k^\dagger - \mathbf{B}_{i,j,k}\right\|_2 \leq \varepsilon + \|\mathbf{A}_{i,j,k}\|_2 \times \frac{\varepsilon}{\delta-\varepsilon} \leq \left(1 + \frac{2\lambda}{\delta}\right)\varepsilon$$

$$\implies \left\|\mathsf{diag}(\mathbf{U}'_j, \mathbf{U}''_j)\left(\mathbf{U}_R \mathbf{A}_{i,j,k}\right)\mathbf{U}_k^\dagger - \mathbf{U}_L^\dagger \mathbf{B}_{i,j,k}\right\|_2 \leq \left(1 + \frac{2\lambda}{\delta}\right)\varepsilon \ ,$$

and for every $i \in [m]$:

$$\left\|\mathbf{U}_L \mathsf{diag}(\mathbf{U}'_j, \mathbf{U}''_j)\mathbf{U}_R \mathbf{A}_{i,j,j}\mathbf{U}_R^\dagger \mathsf{diag}(\mathbf{U}'_j, \mathbf{U}''_j)^\dagger \mathbf{U}_L^\dagger - \mathbf{B}_{i,j,j}\right\|_2 \leq \varepsilon + \|\mathbf{A}_{i,j,j}\|_2 \times \frac{3\varepsilon}{\delta-\varepsilon} \leq \left(1 + \frac{6\lambda}{\delta}\right)\varepsilon$$

$$\implies \left\|\mathsf{diag}(\mathbf{U}'_j, \mathbf{U}''_j)\left(\mathbf{U}_R \mathbf{A}_{i,j,j}\mathbf{U}_R^\dagger\right)\mathsf{diag}(\mathbf{U}'_j, \mathbf{U}''_j)^\dagger - \mathbf{U}_L^\dagger \mathbf{B}_{i,j,j}\mathbf{U}_L\right\|_2 \leq \left(1 + \frac{6\lambda}{\delta}\right)\varepsilon \ .$$

Thus, we can replace $\mathbf{U}_j$ by $\mathbf{U}'_j, \mathbf{U}''_j$ and keep $\mathbf{U}_\ell$ unchanged for any $\ell \neq j$. We can also set $\mathbf{s} = (r_1, \ldots, r_{j-1}, r'_j, r''_j, r_{j+1}, \ldots, r_p)$. Let $a = r_1 + \cdots + r_{j-1}$ and $b = r_{j+1} + \cdots + r_p$ and define $\mathbf{L} = \mathsf{diag}(\mathbf{I}_a, \mathbf{U}_L, \mathbf{I}_b)$ and $\mathbf{R} = \mathsf{diag}(\mathbf{I}_a, \mathbf{U}_R, \mathbf{I}_b)$. Then, for all $i \in [m]$ make

$$\mathbf{A}'_i = \mathbf{R}\mathbf{A}_i\mathbf{R}^\dagger, \quad \text{and} \quad \mathbf{B}'_i = \mathbf{L}^\dagger \mathbf{B}_i \mathbf{L}.$$

With the change above, the max spectral norm of the new blocks of the matrices $\mathbf{A}_i$ and $\mathbf{B}_i$ are still bounded by $\lambda$, since $\|\mathbf{U}_L\|_2, \|\mathbf{U}_R\|_2 = 1$.

We have proved that if $(\overrightarrow{\mathbf{A}}, \overrightarrow{\mathbf{B}}, \varepsilon, \mathbf{r})$ has a solution, then $(\overrightarrow{\mathbf{A}}', \overrightarrow{\mathbf{B}}', (1+6\lambda/\delta)\varepsilon, \mathbf{s})$ also has a solution. Now, we need to show that any solution to $(\overrightarrow{\mathbf{A}}', \overrightarrow{\mathbf{B}}', \tau, \mathbf{s})$ can be converted into a solution to $(\overrightarrow{\mathbf{A}}, \overrightarrow{\mathbf{B}}, \tau, \mathbf{s})$.

To do this, suppose we get matrices $\mathbf{U}'_1, \cdots, \mathbf{U}'_p, \mathbf{U}'_{p+1}$ as solutions to $(\overrightarrow{\mathbf{A}}', \overrightarrow{\mathbf{B}}', \tau, \mathbf{s})$. We can just let $\mathbf{U}_\ell = \mathbf{U}'_\ell$ for $\ell < j$, $\mathbf{U}_\ell = \mathbf{U}_{\ell'+1}$ for $\ell > j$ and $\mathbf{U}_j = \mathbf{U}_L \mathsf{diag}(\mathbf{U}'_j, \mathbf{U}'_{j+1})\mathbf{U}_R$. This completes the proof. □

### F.2.3 Proof of Lemma F.13

*Proof of Lemma F.13.* Recall all blocks $\mathbf{A}_{i,j,k}, \mathbf{B}_{i,j,k}$ are in $\mathsf{Mat}_r(\mathbb{C})$. Since $G_\delta$ is connected, for each $t \in [p]$, we can find a path from 1 to $t$ in $G_\delta$. Without loss of generality, suppose this path is $1 \to 2 \to 3 \to \cdots \to j$ (by renaming the indices in the path).

By the assumption, we have $\sigma_{\min}(\mathbf{A}_{i_\ell,\ell,\ell+1}) \geq \delta$ for all $\ell \in [j-1]$. (If $\sigma_{\min}(\mathbf{A}_{i_\ell,\ell,\ell+1}) = 0$, since $G_\delta$ is connected we must have that $\sigma_{\min}(\mathbf{A}_{i_\ell,\ell+1,\ell}) \geq \delta$ and thus we could replace $\mathbf{A}_{i_\ell,\ell,\ell+1}$ by $\mathbf{A}^\dagger_{i_\ell,\ell+1,\ell}$ in the argument above. For simplicity, we will assume all $\mathbf{A}_{i_\ell,\ell,\ell+1}$ are invertible.)

Let $\mathsf{diag}(\mathbf{U}_i)_{i\in[p]}$ be a solution to $(\overrightarrow{\mathbf{A}}, \overrightarrow{\mathbf{B}}, \varepsilon, \mathbf{r})$. For all $\ell \in [j-1]$ we have

$$\left\|\mathbf{U}_\ell \mathbf{A}_{i_\ell,\ell,\ell+1}\mathbf{U}_{\ell+1}^\dagger - \mathbf{B}_{i_\ell,\ell,\ell+1}\right\|_2 \leq \varepsilon \ .$$

This implies $\mathbf{R}_\ell \stackrel{\text{def}}{=} \mathbf{U}_\ell \mathbf{A}_{i_\ell,\ell,\ell+1}\mathbf{U}_{\ell+1}^\dagger - \mathbf{B}_{i_\ell,\ell,\ell+1}$ satisfies $\|\mathbf{R}_\ell\|_2 \le \varepsilon$. Therefore,

$$\mathbf{U}_1 \prod_{\ell=1}^{j-1} \mathbf{A}_{i_\ell,\ell,\ell+1}\mathbf{U}_j^\dagger - \prod_{\ell=1}^{j-1} \mathbf{B}_{i_\ell,\ell,\ell+1} = \prod_{\ell=1}^{j-1}(\mathbf{B}_{i_\ell,\ell,\ell+1} + \mathbf{R}_\ell) - \prod_{\ell=1}^{j-1} \mathbf{B}_{i_\ell,\ell,\ell+1}$$

Using the assumption $\|\mathbf{B}_{i_\ell,\ell,\ell+1}\|_2 \le \lambda$, we have:

$$\left\| \mathbf{U}_1 \prod_{\ell=1}^{j-1} \mathbf{A}_{i_\ell,\ell,\ell+1}\mathbf{U}_j^\dagger - \prod_{\ell=1}^{j-1} \mathbf{B}_{i_\ell,\ell,\ell+1} \right\|_2 \le (2\lambda)^{j-1}\varepsilon \le (2\lambda)^p \varepsilon$$

Let $\mathbf{C}_j = \delta^{-j+1} \cdot \prod_{\ell=1}^{j-1} \mathbf{A}_{i_\ell,\ell,\ell+1}$ and $\mathbf{D}_j = \delta^{-j+1} \cdot \prod_{\ell=1}^{j-1} \mathbf{B}_{i_\ell,\ell,\ell+1}$. Then, $\sigma_{\min}(\mathbf{C}_j), \sigma_{\min}(\mathbf{D}_j) \ge 1$, which together with inequality above imply

$$\|\mathbf{U}_j^\dagger - \mathbf{C}_j^{-1}\mathbf{U}_1^\dagger \mathbf{D}_j\|_2 \le \left(\frac{2\lambda}{\delta}\right)^{j-1} \varepsilon \le \left(\frac{2\lambda}{\delta}\right)^p \varepsilon \ .$$

Now, for each $j \in [p]$, let us write $\mathbf{U}_j \stackrel{\text{def}}{=} \mathbf{D}_j^\dagger \mathbf{U}_1 \mathbf{C}_j^{-\dagger} + \mathbf{X}_j$ where $\|\mathbf{X}_j\|_2 \le \left(\frac{2\lambda}{\delta}\right)^p \cdot \varepsilon$ (given by the inequality above). Since $\mathbf{U}_j$ is unitary, we have:

$$\left\|(\mathbf{C}_j^{-1}\mathbf{U}_1^\dagger \mathbf{D}_j)^\dagger \mathbf{C}_j^{-1}\mathbf{U}_1^\dagger \mathbf{D}_j - \mathbf{I}\right\|_2 \le 3\left(\frac{2\lambda}{\delta}\right)^p \varepsilon \tag{F.1}$$

We are now ready to define $\overrightarrow{\mathbf{A}}'$ and $\overrightarrow{\mathbf{B}}'$. Let

$$\mathbf{A}'_{i,j,k} = \mathbf{C}_j^{-\dagger}\mathbf{A}_{i,j,k}\mathbf{C}_k^{-1} \text{ and } \mathbf{B}'_{i,j,k} = \mathbf{D}_j^{-\dagger}\mathbf{B}_{i,j,k}\mathbf{D}_k^{-1}.$$

In this case, using $\sigma_{\min}(\mathbf{C}_j), \sigma_{\min}(\mathbf{D}_j) \ge 1$, their spectral norms are bounded by

$$\left\|\mathbf{A}'_{i,j,k}\right\|_2 = \left\|\mathbf{C}_j^{-\dagger}\mathbf{A}_{i,j,k}\mathbf{C}_k^{-1}\right\|_2, \quad \left\|\mathbf{B}'_{i,j,k}\right\|_2 = \left\|\mathbf{D}_j^{-\dagger}\mathbf{B}_{i,j,k}\mathbf{D}_k^{-1}\right\|_2 \le \lambda. \tag{F.2}$$

For every $i \in [m], j, k \in [p]$, we have

$$\begin{aligned}
\left\|\mathbf{U}_1\mathbf{A}'_{i,j,k}\mathbf{U}_1^\dagger - \mathbf{B}'_{i,j,k}\right\|_2 &= \left\|\mathbf{U}_1\mathbf{C}_j^{-\dagger}\mathbf{A}_{i,j,k}\mathbf{C}_k^{-1}\mathbf{U}_1^\dagger - \mathbf{D}_j^{-\dagger}\mathbf{B}_{i,j,k}\mathbf{D}_k^{-1}\right\|_2 \\
&\le \left\|\mathbf{D}_j^{-\dagger}\right\|_2 \cdot \|\mathbf{D}_k^{-1}\|_2 \cdot \left\|\mathbf{D}_j^\dagger\mathbf{U}_1\mathbf{C}_j^{-\dagger}\mathbf{A}_{i,j,k}\mathbf{C}_k^{-1}\mathbf{U}_1^\dagger\mathbf{D}_k - \mathbf{B}_{i,j,k}\right\|_2 \\
&\le \left\|(\mathbf{U}_j - \mathbf{X}_j)\mathbf{A}_{i,j,k}(\mathbf{U}_k - \mathbf{X}_k)^\dagger - \mathbf{B}_{i,j,k}\right\|_2 \\
&\le \left\|\mathbf{U}_j\mathbf{A}_{i,j,k}\mathbf{U}_k^\dagger - \mathbf{B}_{i,j,k}\right\|_2 + \left\|\mathbf{X}_j\mathbf{A}_{i,j,k}\mathbf{U}_k^\dagger\right\|_2 + \left\|\mathbf{X}_j\mathbf{A}_{i,j,k}\mathbf{X}_k^\dagger\right\|_2 + \left\|\mathbf{U}_j\mathbf{A}_{i,j,k}\mathbf{X}_k^\dagger\right\|_2 \\
&\le \varepsilon + 2(2\lambda/\delta)^p\lambda\varepsilon + (2\lambda/\delta)^{2p}\lambda^2\varepsilon^2 \le \left(\frac{2\lambda}{\delta}\right)^{3p} \cdot \varepsilon.
\end{aligned} \tag{F.3}$$

To be able to convert any solution $\mathbf{U}_1$ to the matrices $\mathbf{A}'_{i,j,k}$ and $\mathbf{B}'_{i,j,k}$ above, we will need to add the matrices $\mathbf{E}_j = (\mathbf{C}_j\mathbf{C}_j^\dagger)^{-1}$ to the tuple of matrices $\overrightarrow{\mathbf{A}}'$ and the matrices $\mathbf{F}_j = (\mathbf{D}_j\mathbf{D}_j^\dagger)^{-1}$ to the tuple of matrices $\overrightarrow{\mathbf{B}}'$. We have

$$\begin{aligned}
\left\|\mathbf{U}_1\mathbf{E}_j\mathbf{U}_1^\dagger - \mathbf{F}_j\right\|_2 &= \left\|\mathbf{U}_1\mathbf{C}_j^{-\dagger}\mathbf{C}_j^{-1}\mathbf{U}_1^\dagger - \mathbf{D}_j^{-\dagger}\mathbf{D}_k^{-1}\right\|_2 \\
&\le \left\|\mathbf{D}_j^{-\dagger}\right\|_2 \cdot \|\mathbf{D}_k^{-1}\|_2 \cdot \left\|\mathbf{D}_j^\dagger\mathbf{U}_1\mathbf{C}_j^{-\dagger}\mathbf{C}_k^{-1}\mathbf{U}_1^\dagger\mathbf{D}_k - \mathbf{I}\right\|_2 \le 3\left(\frac{2\lambda}{\delta}\right)^p \cdot \varepsilon,
\end{aligned} \tag{F.4}$$

where the last inequality uses (F.1).

With the above facts at hand, let us now prove parts 1 and 2 of the lemma. We begin by noting that the set $\mathbf{A}'_{i,j,k}$ and $\mathbf{E}_j$ (as well as $\mathbf{B}'_{i,j,k}$ and $\mathbf{F}_j$) is a set of $\ell = mp^2 + p$ matrices in $\mathsf{Mat}_r(\mathbb{C})$. Thus, given any $0 < \tau < 1$ we will define our new problem $(\overrightarrow{\mathbf{A}}', \overrightarrow{\mathbf{B}}', \tau, r)$ by the matrices just described.

Inequality (F.2) and the fact that $\sigma_{\min}(\mathbf{C}_j), \sigma_{\min}(\mathbf{D}_j) \geq 1$ imply part 1.

To prove part 2, note that inequalities (F.3) and (F.4) imply that any solution $\mathsf{diag}(\mathbf{U}_1, \cdots, \mathbf{U}_p)$ to $(\overrightarrow{\mathbf{A}}, \overrightarrow{\mathbf{B}}, \varepsilon, \mathbf{r})$ yield the solution $\mathbf{U}_1$ to $\left(\overrightarrow{\mathbf{A}}', \overrightarrow{\mathbf{B}}', 3\left(\frac{2\lambda}{\delta}\right)^{3p} \cdot \varepsilon, r\right)$.

Let us prove part 3. Let $\mathbf{U}_1$ be a solution to $(\overrightarrow{\mathbf{A}}', \overrightarrow{\mathbf{B}}', \tau, r)$. Thus, $\left\|\mathbf{U}_1\mathbf{A}'_{i,j,k}\mathbf{U}_1^\dagger - \mathbf{B}'_{i,j,k}\right\|_2 \leq \tau$ and $\left\|\mathbf{U}_1\mathbf{E}_j\mathbf{U}_1^\dagger - \mathbf{F}_j\right\|_2 \leq \tau$. From the latter inequality we obtain

$$\left\|\mathbf{D}_j^\dagger\mathbf{U}_1\mathbf{C}_j^{-\dagger}\mathbf{C}_j^{-1}\mathbf{U}_1^\dagger\mathbf{D}_j - \mathbf{I}\right\|_2 \leq \left\|\mathbf{D}_j^\dagger\right\|_2 \cdot \|\mathbf{D}_j\|_2 \cdot \left\|\mathbf{U}_1\mathbf{E}_j\mathbf{U}_1^\dagger - \mathbf{F}_j\right\|_2 \leq \left(\frac{\lambda}{\delta}\right)^{2p} \cdot \tau.$$

From equation above, we know that $\mathbf{D}_j^\dagger\mathbf{U}_1\mathbf{C}_j^{-\dagger}$ is approximately a unitary matrix. Let $\mathbf{U}_j$ be the unitary matrix obtained by running SVD on $\mathbf{D}_j^\dagger\mathbf{U}_1\mathbf{C}_j^{-\dagger}$ and setting all singular values to be one (this can be done in deterministic time $\mathsf{poly}(r\log\lambda/\tau)$). We then have $\left\|\mathbf{U}_j - \mathbf{D}_j^\dagger\mathbf{U}_1\mathbf{C}_j^{-\dagger}\right\|_2 \leq \left(\frac{\lambda}{\delta}\right)^{2p} \cdot \tau$, and therefore can write $\mathbf{U}_j = \mathbf{D}_j^\dagger\mathbf{U}_1\mathbf{C}_j^{-\dagger} + \mathbf{Z}_j$, where $\|\mathbf{Z}_j\|_2 \leq \left(\frac{\lambda}{\delta}\right)^{2p} \cdot \tau$.

Thus,

$$\begin{aligned}
\left\|\mathbf{U}_j\mathbf{A}_{i,j,k}\mathbf{U}_k^\dagger - \mathbf{B}_{i,j,k}\right\|_2 &= \left\|(\mathbf{D}_j^\dagger\mathbf{U}_1\mathbf{C}_j^{-\dagger} + \mathbf{Z}_j)\mathbf{A}_{i,j,k}(\mathbf{C}_k^{-1}\mathbf{U}_1^\dagger\mathbf{D}_k + \mathbf{Z}_k^\dagger) - \mathbf{B}_{i,j,k}\right\|_2 \\
&\leq \left\|\mathbf{D}_j^\dagger\mathbf{U}_1\mathbf{C}_j^{-\dagger}\mathbf{A}_{i,j,k}\mathbf{C}_k^{-1}\mathbf{U}_1^\dagger\mathbf{D}_k - \mathbf{B}_{i,j,k}\right\|_2 + \left\|\mathbf{D}_j^\dagger\mathbf{U}_1\mathbf{C}_j^{-\dagger}\mathbf{A}_{i,j,k}\mathbf{Z}_k^\dagger)\right\|_2 \\
&+ \left\|\mathbf{Z}_j\mathbf{A}_{i,j,k}\mathbf{C}_k^{-1}\mathbf{U}_1^\dagger\mathbf{D}_k\right\|_2 + \left\|\mathbf{Z}_j\mathbf{A}_{i,j,k}\mathbf{Z}_k^\dagger\right\|_2 \\
&\leq \left\|\mathbf{D}_j^\dagger\right\|_2 \cdot \|\mathbf{D}_k\|_2 \cdot \left\|\mathbf{U}_1\mathbf{A}'_{i,j,k}\mathbf{U}_1^\dagger - \mathbf{D}_j^{-\dagger}\mathbf{B}_{i,j,k}\mathbf{D}_k^{-1}\right\|_2 + 2(\lambda/\delta)^{2p+1}\tau + (\lambda/\delta)^{4p+1}\tau^2 \\
&\leq (\lambda/\delta)^{2p} \cdot \left\|\mathbf{U}_1\mathbf{A}'_{i,j,k}\mathbf{U}_1^\dagger - \mathbf{B}'_{i,j,k}\right\|_2 + 2(\lambda/\delta)^{2p+1}\tau + (\lambda/\delta)^{4p+1}\tau^2 \\
&\leq 3(\lambda/\delta)^{2p+1}\tau + (\lambda/\delta)^{4p+1}\tau^2 \leq 4(\lambda/\delta)^{4p+1}\tau
\end{aligned}$$

Part 4 follows from the calculation below, where we use the facts $n = rp$, $p > 1$.

$$P((\overrightarrow{\mathbf{A}}, \overrightarrow{\mathbf{B}}, \varepsilon, \mathbf{r})) = mn^2 + n^3 = m(rp)^2 + (rp)^3 > (mp^2 + p)r^2 + r^3 = P((\overrightarrow{\mathbf{A}}', \overrightarrow{\mathbf{B}}', \varepsilon, r)).$$

□

### F.2.4 Proof of Lemma F.14

*Proof of Lemma F.14.* We can compute the eigenvalues of each $\mathbf{A}_i, \mathbf{B}_i$ (i.e., roots of their characteristic polynomials) in time $\mathsf{poly}(n, m\log(\lambda/\varepsilon))$. Suppose that there exists $\mathbf{U} \in \mathsf{U}(r)$ such that $\left\|\mathbf{U}\overrightarrow{\mathbf{A}}\mathbf{U}^\dagger - \overrightarrow{\mathbf{B}}\right\|_2 \leq \varepsilon$. Without loss of generality, suppose $\mathbf{A}_1$ has a pair of eigenvalues which are $r \geq \delta$ apart.

We can use Lemma F.5 to find unitary matrices $\mathbf{U}_L, \mathbf{U}_R$ and positive integers $r_1, r_2$ with $r_1 + r_2 = r$ such that there exist unitary matrices $\mathbf{U}_1, \mathbf{U}_2$ with dimensions $r_1, r_2$ and

$$\|\mathbf{U}_L\mathsf{diag}(\mathbf{U}_1, \mathbf{U}_2)\mathbf{U}_R - \mathbf{U}\|_2 \leq \frac{\varepsilon}{\delta - \varepsilon}.$$

Thus, the following holds for every $i \in [m]$:

$$\left\|\mathbf{U}_L\mathsf{diag}(\mathbf{U}_1, \mathbf{U}_2)\mathbf{U}_R\mathbf{A}_i\mathbf{U}_R^\dagger\mathsf{diag}(\mathbf{U}_1, \mathbf{U}_2)^\dagger\mathbf{U}_L^\dagger - \mathbf{B}_i\right\|_2 \leq \varepsilon + \|\mathbf{A}_i\|_2 \times \frac{2\varepsilon}{\delta - \varepsilon} \leq \left(1 + \frac{2\lambda}{\delta}\right)\varepsilon \ ,$$

and as a consequence,

$$\left\|\mathsf{diag}(\mathbf{U}_1, \mathbf{U}_2)\mathbf{U}_R\mathbf{A}_i\mathbf{U}_R^\dagger\mathsf{diag}(\mathbf{U}_1, \mathbf{U}_2)^\dagger - \mathbf{U}_L^\dagger\mathbf{B}_i\mathbf{U}_L\right\|_2 \le \left(1 + \frac{2\lambda}{\delta}\right)\varepsilon \ .$$

This means, we can replace $\mathbf{U}$ by $\mathbf{U}_1, \mathbf{U}_2$, set $\mathbf{r} = (r_1, r_2)$, and set $\mathbf{A}_i' = \mathbf{U}_R\mathbf{A}_i\mathbf{U}_R^\dagger$ and $\mathbf{B}_i' = \mathbf{U}_L^\dagger\mathbf{B}_i\mathbf{U}_L$.

With the change above, it is easy to see that the max spectral norm of the new blocks of the matrices $\mathbf{A}_i'$ and $\mathbf{B}_i'$ are still bounded by $\lambda$, since $\|\mathbf{U}_L\|_2, \|\mathbf{U}_R\|_2 = 1$.

We have proved that if $(\overrightarrow{\mathbf{A}}, \overrightarrow{\mathbf{B}}, \varepsilon, r)$ has a solution, then $(\overrightarrow{\mathbf{A}}', \overrightarrow{\mathbf{B}}', (1+2\lambda/\delta)\varepsilon, (r_1, r_2))$ also has a solution. Now, we need to show that any solution to $(\overrightarrow{\mathbf{A}}', \overrightarrow{\mathbf{B}}', \tau, \mathbf{s})$ can be converted into a solution to $(\overrightarrow{\mathbf{A}}, \overrightarrow{\mathbf{B}}, \tau, \mathbf{s})$.

To do this, suppose we get matrices $\mathsf{diag}(\mathbf{U}_1, \mathbf{U}_2)$ as a solution to $(\overrightarrow{\mathbf{A}}', \overrightarrow{\mathbf{B}}', \tau, (r_1, r_2))$. We can just let $\mathbf{U} = \mathbf{U}_L\mathsf{diag}(\mathbf{U}_1, \mathbf{U}_2)\mathbf{U}_R$. This completes the proof. □

### F.3 Final Algorithms

With lemmas introduced in Section F.2, we can now describe an algorithm to decompose an instance of Problem (F.2) into small "near identity" cases. It keeps applying Lemmas F.11, F.12, F.13 and F.14 until we are left with "near-identity" instances. If in any step we find out that the singular values of $\mathbf{A}_{i,j,k}$ are far away from $\mathbf{B}_{i,j,k}$, we return **Fail** to indicate that the problem has no solution.

**Algorithm 4** Dimension Reduction Algorithm

**Input:** $(\overrightarrow{\mathbf{A}}, \overrightarrow{\mathbf{B}}, \varepsilon, \mathbf{r})$ instance of Problem (F.2), where $\mathbf{r} = (r_1, \ldots, r_p)$, a positive spectral parameter $\delta \in [2\varepsilon, 1)$, and a bound $\lambda \ge 2$ on $\|\overrightarrow{\mathbf{A}}\|_2, \|\overrightarrow{\mathbf{B}}\|_2$.

**Output:** Either **Fail**, indicating there cannot be any solution to $(\overrightarrow{\mathbf{A}}, \overrightarrow{\mathbf{B}}, \varepsilon, \mathbf{r})$, or **Success** with a decomposition of $(\overrightarrow{\mathbf{A}}, \overrightarrow{\mathbf{B}}, \varepsilon, \mathbf{r})$ into "near identity" instances.

```
Σ_ijk, Σ'_ijk ← vectors of ordered singular values of A_ijk and B_ijk.
 if ‖Σ_ijk − Σ'_ijk‖_∞ > ε^{1/2} for any i, j, k then return Fail.
Define the spectral indicator graph G_δ = G_δ(A⃗, B⃗, ε, r) according to Definition F.10.
if G_δ is disconnected then
    Let S ⊂ [p] be a connected component and T = [p] \ S.
    Apply Algorithm 4 on (A⃗^S, B⃗^S, ε, r^S) and (A⃗^T, B⃗^T, ε, r^T) with same δ, using Lemma F.11.
    If either of them fails, then return Fail.
    Otherwise, return the union of subproblems generated from Algorithm 4.
else if ∃ some edge {i, k} of G_δ where A_ijk or A_ikj has singular value gap > δ then
    return Algorithm 4 applied on (A⃗', B⃗', (1+2λ/δ)ε, s) with the same δ, using Lemma F.12.
else if p > 1 then                                    ◇ we must have r_1 = r_2 = ··· = r_p = r
    return Algorithm 4 applied on (A⃗', B⃗', (3λ/δ)^{4p}·ε, r) with new δ ← δ^{6p}, using Lemma F.13.
                                                      ◇ see Remark F.15 for why we decrease δ to δ^{6p}
else if ∃ some A_i or B_i with eigenvalue gap > rδ then    ◇ we must have p = 1
    return Algorithm 4 applied on (A⃗', B⃗', (1 + 2λ/δ) · ε, (r_1, r_2)) with δ, using Lemma F.14.
end if
 return (A⃗, B⃗, ε, r) which is already a "near identity" instance.
```

**Remark F.15.** *We emphasize that whenever Lemma F.12 is applied, we decrease $\delta$ to $\delta^{6p}$ in the subproblem. This is a crucial step for our decomposition, in order to make sure that error does not blow up when we reconstruct a final solution from the leaf instances of this recursive tree.*

*Indeed, suppose we fix the value of $\delta > 0$ throughout the recursion. Then, whenever Lemma F.11 is applied and we obtain a solution to its subproblems with error $\tau$, this error blows up to $\Omega(\tau + \mathsf{poly}(n)\delta)$. If somewhere higher in this recursion tree, we have also applied Lemma F.12 once, this error will further blow up to $\Omega(\tau + \mathsf{poly}(n)\delta) \cdot 4(\lambda/\delta)^{4p+1} \geq \Omega(1)$, which is* independent *of $\tau$.*

*Instead, we decrease $\delta > 0$ sufficiently each time Lemma F.12 is applied, this issue will go away.*

We next provide Algorithm 5 for checking unitary equivalence. Algorithm 5 recursively applies Algorithm 4 to decompose the problem into "near identity" instances, and then piece together the solutions.

**Algorithm 5** Algorithm for checking unitary equivalence

**Input:** $\overrightarrow{\mathbf{A}}, \overrightarrow{\mathbf{B}}$ in $\mathsf{Mat}_n(\mathbb{C})^m$ where $\|\overrightarrow{\mathbf{A}}\|_2, \|\overrightarrow{\mathbf{B}}\|_2$ are upper bounded by $\lambda \geq 2$, and error parameters $0 < \varepsilon < \lambda^{-\mathsf{poly}(n,m)}$.

**Output:** either **No** or "**Yes** with $\mathbf{U}, \mathbf{V} \in \mathsf{U}_n(\mathbb{C})$ s.t. $\|\mathbf{U}\overrightarrow{\mathbf{A}}\mathbf{V} - \overrightarrow{\mathbf{B}}\| \leq \varepsilon' \stackrel{\text{def}}{=} 2^{8n}(6\lambda)^{\frac{1}{n^5}}\varepsilon^{\frac{1}{20mn^{10}}}$."

◇ *If $\Delta_U(\overrightarrow{\mathbf{A}}, \overrightarrow{\mathbf{B}}) \leq \varepsilon$, then will always output* **Yes**.

$\delta \leftarrow (6\lambda)^{\frac{1}{n^5}} \cdot \varepsilon^{\frac{1}{20mn^{10}}}$.

$\mathbf{A}'_i \leftarrow \begin{pmatrix} 0 & \mathbf{A}_i \\ 0 & 0 \end{pmatrix}$ and $\mathbf{B}'_i \leftarrow \begin{pmatrix} 0 & \mathbf{B}_i \\ 0 & 0 \end{pmatrix}$ so we have instance $(\overrightarrow{\mathbf{A}}', \overrightarrow{\mathbf{B}}', \varepsilon, (n,n))$ for Problem (F.2).

Recursively apply Algorithm 4 to reduce $(\overrightarrow{\mathbf{A}}', \overrightarrow{\mathbf{B}}', \varepsilon, (n,n))$ to subproblems, updating the parameter $\delta$ accordingly.

If at any point Algorithm 4 fails, then **return No**.

Otherwise, the leaves of the recursive tree of Algorithm 4 must consist of "near identity" instances. Solve each of them by Lemma F.6.

Once all subproblems at the leaves are solved, piece together the solutions to form a candidate solution to $(\overrightarrow{\mathbf{A}}', \overrightarrow{\mathbf{B}}', \varepsilon', (n,n))$. Denote the final solution by $\mathbf{U}' = \mathsf{diag}(\mathbf{U}, \mathbf{V})$.

If $\|\mathbf{U}'\overrightarrow{\mathbf{A}}'(\mathbf{U}')^\dagger - \overrightarrow{\mathbf{B}}'\|_2 \leq \varepsilon'$, we **return** "**Yes** with $\mathbf{U}, \mathbf{V}$;" otherwise, **return No**.

## F.4 Proof of Theorem M3

*Proof of Theorem M3.*

**Size of recursion tree.** We bound the number of calls to Algorithm 4, and the maximum size of any subproblem. Recall the total sum of the potential function $P((\overrightarrow{\mathbf{A}}, \overrightarrow{\mathbf{B}}, \varepsilon, \mathbf{r}))$ from Definition F.9 never increases when we perform reductions using Lemma F.11, F.12, F.13, and F.14. Thus,

- Algorithm 4 applies reductions at most $2(mn^2 + n^3)$ times;
- the number of leaves in the recursion tree of Algorithm 4 is at most $mn^2 + n^3$;
- the total number of matrices appeared in all reduction steps is at most $2(mn^2 + n^3)$.

**Bound on spectral parameter $\boldsymbol{\delta}$.** We first derive a lower bound on $\delta$, which is our spectral parameter that is updated in the recursion of Algorithm 4 (in particular, whenever Lemma F.13 is applied). Take any path in the recursion tree, going from the original problem to a leaf (i.e., a near-identity instance). Let $t$ be the number of times Lemma F.13 is applied on this path, and $\delta_1, \ldots, \delta_t$ be the values of the spectral parameter $\delta$ on this path after each Lemma F.13 is applied, and $\delta_0 = \delta$. For each $i \in [t]$, suppose the $i$-th time we apply Lemma F.13, we have $\mathbf{r} = (r_i, \ldots, r_i)$ and $p = p_i$. We have $\delta_i = \delta_{i-1}^{6p_i}$ for $i \in [t]$.

Since it is easy to verify that $r_i \geq p_{i+1}r_{i+1}$ for $i \in [t-1]$ and $n \geq p_1 r_1$, we have

$$\delta_t = \delta_0^{6^t p_1 \cdots p_t} \geq \delta^{n^5} \tag{F.5}$$

where the inequality follows from the fact that $t \leq \log_2 n$ (as $r_{i+1} \leq r_i/p_{i+1} \leq r_i/2$) and from the inequality $r_t \cdot \prod_{i=1}^{t} p_i \leq r_{t-1} \cdot \prod_{i=1}^{t-1} p_i \leq \cdots \leq r_1 p_1 \leq n$.

**Forward Error Propagation.** In the forward direction, we bound if there is a solution to $(\overrightarrow{\mathbf{A}}', \overrightarrow{\mathbf{B}}', \varepsilon, (n,n))$ at the root, then how this error parameter $\varepsilon$ increases when we perform reductions using Algorithm 4.

When Lemma F.11 is applied, $\varepsilon$ does not change. When Lemma F.12 or Lemma F.14 is applied, the error goes from $\varepsilon$ to $\leq (1 + 6\lambda/\delta^{n^5}) \cdot \varepsilon$. When Lemma F.13 is applied, the error goes from $\varepsilon$ to $\leq (3\lambda/\delta^{n^5})^{4p}\varepsilon$.

Overall, in each reduction $\varepsilon$ grows at most to $(6\lambda/\delta^{n^5})^{4n}\varepsilon$, but the depth of our recursion tree is at most $2(mn^2 + n^3)$. Therefore, at the leaves of the recursion tree, the error is at most

$$\leq (6\lambda/\delta^{n^5})^{8n(mn^2+n^3)} \cdot \varepsilon \leq \varepsilon^{1-1/n^2} \leq \varepsilon^{1/2} \ .$$

In sum, we have just shown that, if there is a solution to $(\overrightarrow{\mathbf{A}}', \overrightarrow{\mathbf{B}}', \varepsilon, (n,n))$ at the root, then Algorithm 4 never fails (thus Line 2 will not terminate the algorithm).

**Backward Error Propagation.** Suppose Algorithm 4 does not fail. We need to bound how error propagates when reconstructing the solution from the leaves back to the original problem at the root.

Recall whenever we apply Lemma F.12 or Lemma F.14, and whenever we get solutions to the subproblem with error $\tau$, we also have $\tau$ error for the original problem after reconstruction. Therefore, we only need to carefully bound how error propagates when Lemma F.11 or Lemma F.13 is applied.

Using the notation before, take any path in the recursion tree, going from the original problem to a leaf (i.e., a near-identity instance). Let $t$ be the number of times Lemma F.13 is applied on this path, and $\delta_1, \ldots, \delta_t$ be the values of the spectral parameter on this path after each Lemma F.13 is applied. Again, for each $i \in [t]$, suppose the $i$-th time we apply Lemma F.13, we have $\mathbf{r} = (r_i, \ldots, r_i)$ and $p = p_i$. We have $\delta_0 = \delta$ and $\delta_i = \delta_{i-1}^{6p_i}$ for $i \in [t]$.

Suppose for each $i \in [t]$, we have that when the $i$-th time Lemma F.13 is applied on this path, the error $\tau_i'$ of the subproblem propagates in error $\tau_i$ of the parent problem. We also assume that for each $i = 0, 1, \ldots, t$, between the $i$-th and $(i+1)$-st calls calls of Lemma F.13 on this path, the error $\tau_{i+1}$ propagates to $\tau_i'$ (and this must be due to Lemma F.11). Note that $\tau_{t+1}$ is the error from the leave (i.e., for solving the near-identity case), and $\tau_0'$ is the error for the root (i.e., for solving the original problem). We have

- $\tau_{t+1} \leq 11n^2\delta_t$.

  By the definition of leaf, we know that Line 16 of Algorithm 4 is reached. Therefore, all the matrices $\mathbf{A}_i$ and $\mathbf{B}_i$ must have (1) singular value gaps $\leq \delta_t$ and (2) eigenvalues being $\leq n\delta_t$ far apart. Lemma F.6 implies that we can solve this "near identity" problem to error $11n^2\delta_t$.

- $\tau_i' \leq 2^n \tau_{i+1} + 2^n(n+1)^4\delta_i$ for each $i = 0, 1, \ldots, t$.

  This is because, whenever Lemma F.11 is applied and some $\eta$-error is obtained for its subproblems $(\overrightarrow{\mathbf{A}}^S, \overrightarrow{\mathbf{B}}^S)$ and $(\overrightarrow{\mathbf{A}}^T, \overrightarrow{\mathbf{B}}^T)$, we can reconstruct a solution for the parent problem $(\overrightarrow{\mathbf{A}}, \overrightarrow{\mathbf{B}})$ with error at most $2\eta + (n+1)^3\delta_i$. There are at most $n$ applications of Lemma F.11 in total.

- $\tau_i \le 4(\lambda/\delta_{i-1})^{4p_i+1}\tau_i'$ for each $i = 1, 2, \dots, t$.

  This is a direct consequence of Lemma F.13.

Combining the second and third inequalities, we have for every $i \in [t]$,

$$\begin{aligned}
\tau_i &\le 4(\lambda/\delta_{i-1})^{4p_i+1} \cdot \big(2^n \tau_{i+1} + 2^n (n+1)^4 \delta_i\big) \\
&\le 4\lambda^{5n} 2^n \cdot \frac{1}{\delta_{i-1}^{4p_i+1}} \cdot \big(\tau_{i+1} + (n+1)^4 \delta_i\big) \\
&\overset{①}{=} 4\lambda^{5n} 2^n \cdot \frac{\delta_{i-1}^{6p_i}}{\delta_{i-1}^{4p_i+1}} \cdot \big(\frac{\tau_{i+1}}{\delta_i} + (n+1)^4\big) \\
&\le \big(4\lambda^{5n} 2^n \cdot \delta_{i-1}\big) \cdot \delta_{i-1} \cdot \big(\frac{\tau_{i+1}}{\delta_i} + (n+1)^4\big) \\
&\overset{②}{\le} \delta_{i-1} \cdot \big(\frac{\tau_{i+1}}{\delta_i} + (n+1)^4\big) \enspace .
\end{aligned}$$

Above, in equality ① we used $\delta_i = \delta_{i-1}^{6p_i}$, and in inequality ② we used $\lambda^{10n}\delta_{i-1} \le \lambda^{10n}\delta < 1$. Recursively applying this inequality $t$ times, we have

$$\tau_1 \le \frac{\delta_0}{\delta_t}\tau_{t+1} + (n+1)^5 \delta_0 \le 11n^2 \delta_0 + (n+1)^5 \delta_0 \le 3(n+1)^5 \delta_0$$

and accordingly

$$\tau_0' \le 2^n \tau_1 + 2^n (n+1)^4 \delta_0 \le 2^{n+2}(n+1)^5 \delta_0 = 2^{n+2}(n+1)^5 \delta \le 2^{8n}\delta = \varepsilon' \enspace .$$

To sum up, if $\Delta_U(\overrightarrow{\mathbf{A}}, \overrightarrow{\mathbf{B}}) \le \varepsilon$, Algorithm 5 outputs a solution which has unitary distance $< \varepsilon'$, and this guarantees that the algorithm outputs yes. This proves correctness.

**Numerical Error and Time Complexity.** Since we only need to distinguish between $\Delta_U(\overrightarrow{\mathbf{A}}, \overrightarrow{\mathbf{B}}) \le \varepsilon$ or $\Delta_U(\overrightarrow{\mathbf{A}}, \overrightarrow{\mathbf{B}}) > \varepsilon'$, we can assume the bit complexities of the matrices obtained throughout the algorithm are bounded by $\mathsf{poly}(n, m, \log(\lambda/\varepsilon))$. This is a consequence of the following facts:

- the spectral norms of all matrices obtained by reductions are no more than $\lambda$;
- the spectral parameter $\delta$ is always polynomially bounded, see (F.5);
- the error propagation factor is only $\exp(\mathsf{poly}(n, m, \log(\lambda/\varepsilon)))$.

Therefore, if we encounter any matrix with higher bit complexity, we can simply truncate its entries to have proper bit complexity. Such truncation does not affect the approximation of the algorithm. □